%% file: pwnpop.tex
\documentclass[longauth,traditabstract]{aa} 
\usepackage{natbib}
\usepackage{amssymb}
\usepackage{amsmath}
\usepackage{deluxetable}
\usepackage[colorlinks=false,breaklinks=true]{hyperref}
\bibpunct{(}{)}{;}{a}{}{,}    
\usepackage{graphicx}
\usepackage{pifont}
\usepackage{stmaryrd}

\newcommand{\eh}[1]{\,\mathrm{#1}}
\newcommand{\asurv}{\textsc{asurv}}
\newcommand{\cox}{the Cox proportional hazards model}
\newcommand{\emalg}{EM algorithm}

\newcommand{\tev}{\eh{TeV}}

\newcommand{\kyr}{\eh{kyr}}

\newcommand{\pc}{\eh{pc}}

\newcommand{\kpc}{\eh{kpc}}

\newcommand{\erg}{\eh{erg}}

\newcommand{\ergs}{\eh{erg\,s^{-1}}}

\newcommand{\kms}{\eh{km\, s^{-1}}}

\newcommand{\ergskpc}{\eh{erg\,s^{-1}\,kpc^{-2}}}

\newcommand{\ttt}[1]{\times10^{#1}}

\newcommand{\tin}[1]{_\mathrm{#1}}

\newcommand{\dg}{^{\circ}}

\newcommand{\pct}{\eh{\%}}

\newcommand{\cl}{confidence level}

\newcommand{\lgt}{\lg}

\newcommand{\mr}[1]{\mathrm{#1}}

\newcommand{\ind}[1]{_{\mr{#1}}}

\newcommand{\che}[1]{}

\renewcommand{\epsilon}{\varepsilon}

\newcommand{\hess}{H.E.S.S.}

\newcommand{\hgps}{H.E.S.S. Galactic Plane Survey}
\newcommand{\model}{\emph{baseline model}}
\newcommand{\scatter}{\emph{varied model}}

\newcommand{\fermi}{\emph{Fermi} Large Area Telescope}

\let\seriesfb\bfseries\def\bfseries{\boldmath\seriesfb}
\let\seriesdm\mdseries\def\mdseries{\unboldmath\seriesdm}

\newcommand{\velax}{{Vela\,X}}

\newcommand{\jxviii}{HESS\,J1825$\!-\!$137}
\newcommand{\lmc}{{N\,157B}}
\newcommand{\ctaone}{{CTA\,1}}
\newcommand{\threec}{3C\,58}

\newcommand{\edot}{\dot{E}}
\newcommand{\pdot}{\dot{P}}
\newcommand{\edotdsq}{\dot{E}/d^2}
\newcommand{\age}{\tau\tin{c}}
\newcommand{\lumi}{L_{1-10\,\mathrm{TeV}}}

\newcommand{\aref}[1]{Appendix~\ref{#1}}
\newcommand{\tref}[1]{Table~\ref{#1}}
\newcommand{\fref}[1]{Fig.~\ref{#1}}
\newcommand{\figref}[1]{Figure~\ref{#1}}        
\newcommand{\eref}[1]{Eq.~\ref{#1}}
\newcommand{\sref}[1]{Sect.~\ref{#1}}

\begin{document}


\title{The population of TeV pulsar wind nebulae in\\ the \hess\
Galactic Plane Survey}

\authorrunning{\hess\ Collaboration}
\titlerunning{Population of TeV pulsar wind nebulae}


%
%
%
%

\author{\tiny{ H.E.S.S. Collaboration
\and H.~Abdalla \inst{1}
\and A.~Abramowski \inst{2}
\and F.~Aharonian \inst{3,4,5}
\and F.~Ait Benkhali \inst{3}
\and A.G.~Akhperjanian$^\dagger$ \inst{6,5} 
\and T.~Andersson \inst{10}
\and E.O.~Ang\"uner \inst{7}
\and M.~Arrieta \inst{15}
\and P.~Aubert \inst{24}
% ...
}}

\institute{
Centre for Space Research, North-West University, Potchefstroom 2520, South Africa \and 
Universit\"at Hamburg, Institut f\"ur Experimentalphysik, Luruper Chaussee 149, D 22761 Hamburg, Germany \and 
Max-Planck-Institut f\"ur Kernphysik, P.O. Box 103980, D 69029 Heidelberg, Germany \and 
Dublin Institute for Advanced Studies, 31 Fitzwilliam Place, Dublin 2, Ireland \and 
National Academy of Sciences of the Republic of Armenia,  Marshall Baghramian Avenue, 24, 0019 Yerevan, Republic of Armenia  \and
Yerevan Physics Institute, 2 Alikhanian Brothers St., 375036 Yerevan, Armenia \and
Institut f\"ur Physik, Humboldt-Universit\"at zu Berlin, Newtonstr. 15, D 12489 Berlin, Germany \and
University of Namibia, Department of Physics, Private Bag 13301, Windhoek, Namibia \and
GRAPPA, Anton Pannekoek Institute for Astronomy, University of Amsterdam,  Science Park 904, 1098 XH Amsterdam, The Netherlands \and
Department of Physics and Electrical Engineering, Linnaeus University,  351 95 V\"axj\"o, Sweden \and
Institut f\"ur Theoretische Physik, Lehrstuhl IV: Weltraum und Astrophysik, Ruhr-Universit\"at Bochum, D 44780 Bochum, Germany \and
GRAPPA, Anton Pannekoek Institute for Astronomy and Institute of High-Energy Physics, University of Amsterdam,  Science Park 904, 1098 XH Amsterdam, The Netherlands \and
Institut f\"ur Astro- und Teilchenphysik, Leopold-Franzens-Universit\"at Innsbruck, A-6020 Innsbruck, Austria \and
School of Physical Sciences, University of Adelaide, Adelaide 5005, Australia \and
LUTH, Observatoire de Paris, PSL Research University, CNRS, Universit\'e Paris Diderot, 5 Place Jules Janssen, 92190 Meudon, France \and
Sorbonne Universit\'es, UPMC Universit\'e Paris 06, Universit\'e Paris Diderot, Sorbonne Paris Cit\'e, CNRS, Laboratoire de Physique Nucl\'eaire et de Hautes Energies (LPNHE), 4 place Jussieu, F-75252, Paris Cedex 5, France \and
Laboratoire Univers et Particules de Montpellier, Universit\'e Montpellier, CNRS/IN2P3,  CC 72, Place Eug\`ene Bataillon, F-34095 Montpellier Cedex 5, France \and
DSM/Irfu, CEA Saclay, F-91191 Gif-Sur-Yvette Cedex, France \and
Astronomical Observatory, The University of Warsaw, Al. Ujazdowskie 4, 00-478 Warsaw, Poland \and
Aix Marseille Universit\'e, CNRS/IN2P3, CPPM UMR 7346,  13288 Marseille, France \and
Instytut Fizyki J\c{a}drowej PAN, ul. Radzikowskiego 152, 31-342 Krak{\'o}w, Poland \and
Funded by EU FP7 Marie Curie, grant agreement No. PIEF-GA-2012-332350,  \and
School of Physics, University of the Witwatersrand, 1 Jan Smuts Avenue, Braamfontein, Johannesburg, 2050 South Africa \and
Laboratoire d'Annecy-le-Vieux de Physique des Particules, Universit\'{e} Savoie Mont-Blanc, CNRS/IN2P3, F-74941 Annecy-le-Vieux, France
% ...
}

\offprints{H.E.S.S.~collaboration,
\\\email{\href{mailto:contact.hess@hess-experiment.eu}{contact.hess@hess-experiment.eu}}
\\$^\textrm{\ding{72}}$Corresponding authors
\\$^\dagger$Deceased
\\$^{\ddagger}$Now at: Technische Universit\"at Kaiserslautern, Fachgebiet Bauphysik/Energetische Geb\"audeoptimierung, Paul-Ehrlich-Str., Geb\"aude 29, Raum 214, D-67663 Kaiserslautern
\\$^\S$Now at: Institute for Space Sciences (CSIC−IEEC), Campus UAB, Carrer de Can Magrans s/n, 08193 Barcelona, Spain
\\$^\parallel$Now at: Karlsruhe Institute of Technology, Institute for Nuclear Physics, P.O. Box 3640, D-76021 Karlsruhe, Germany}



\abstract{
The nine-year H.E.S.S. Galactic Plane Survey (HGPS) has yielded the most uniform observation scan of the
inner Milky Way in the TeV gamma-ray band to date. The sky maps
and source catalogue of the HGPS allow for a systematic
study of the population of TeV pulsar wind nebulae found throughout the
last decade. To investigate 
the nature and evolution of pulsar wind nebulae, for the first time we also present several upper limits for
regions around pulsars without a detected TeV wind nebula. 
Our data exhibit a correlation of TeV surface brightness with pulsar
spin-down power $\edot$. This seems to be caused both by an increase of extension
with decreasing $\edot$, and hence with time, compatible with a power law
$R\tin{PWN}(\edot) \sim \edot^{-0.65 \pm 0.20}$\che{2016/06/09}, and by a mild
decrease of TeV gamma-ray luminosity with decreasing $\edot$, compatible
with $\lumi \sim \edot^{0.59 \pm 0.21}$\che{2016/06/09}. We also find
that the offsets of pulsars
with respect to the wind nebula centre with ages around $10\kyr$ are
frequently larger than can be plausibly explained by pulsar proper motion and
could be due to an asymmetric environment. In the present data, it
seems that a large pulsar offset is correlated with a high apparent TeV
efficiency $\lumi/\edot$. 
In addition to 14\che{2016/06/09} HGPS sources considered firmly identified
pulsar wind nebulae
and 5\che{2016/06/09} additional pulsar wind nebulae taken from literature, we find
10\che{2016/06/09} HGPS sources that are likely TeV
pulsar wind nebula candidates.
Using a model that subsumes the present common understanding of
the very high-energy radiative evolution of pulsar wind nebulae, we find that
the trends and variations of the TeV
observables and limits can be reproduced to a good level,
drawing a consistent
picture of present-day TeV data and theory.
}


%



\maketitle


\onecolumn
\tableofcontents
\twocolumn
\newpage

\section{Introduction}\label{sec:intro}

Pulsar wind nebulae (PWNe) are clouds of magnetised
electron-positron plasma that
can span many parsecs and are observed via their synchrotron
or inverse Compton (IC) radiation \citep[see][for a comprehensive review on
the subject]{gs}. They are created inside supernova remnants (SNRs) by the energetic
outflow (``wind") of a pulsar, which is a swiftly rotating neutron star that is the compact leftover of the supernova explosion.
The pulsar wind runs into the supernova ejecta and develops a standing
shock wave beyond which the PWN builds up as an expanding bubble of diffuse
plasma. 
Pulsars can live for up to $10^{5-6}\kyr$, but their magnetic and particle
outflow is decreasing steadily. Therefore, most of the observed PWNe are associated
with pulsars that are less than a few $100\kyr$ old \citep{mcgill}.

It is instructive to consider the energetics of a typical PWN system. A pulsar
releases a total amount of energy of up to
$10^{49}$--$10^{50}\erg$ over its lifetime, but only $\lesssim 10\pct$ of this energy is
emitted as pulsed electromagnetic radiation \citep{fermi_pulsars}. Most of the
pulsar outflow consists of high-energy
particles and magnetic fields that feed into the growing PWN plasma. This plasma is dynamically inferior to the
$\sim10^{51}\erg$ carried away by the supernova blast wave around it. A good
portion
 of the PWN energy is radiated off, predominantly
through synchrotron emission in the first few thousand years, which can be observed in the
X-ray and radio bands. 
Besides that, a few percent of the
PWN energy are converted to IC radiation in the TeV regime. In
\citet{gould65}, but also in later works
\citep{okkie95, msh95, cmb_aharonian_95}, it was already suggested that this could allow 
for the detection of TeV emission. And even though the IC photons are an energetically subdominant emission
component, they carry important information that the synchrotron
emission, albeit much higher in flux and energy transport, does
not give access to; they emerge predominantly from homogeneous, time-constant CMB and IR photon seed fields, and
therefore trace the electron plasma independent of the time- and space-varying magnetic fields. 
In \citet{aharonian1997}, it was suggested that the TeV nebulae could be much
larger nebulae than those observed in
the radio or X-ray regimes. So in general, the IC image gives a more accurate and complete picture of the electron
population than the synchrotron photons.

Indeed, since the TeV detection of the Crab PWN in 1989 with the Whipple telescope
\citep{whipple_crab}, tens of Galactic sources have meanwhile been associated
with TeV pulsar wind nebulae.
Most of these objects are situated in the inner Galaxy; many were
therefore discovered and extensively
investigated from the southern hemisphere using the \hess\ Imaging Atmospheric Cherenkov Telescope (IACT) array 
\citep[e.g.][]{hess_j1825_edep}, which can
observe the inner Milky Way at low zenith angles and high sensitivity. The northern IACT systems
MAGIC \citep[e.g.][]{3c58_magic} and
VERITAS \citep[e.g.][]{veritas_cta1}, and arrays of air shower detectors
such as MILAGRO \citep{milagro_fermilist},
have also observed PWNe and contributed very valuable case
studies, mostly of systems evolving in the less dense outer Milky Way regions. Also HAWC shows promising potential to contribute new data soon \citep{hawc_icrc} but has not provided a
major data release yet. In the $1$--$10\tev$ regime, IACTs generally have a
better angular resolution and sensitivity than air shower arrays, even
though their fields of view (FOV) are limited to one or few objects, and their duty
cycle is restricted to dark, cloudless nights.

A systematic search with the \fermi\ for GeV pulsar wind nebulae in the vicinity of 
TeV-detected sources \citep{fermi_pwne} yielded 5 firmly identified
high-energy gamma-ray PWNe and 11 further candidates.
The PWN detections were
also often complemented by multi-wavelength observations in the X-ray 
or radio bands \citep[see e.g.][]{kargaltsev2013}.

In this paper, we proceed along the lines of previous work that aimed at a uniform
analysis of the whole population of TeV pulsar wind nebulae, such as
\citet{thesis_carrigan}, \citet{svenja_merida07}, \citet{thesis_marandon}, and
\citet{thesis_mayer}.
To do so, we take
advantage of the newly released TeV source catalogue
\citep{hgps_catalog_paper}\che{2016/06/09},
which is based on the nine-year \hess\ Galactic Plane Survey (HGPS). 
It provides a uniform
analysis of source sizes, positions, and spectra based on data taken during
nearly $3000\eh{h}$ of observations. It covers the Galactic plane at
longitudes $\ell=250\degr$ to $65\degr$ and latitudes $|b| \la 3.5^{\circ}$. We undertake a census of all the firmly
identified PWNe detected with \hess\ and other IACTs, and for the first time
complement this sample with HGPS flux upper limits for all covered pulsar locations
without a corresponding TeV detection. This allows for a less biased judgement of the
whole population. We compare the common
properties and trends of this population to those found in the numerous efforts
to theoretically describe the nature of pulsar wind nebulae.

\section{Observational data}\label{ec:data}

\subsection{HGPS and ATNF catalogues as data sources}\label{subsec:datasources}

We use two different sets of astronomical tables: the
\hess\ Galactic Plane Survey\footnote{\url{http://www.mpi-hd.mpg.de/HESS/hgps}} \citep[HGPS;][]{hgps_catalog_paper} and the ATNF pulsar
catalogue\footnote{\url{http://www.atnf.csiro.au/research/pulsar/psrcat}} \citep[version
1.54]{atnf}. For most purposes in this paper, the HGPS source catalogue and
the full ATNF listing are
used. Only the TeV-PSR spatial correlation study in
\sref{subsec:psr_pwn_correlations} makes use of less biased
listings, namely the HGPS components
list (HGPSC) and Parkes Multibeam Pulsar Survey
\citep[PMPS;][and references therein]{pmps, pmps_update}, which is a
subset\footnote{The difference between the two is that the
ATNF pulsar catalogue is a full listing of different surveys and targeted
observations, including, for instance, \emph{Fermi}-LAT detected gamma-ray
pulsars, whereas the PMPS is a comparably uniform survey of one particular
radio instrument and hence it is less prone to observational biases.} of the ATNF
pulsar catalogue. The HGPSC components list is an unbiased representation
of the TeV objects in terms of Gaussian components, which
does not invoke a priori knowledge of source associations or other prejudiced
assumptions. 

\input{table2d2}

For the pulsar distances, we choose the distance estimates of
\citet{cordes_lazio} provided by the ATNF team. Their uncertainty, however, is not very well defined and
can be as large as a factor of $2$. For the few cases in which pulsar
distance estimations were added or replaced from references other than the ATNF pulsar catalogue, these values are listed
in \tref{tab:atnf_exceptions}.

\begin{table*}
\centering
\caption{List of ATNF pulsar distance estimates that were modified.}
\label{tab:atnf_exceptions}
\tiny
\begin{tabular}{llllc}
\hline\hline
PSR & Distance & Method/Adjacent object & Reference \\
 & (kpc) & & \\
\hline
J0205+6449 (3C58)       & 2.0   & H{\sc i} absorption & \citet{3c58_dist_kothes} \\
J1023$-$5746 (Westerlund 2) & 8.0 & Westerlund 2 open cluster & \citet{westerlund_distance} \\
J1418$-$6058 (Rabbit) & 5.0 & Fiducial distance to Rabbit PWN & \citet{rabbit_distance} \\        
J1849$-$000             & 7.0 & Scutum arm tangent region & \citet{j18490_distance} \\
\hline                                   
\end{tabular}
\end{table*}

\subsection{Firmly identified TeV pulsar wind nebulae}

To determine which of the known TeV sources should
be considered as firmly identified PWNe, we use the identification criteria
discussed in the HGPS paper and take as a starting point the list of all
12 identified PWNe and the 8 identified composite SNRs \citep[][Table 
3]{hgps_catalog_paper}\che{2016/06/09}. Most PWNe in the HGPS are
identified by positional and/or morphological coincidence with a PWN identified in other
wavelengths, or by their specific (mostly energy dependent) TeV morphology.
Our selection for this paper also requires that the
corresponding pulsar has been detected and timed; if this is not the case, the properties
of the source cannot be put into the physics context of this
study, despite its identified PWN nature. This excludes the PWNe in SNRs G327.1$-$1.1 and G15.4+0.1,
and the identified composite SNRs CTB 37A and W41 (see
\citet{hgps_catalog_paper}, Table~3 and references therein). In composite SNRs,
the PWN component is mostly believed to outshine the potential
contribution from the SNR shell in TeV gamma-rays, and we
assume here that this is the case for TeV sources identified
as composite SNRs with the exception of HESS J1640$-$465. For
this object, detailed observations with H.E.S.S. suggest
that a significant part of the TeV emission may originate from
the SNR shell \citep{j1640}.
Therefore, we exclude HESS J1640$-$465 from  firm identification
and consider it a PWN candidate.
The sample we arrive at is listed
in \tref{tab:rePAPd2}.

In addition to the firmly identified objects found in the HGPS, we
include five\che{2016/06/09} HGPS-external PWNe, among them G0.9+0.1, which is
inside the plane scan, but was not re-analysed with the HGPS pipeline. These
PWNe are displayed using distinct symbols in the figures throughout this work. This latter group, listed in \tref{tab:rePORGd2}, is based both on  dedicated \hess\ observations outside of the scope of the HGPS and on data from other IACTs.

\input{table1d2}

We do not include detections that are only reported from
(direct) air shower detectors, such as MILAGRO, HAWC, or ARGO-YBJ, because their angular and spectral uncertainties are much higher, making the
source resolution and pulsar association more difficult and the spectral
statements more uncertain.

\subsection{Data extracted from the HGPS} \label{subsec:hgps_data_extraction}

The quantities taken from the HGPS catalogue are source position, extension,
integral flux $>1\tev$, and spectral index $\Gamma$ from the power-law fit
of the differential photon flux $\phi_0 \times (E/E_0)^{-\Gamma}$.
The extension measure $\sigma$ is given as the standard deviation of a
circular Gaussian function. Extension upper limits were used as provided in
the catalogue, namely in cases where the extension is not more than two
standard deviations larger than the systematic minimum extension of $0.03\dg$.

Offsets between pulsar and PWN centroid position were calculated and, where necessary, converted to
$2\sigma$ limits following a similar prescription, namely in the cases where the
offset was less than $3\sigma$ above a systematic minimum of $0.0056\dg$, which
is a typical value for the systematic positional uncertainty of \hess 

The integral photon
flux $I_{>1\tev}$ and index $\Gamma$ is converted to a
luminosity between $1$ and $10\tev$ using
\begin{equation}\label{eq:lumi}\begin{split}
\lumi & =  1.92\ttt{44}\frac{I_{>1\tev}}{\mr{cm^{-2}s^{-1}}}\, \\
      &  \times \frac{\Gamma-1}{\Gamma-2} \, (1 - 10^{2-\Gamma})
\left(\frac{d}{\mr{kpc}}\right)^2\eh{erg\,s^{-1}},
\end{split}\end{equation}
where $d$ is the source distance and the integral flux $I_{>1\tev}$ is taken
from the \verb=Flux_Map= column of the catalogue, which is recommended there as
the most reliable estimate of the integral flux.
The errors, propagated from the index errors $\sigma_\Gamma$ and integral flux
errors
$\sigma_I$, are
\begin{equation}\label{eq:lumi_err}\begin{split}
\left(\frac{\sigma_{L}}{\lumi}\right)^2 & = \left[\frac{\sigma_I}{I_{>1\tev}}\right]^2  \\
+ & \left[\left(\frac{1}{(\Gamma-1)(\Gamma-2)} + \frac{\ln{10}}{1-10^{\Gamma-2}}\right)
\sigma_\Gamma\right]^2.
\end{split}\end{equation}

The errors on
flux and index are assumed to be independent because the reference energy of
$1\tev$ is typically very close to the mean pivot energy of the fits. The
errors on the distance estimation of pulsars are not available consistently and are likely not Gaussian in most cases, so they are not treated here and remain a
systematic uncertainty. For uniformity, the power-law integration is also used in the few cases where a
high-energy cut-off is found to be significant, as
the cut-off has very little influence on the
integral\footnote{\velax\ is the only
source where this prescription leads to a significant
deviation from previously published dedicated analyses, both
because of its energy cut-off and its extended emission component up to
$1.2\dg$ away from its centre \citep{hess_velax_2012}. Therefore, we convert its $I_{>1\tev}$ to an
energy flux using its cut-off spectrum ($\Gamma=1.35 \pm 0.08$;
$\lambda=0.0815 \pm 0.0115$ for a flux function $F(E) \sim E^{-\Gamma}
\exp(-\lambda E)$), which leads to a $17\pct$ higher
energy flux than when only using the power-law approximation. Furthermore, the
extended and faint ``ring'' emission component noted in
\citet{hess_velax_2012} is taken into account by applying a correction factor
of $1.31\pm0.16$. This emission component is derived from the ratio of ``Inner''
and ``Total'' integral fluxes presented in \citet{hess_velax_2012}, Table~3.}.

We also extract flux upper limits from the sky maps of the HGPS data release.
The $95\pct$ \cl\ limit $I\tin{>1\,TeV}$ on the flux is converted as above, assuming a spectral
index of $\Gamma=2.3$, which is the typical TeV index also used in several
pipeline analysis steps of the HGPS analysis
\citep{hgps_catalog_paper}.
The flux limits are available for integration radii of $0.1\dg$, $0.2\dg$, and
$0.4\dg$; the latter of which is only available internally and will not be
part of the public HGPS data release. For pulsars that qualify for an upper limit, we use the \model\
(\aref{app:modelling}) to
estimate the PWN extension. Assuming $1000\eh{km\,s^{-1}}$
for the offset speed (see \sref{subsubsec:offset}), a required flux limit radius
$R\tin{lim}=R\tin{PWN}+R\tin{off}$ is derived and a corresponding angular
extent $\theta_{\mathrm{pred}}$ as seen from Earth is calculated. If this extension is
below $0.4\dg$, the value is rounded up to the next available correlation
radius and a flux limit is looked up in the respective limit map. In the
case of $0.4\dg<\theta_{\mathrm{pred}}<0.6\dg$, we assume that the source
could have been detected, and calculate a limit from the $0.4\dg$ map, scaling
it up by $(\theta_{\mathrm{pred}}/0.4\dg)^2$ to account for the uncontained
part of the PWN. If $\theta_{\mathrm{pred}}>0.6\dg$, no limit is calculated since
one cannot exclude that a potential weak and undetected PWN emission would have been confused with
background in the background subtraction of the HGPS pipeline.

\subsection{Caveats of the HGPS} \label{subsec:caveats}

The HGPS data contain unbiased observations, a priori targeted observations, and re-observations of hotspots.
It is therefore impossible to raise truly objective and
statistically robust statements on chance coincidence detections of TeV
objects near energetic pulsars. A way to unbias the
data would be to remove all deep and targeted observations from the catalogue construction  pipeline, which would obviously discard very interesting parts of
the data set and lead to a different catalogue content. We refrain from this exercise
here, trying to make use of the richness that is present in the full
data set and catalogue. 

A uniform source analysis, as provided in the HGPS and exploited here,
has many advantages with regard to a population study. The fluxes and
extensions are determined with one software version, data quality cut, analysis
algorithm, and event selection cut set, leading to values that are comparable and consistently
defined among all sources.
The disadvantage of uniformity is that it comes with a lack of adjustment. Customised data
quality cuts can allow for the detection of weaker sources or for lower systematic uncertainties for very strong sources.
This is deliberately not done here.

Besides this, the energy threshold and sensitivity of Cherenkov telescopes vary with the zenith angle of observation, and
therefore with the declination of a given sky region. The IACT data thus are
intrinsically not completely uniform across different sky regions.

\section{Correlation of TeV sources and pulsars} \label{sec:psr_pwn_correlations}

The total energy output of a pulsar at a given time is characterised by its
spin-down power $\edot$, which can be observationally determined from its
period $P$ and period derivative $\pdot$, assuming a
neutron star moment of inertia of $I = 10^{45}\eh{g\, cm^2}$ (see also
\aref{app:formulae_derivation} for the basic formulae of pulsar evolution). Pulsars deploy most of their
spin-down energy within few tens of kiloyears. The
pulsar wind nebulae thereby created are loss-dominated ever after that period, when
the electrons are diffused and lose their energy through radiative or
adiabatic cooling with cooling times of $\mathcal{O}(10\kyr)$ (see
\sref{subsec:model-cooling}). Therefore, the natural expectation
for a bright PWN is that it has to have an accordingly young
($\mathcal{O}(<10\kyr)$) and still energetic pulsar nearby.

Observationally, this
is supported by the fact that most TeV pulsar wind nebulae (and sources in general) are found
at Galactic latitudes $<0.5\dg$; if pulsars were to grow TeV nebulae in
their late stage of evolution, then TeV
sources should also be more numerous at higher latitudes, where many old
pulsars drift off to.

\subsection{Spatial correlation}\label{subsec:psr_pwn_correlations}

A way to find general support for the association of energetic pulsars
and TeV sources was explored by \citet{svenja_merida07}, where the whole HGPS
sky map of that time was used
along with the PMPS pulsar catalogue to evaluate a detection fraction
$N\tin{detected}/N\tin{pulsars}$ for pulsars in different bands in $\edotdsq$.

\begin{figure*}
\includegraphics[width=0.494\textwidth]{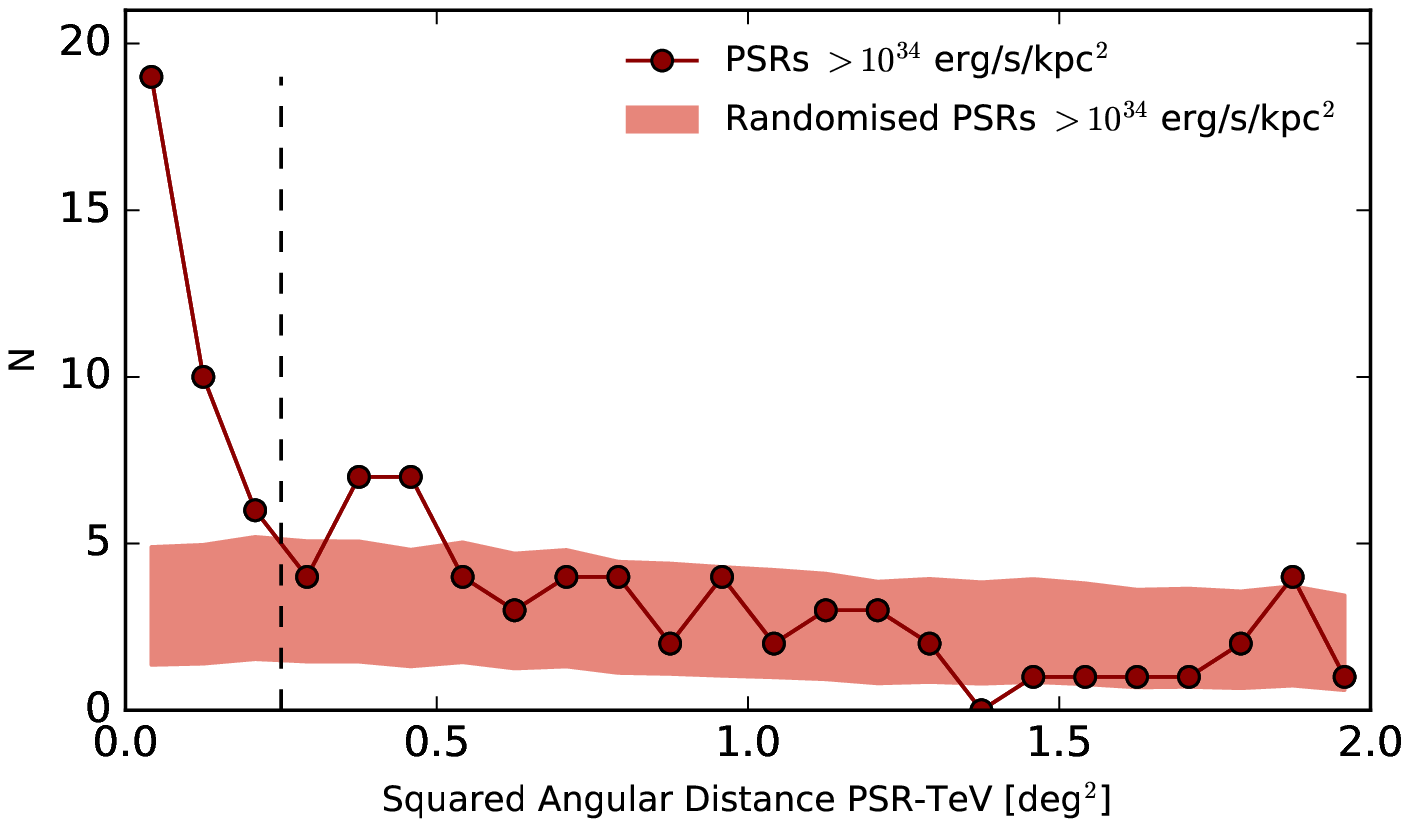}
\includegraphics[width=0.494\textwidth]{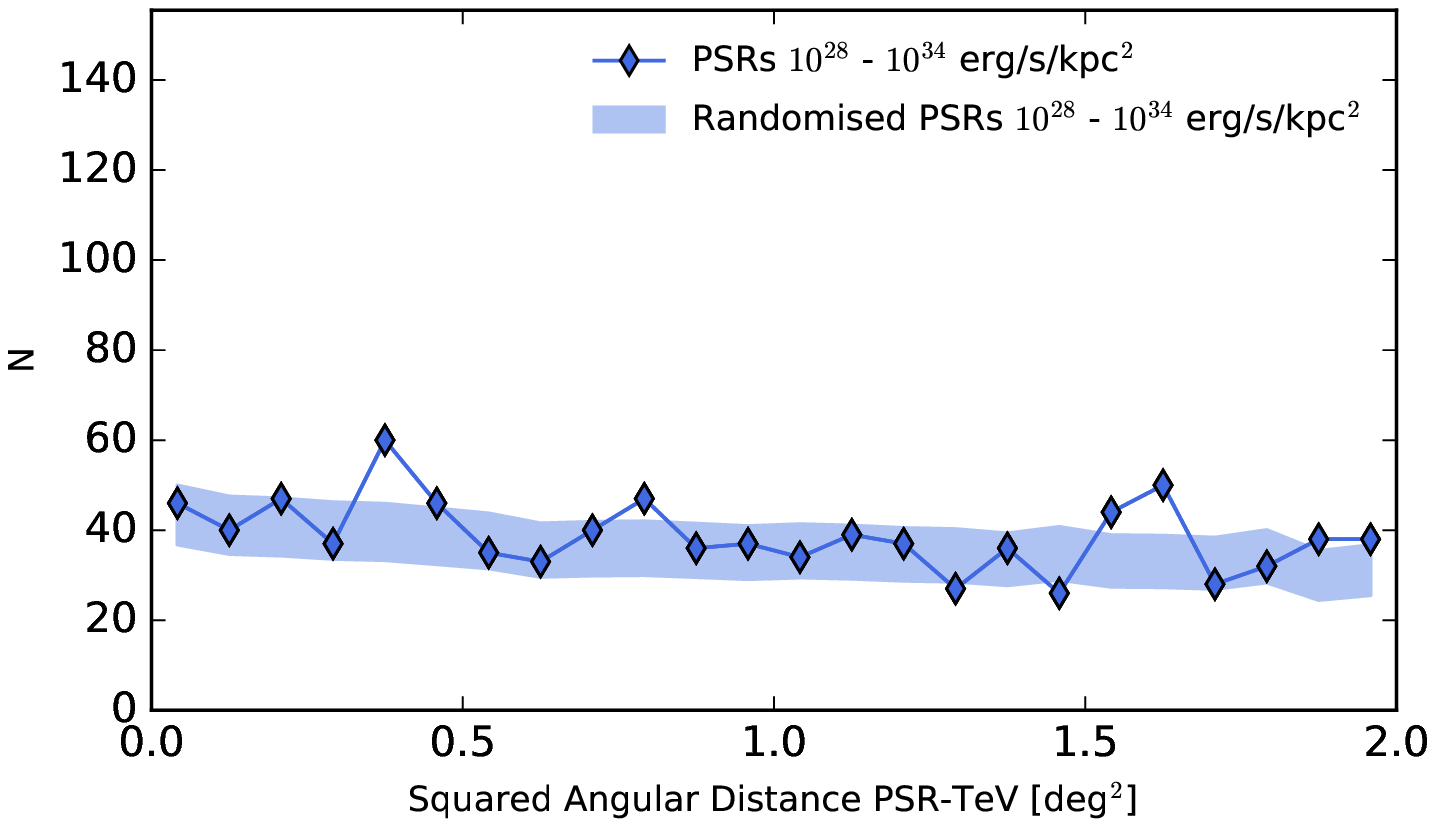}
\caption{
Histograms of spatial separation between PMPS pulsars and TeV source components from the HGPSC list. 
In the high-$\edot$ pulsar sample (left), a clear correlation
is seen as a peak at small squared angular distances,
whereas the low-$\edot$ associations, if present, are not significant beyond
the expected rate of chance coincidences (right). The angular separation cut
of $\theta < 0.5\dg$\che{2016/06/09} applied in the preselection of PWN candidates (\sref{subsec:candidate_selection}) is indicated by a dashed vertical line in the left panel.
\label{fig:theta2}}
\end{figure*}

To investigate whether this spatial correlation is still manifest in the data,
\fref{fig:theta2} shows the distribution of angular distances between all pulsars of a given
range in $\edotdsq$ and all ``Gaussian components'' listed in the unbiased HGPSC component
list\footnote{We use $\edotdsq$ as an estimator for
detectability for consistency with previous works.
This is optimal under the assumptions that (a) the TeV luminosity
scales linearly with $\edot$, and (b) the sources appear small compared to the correlation
radius. Both of these assumptions are questionable, given the large extension of
some objects and the weak correlation between $\edot$ and TeV luminosity. For this
reason, we cross-checked the study with just $\edot$ as the estimator, and we find
very similar results. Presumably, the fact that $d$ only varies by a factor of
$10$ throughout the
population makes the
distance correction a subdominant effect against intrinsic luminosity
variations.}.
The shaded band shows the
expectation derived from simulated pulsar samples. It is derived for the same
band of $\edotdsq$, calculating
30000 randomisations of the PMPS pulsar sample. The observed Galactic latitude
and longitude distributions of the pulsars are preserved in the reshuffling. A significant
correlation beyond chance coincidences is found for pulsars with
$\edotdsq>10^{34}\ergskpc$
and is absent for less energetic pulsars. An estimate for the number of chance coincidences for
a cut of $0.5\dg$\che{2016/06/09} yields a value of 
$9.7$\che{2016/06/09}, while 35\che{2016/06/09} HGPSC components are actually
found. Using the full ATNF catalogue instead of PMPS and the
HGPS source
catalogue instead of the components list, the study is more similar to the
source selection
we do in the following, but involves statistically
less unbiased samples. The estimated number of chance coincidences derived in this case
is $11.5$. 

\subsection{Pulsar wind nebulae preselection candidates and flux limits} \label{subsec:candidate_selection}

The strategy employed to select and evaluate unconfirmed PWN candidates in
this paper
is a two-step procedure: First, a loose \textit{preselection} of candidates
has been carried out.
Secondly, these candidates are distinctly marked in the various observables correlation plots
of \sref{sec:pwn_properties}, leading to a subsequent judgement on their physical
plausibility to be a PWN in the \textit{post-selection} of \sref{sec:candidate_rating}.

The criteria we impose for the preselection are that a pulsar should be more
energetic than $\edotdsq=10^{34}\ergskpc$\che{2016/06/09} and have an angular separation $\theta$ from an
HGPS source of less than $0.5\dg$\che{2016/06/09}. We also require a
characteristic age $\age <10^7\eh{yr}$ to prevent millisecond pulsars,
which are different concerning their nature and physics of emission, from entering
the PWN candidate sample\footnote{There is only one case of such a
coincidence, PSR~J1832$-$0836, which correlates with HESS~J1832$-$085 along with the much more likely ordinary
 PSR~B1830$-$08.}.
While these criteria are arbitrary to some
extent, we note that, as a preselection, they were chosen to be relatively loose and
amply include all firmly identified PWNe. 

Energetic pulsars that do not have an HGPS source
nearby or that coincide with an HGPS source that is already firmly associated to
another pulsar are selected for the calculation of a flux upper limit. In the
latter case, the flux of the established source is not subtracted,
since one cannot isolate one from the other and the conservative flux limit
is therefore on top of the emission of the main source.
In the limit calculation step, we include all pulsars with
$\edot>10^{35}\ergs$\che{2016/06/09}, independent of
their distance. For very old and extended objects, a large distance can even be favourable because
only then can their full supposed extent be covered within the \hess\ FOV,
leading to a meaningful flux limit (see also
\sref{subsec:hgps_data_extraction}).

For the same reason as in the selection of firmly identified PWNe, we deliberately choose not to  treat
pulsar systems in which the pulsar is not clearly identified in terms of
period, derivative (presumably because the pulsar beam does not intersect
Earth), and distance. We require a known pulsar distance so as to be able to
quantify
TeV properties, such as luminosity and extension, and compare
them with the firmly identified population.  But we should note
that this implies that we cannot consider among PWN candidates the TeV sources coincident with PSR~J1459$-$6053, PSR~J1813$-$1246 and
PSR~J1826$-$1256 \citep[see][]{hgps_catalog_paper}, 
which are pulsars that are detected in high-energy gamma-rays but not in
the radio domain. 

As a caveat of our cut in $\edotdsq$, we note that potential ancient nebulae
from very old pulsars cannot make it into our selection and are not be considered in this work
(except for being included in terms of a flux limit). Figure \ref{fig:theta2}
(right) shows that the TeV detection of such ancient nebulae has to be treated as
hypothetical, judging from the global catalogue
point of view we adopt in this paper. 

The result of the preselection is that besides the 14\che{2016/06/09} firmly identified
PWNe we consider here, 18\che{2016/06/09} additional PWN
candidates pass the criteria; two\che{2016/06/09} of
these additional candidates have two pulsars they
could be associated with and four pulsars have two possible TeV
counterparts\che{2016/06/14}. The 5\che{2016/06/09} HGPS-external PWNe also match the
criteria. We exclude the
$\gamma$-ray binary
PSR B1259$-$63 here. While the TeV source is believed to contain the wind
nebula of its pulsar,
the TeV
emission is clearly impacted by the binary nature of the object and therefore
out of the scope of this paper. Also, the obvious TeV shells that were
omitted from the standard HGPS pipeline are excluded here, although
coincident
pulsars are allowed to be included in the limits listing if they qualify. 

Among the pulsars without a matching detected TeV source, 65\che{2016/06/09} with
$\edot>10^{35}\ergs$ are selected for the limit calculation; however the assumed PWN extension and offset are small enough to calculate a flux limit with
the HGPS maps for only 22 of those. Of these limits, 3\che{2016/06/09} appear to be on top of
significant emission for various reasons: PSR~J1837$-$0604 coincides with the PWN
HESS~J1837$-$069. The limit of PSR~J1815$-$1738 is
integrated over $0.4\dg$ and therefore contains parts of the emission of HESS
J1813$-$178. PSR~J1841$-$0524 is situated within the very large HESS~J1841$-$055,
possibly consisting of multiple sources; the $\edotdsq$ of this object is too low for it to qualify as a candidate.

\input{table3d2}

\input{table4d2}

\begin{figure*}
\centering
\includegraphics[width=0.494\textwidth]{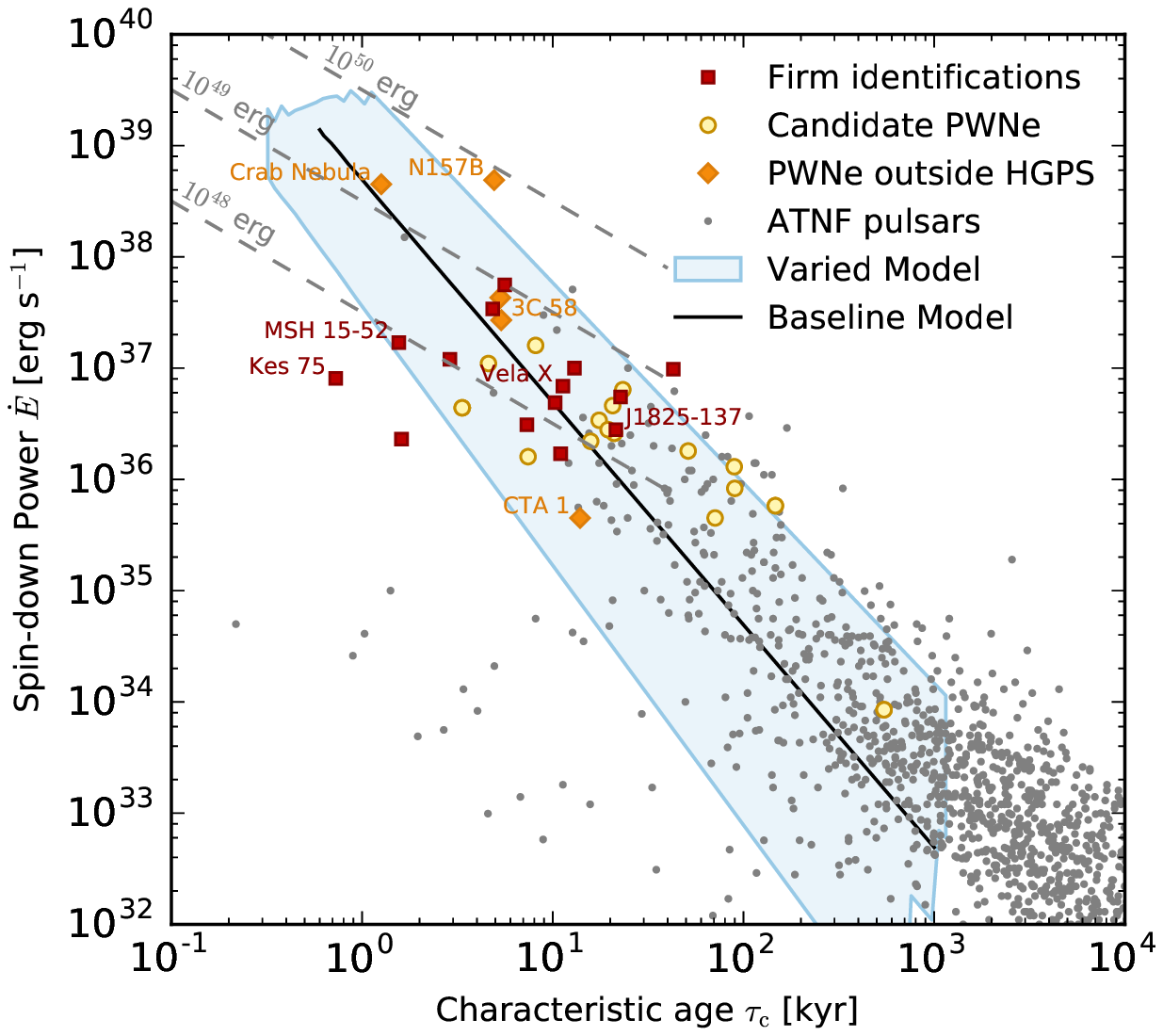}\includegraphics[width=0.494\textwidth]{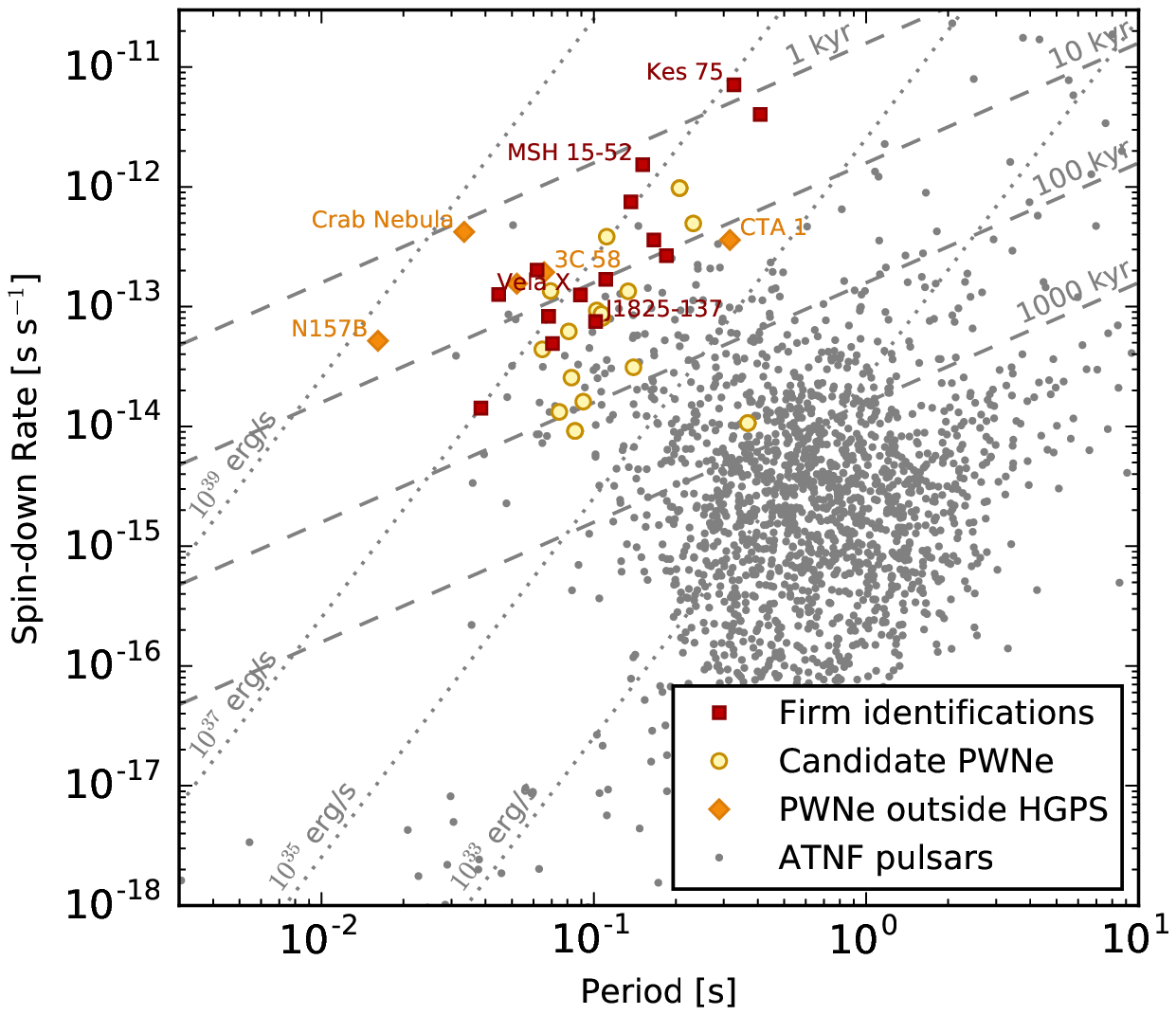}
\caption{
Left: Spin-down power $\edot$ and characteristic age $\age$ of pulsars with a firmly
identified PWN, candidate PWN, and without TeV counterpart (grey dots). The
black line and shaded band show the injection evolution of the
modelling used in this paper. The dashed lines indicate lines of constant
total remaining energy $\edot\tau$; see \aref{app:formulae_derivation}. Hence
a model curve that starts at $\edot_0\tau_0=10^{49}\erg$ represents a pulsar
with total initial rotational energy of $10^{49}\erg$. Since both $\edot$ and $\age$ depend on $P$
and $\pdot$, the axes in this plot do not represent independent quantities. Right: The same data, shown in the commonly used view, using the independently measured $P$ and
$\pdot$.
\label{fig:edot_age}}
\end{figure*}

The pulsars selected as firm PWNe from the HGPS catalogue, as external PWNe, candidate PWNe,
and for flux limits are listed in Tables
\ref{tab:rePAPd2}, \ref{tab:rePORGd2}, \ref{tab:CAPd2}, and \ref{tab:withLIMd2}, respectively.
They are shown in the $\edot$--
$\age$ and $\pdot$--$P$ planes in \fref{fig:edot_age}. The plots also show
 ATNF pulsars without detected TeV wind nebula for comparison and highlight some prominent
or special objects with labels. These are labeled throughout the paper
for orientation.

As expected, the preselection candidates are
young, but on average somewhat older than the already established PWNe.
This is likely because only young wind nebulae have a detectable extended X-ray counterpart,
which allows for a firm identification. Most of the candidates have previously
been hypothesised to be a PWN or to have a PWN component.
The only substantially older pulsar is PSR~B1742$-$30\che{2016/06/09}, which is selected
thanks to its very low distance despite its low $\edot$. We
cannot display this pulsar in all plots of this paper, but we discuss it as a special
case in
\sref{sec:candidate_rating}.

\subsection{Location in the Galaxy}\label{sec:galaxy} 

\begin{figure*}
\centering
\includegraphics[width=0.8\textwidth]{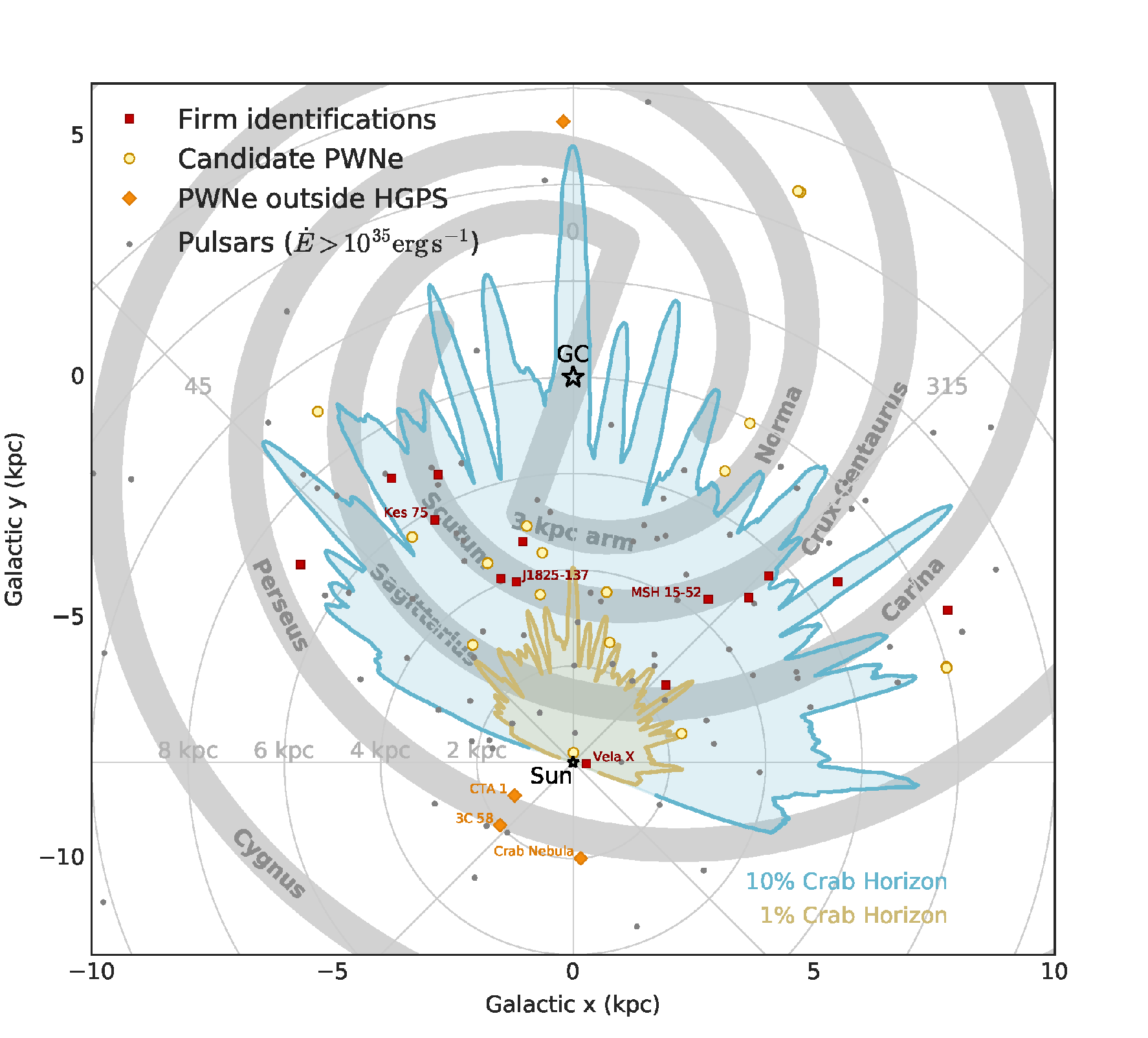}
\caption{
Schematic of the Milky Way and its spiral arms, along with firmly identified
PWNe, candidates, and energetic pulsars ($\edot > 10^{35}\ergs$) without
detected TeV wind nebula.
The yellow and blue curves outline the sensitivity horizon of the HGPS for 
point-like sources with an integrated gamma-ray luminosity ($1$--$10\tev$) 
of 1\% and 10\% of the Crab luminosity, respectively (see \citet{hgps_catalog_paper} for details).
\label{fig:galaxy-topview}}
\end{figure*}

In order to assess the reach of the population study presented in this work 
it is instructive to display the positions of Galactic PWNe together with the
sensitivity (or depth) of the \hgps. 
The map in \fref{fig:galaxy-topview} visualises the 2D
projection of the Galactic distribution of very energetic pulsars 
($\edot > 10^{35}\ergs$)\che{2016/06/09}.
The symbols distinguish between pulsars with firmly identified wind nebulae,
candidate PWNe, and pulsars at $>$\,$10^{35}\ergs$
for which no TeV wind nebula has been detected so far.
For reference, the map comprises a schematic representation of the
spiral arms of the Milky Way according to the parametrisation of \citet{vallee_2008}. 
The overlaid blue and yellow curves define the accessible range of the HGPS 
for point-like sources at an integrated luminosity
($1$--$10\tev$) of 1\% and 10\% of the Crab luminosity, respectively
\citep[for details see][]{hgps_catalog_paper}. 

For sources of $10\pct$ Crab luminosity, the HGPS covers approximately one quarter of our Galaxy, and
generally does not reach much farther from Earth than the distance to the Galactic
centre. For extended objects, the horizon can be expected to be closer,
 and for close-by extended sources, the \hess\ FOV
can limit the capability of isolating them from the background.

Most of the detected PWNe are located close to one of the nearby dense spiral
arm structures, where pulsars are expected to be born. In
particular, the Crux Scutum arm hosts half of all HGPS pulsar wind nebulae. Several
high-$\edot$ pulsars are on closer spiral arms but are not detected.

\begin{figure}
\centering
\includegraphics[width=0.494\textwidth]{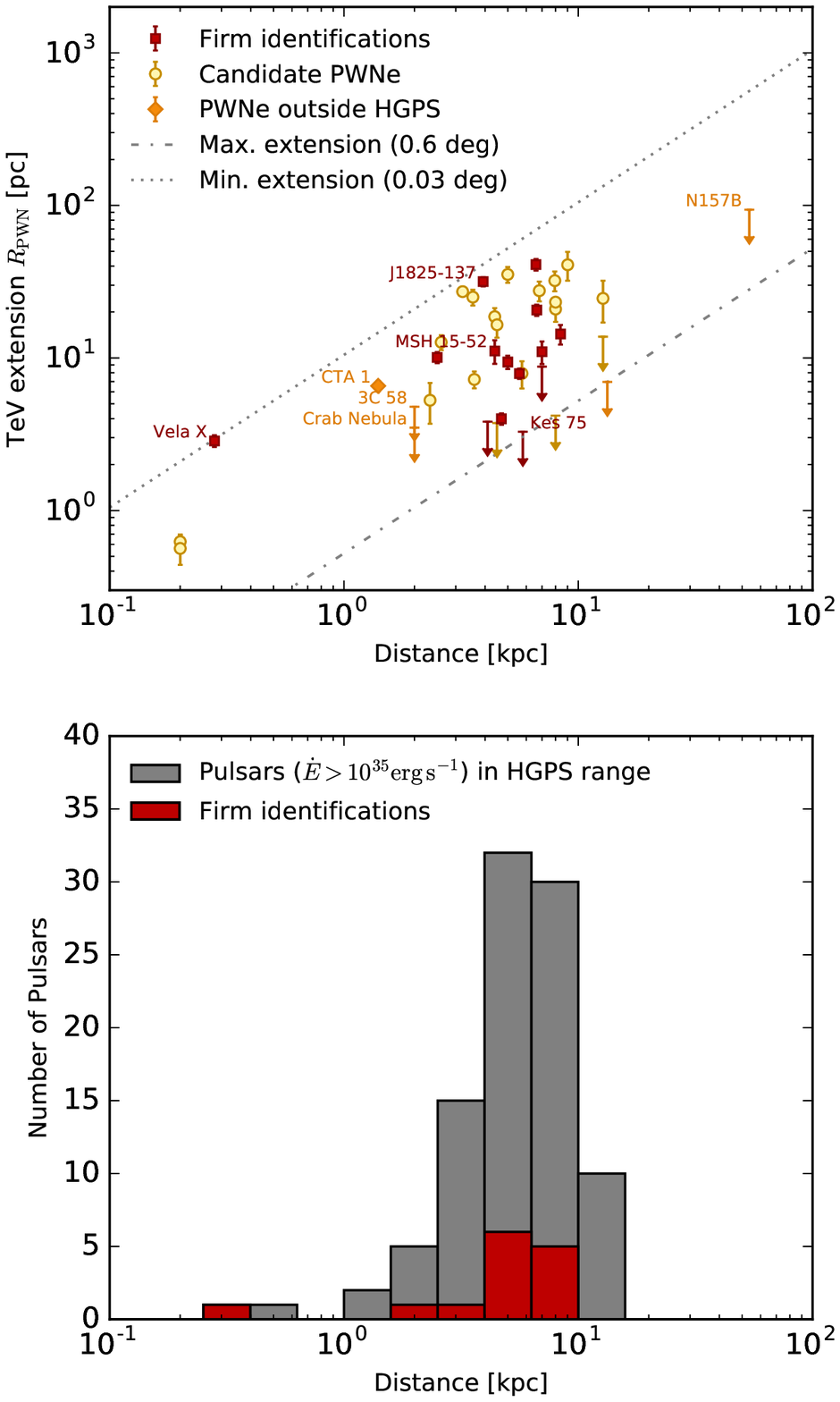}
\caption{
Top: PWN extension occurrences over distance from Earth, in comparison to the
band of extensions that can be expected to be identified in the HGPS analysis
chain. Bottom: Distribution of known distances of energetic pulsars ($\edot >
10^{35}\ergs$)\che{2016/06/09}.
\label{fig:extension_dist}}
\end{figure}

A way to look at the sensitivity to extended PWNe is shown in the upper part of \fref{fig:extension_dist}, where
the extension is plotted against distance from Earth. To guide the eye, two
lines indicate the range of detected extensions between the systematic
minimum of about $0.03\dg$ and the maximum extension in HGPS of $\sim0.6\dg$
(\velax, see
\sref{subsubsec:extension}). 
As can be seen in the lower panel of \fref{fig:extension_dist}, most PWNe
are detected around $5.1\kpc$, which is the average distance of PWNe in
Table \ref{tab:rePAPd2}.
This allows for the determination of radii between $3$ and at
least $60\pc$.

We conclude that both the \hess\ FOV ($5\dg$) and angular resolution
($0.03\dg$) are
adequate to study the wind nebulae of most of the high-$\edot$ pulsars known today.

\section{Theoretical notion of pulsar wind nebulae} \label{sec:pwn_theory}

Before discussing the properties of the PWNe and PWN candidates 
we found, this section recapitulates some concepts of the theoretical
understanding of pulsar wind nebulae.

A PWN is usually considered to be a calorimetrical, dynamical object around a
pulsar. It stores and displays the radiative output of the pulsar during tens of kiloyears
while at the same time undergoing a substantial dynamical
evolution inside the host SNR. Expressed in terms of a diffusion equation,
this means that it is energised by the magnetic and particle
flux from the pulsar, and cooled by radiative (synchrotron emission and IC
scattering), adiabatic, and escape losses \citep[e.g.][and references
therein]{crab_martin,zhang_2008}. In the context of this work, 
acceleration and injection mechanisms are not considered in detail. Pulsars
are regarded as particle-dominated, diffuse injectors of
electrons. Here and in the following, the term ``electrons" always
refers to the full electron and positron outflow.

\subsection{Injection evolution} \label{subsec:injection}

The energy outflow of the pulsar, $\edot$, which determines the energy injection
history of a PWN. This energy outflow is decaying continually at a rate determined by the
so-called spin-down timescale $\tau$, following an evolution similar to that
expected from a dipole (see also \aref{app:formulae_derivation})
\begin{equation}\label{eq:edot}
\edot(t) = \edot_0\left(1+\frac{t}{\tau_0}\right)^{-\frac{n+1}{n-1}}
,\end{equation}
where $\tau_0$ is the initial spin-down timescale, $\edot_0$ is the initial
spin-down
luminosity, $n$ is the so-called ``braking index" \citep[e.g.][]{pacini_salvati},
and $t$ is the time since the birth of the pulsar. Values typically considered are
$\tau_0\sim 10^{2.5-3.5}\,\mathrm{yr}$, $\edot_0\sim 10^{37.5-40}\ergs$, 
and $n\sim3$ \citep{crab_martin, zhang_2008, vorster_2013, gelfand09}. This indicates that most of the pulsar rotational 
energy budget ($E\tin{rot}=\edot_0\tau_0(n-1)/2=I\,\Omega_0^2/2$, typically $\lesssim 10^{50}\erg$;
see \aref{app:formulae_derivation}) is spent in the first few thousand years. 

The present spin-down luminosity can be calculated from the period $P$ and its
time derivative $\pdot$ \citep[Eq. 1]{gs}. Another parameter that can be
derived from the pulsar ephemeris is the so-called characteristic age, which
is defined as
\begin{equation}\label{eq:age}
\age \equiv \frac{P}{2\pdot} = (\tau_0+t)\,\frac{n-1}{2}.
\end{equation}
If $t\gg\tau_0$ and $n=3$, then $\age$ is an estimator for
the true age $t$ of a pulsar.
Independent of this condition, though, \eref{eq:edot} and \eref{eq:age} imply a
straight power-law correlation between $\edot$ and $\age$   , i.e.%
\begin{equation}\label{eq:edotvsage}
\edot = \edot_0\left[\frac{2}{n-1}\cdot\frac{\age}{\tau_0}\right]^{-\frac{n+1}{n-1}},
\end{equation}
or, equivalently, between $\pdot$ and $P$ (see \eref{eq:pdot_vs_p} in
\aref{app:formulae_derivation}), i.e.
\begin{equation}
\pdot(P) = \frac{P_0}{\tau_0} \frac{1}{n-1} \left(\frac{P}{P_0}\right)^{2-n}.
\end{equation}
Consequently, the power indices of the above relations are only
determined by the braking index $n$. Figures~\ref{fig:edot_age} show how real
pulsars populate these diagrams. They are born on the upper left of the plots and
move towards the lower right as their spin-down decays.
Pulsar population synthesis studies have
shown that this distribution can be reproduced assuming
magnetic dipole spindown ($n=3$; e.g. \citet{psr_population}, and references therein).  Some such studies found
evidence for pulsar magnetic field decay, but on timescales
of several Myr \citep[e.g.][]{psr_population2}.  As this is
much longer than the PWN evolution timescales we consider,
in the \model\ of this paper we assume that the injection
evolution is dictated by an average braking index $n=3$, which is a
compromise between theoretical expectation, observed pulsar $\edot$ and $\age$, and
the measured braking indices (see \aref{app:modelling} for more details). 

\subsection{Dynamical evolution} \label{subsec:dynamical_theory}

The dynamical evolution of PWNe can generally be divided into three distinct stages
\citep[and others]{gs, gelfand09, swaluw01, swaluw04}: the free expansion
\mbox{(${<2}$--$6\kyr$)}, reverse shock interaction (until some tens of kyr),
and relic stage. In the free expansion phase, the plasma bubble grows inside the
unshocked ejecta of the SNR, whose forward and reverse shocks do not interact
with the PWN. This phase is comparably well understood because of numerous analytical \citep{rees_gunn, kennel_coroniti_1, kennel_coroniti_2}
and numerical \citep[and references therein]{crab_martin, mhd} works on the subject mostly
focussed on the Crab nebula case, but applicable to other young PWNe. The
PWN is growing fast \citep[$R\sim t^{1.2}$]{chevalier77}, attenuating
the magnetic field strength and synchrotron radiation,
while IC emission from the accumulating
electrons quickly increases in the beginning and then decreases very slowly
\citep{torres2014}. This early stage is the only phase
where the IC scattering on synchrotron photons (synchrotron self-Compton emission) can
also play a dominant role.

The second phase begins after a few thousand years, when the PWN has grown to a
size of the order of $\sim 10\pc$ and encounters the reverse shock of
the SNR, which may be moving spatially inwards \citep{blondin01}. Since the total dynamic energy
in the SNR exceeds that of the PWN by one or two orders of magnitude, the PWN
may be compressed again by up to a factor of 10 \citep{gelfand09} and experiences a series of
contractions and expansions until a steady balance is reached. After that, the
wind nebula continues to grow
at a much slower pace, like $R\sim t^{0.73}$ for $t<\tau_0$ in
\citet{swaluw01} and $R\sim t^{0.3}$ for $t>\tau_0$ in \citet{reynolds84}. In the work of \citet{gelfand09},
where a spherically symmetric case was simulated, the oscillations were found to lead to
dramatic changes in the synchrotron and IC luminosities, making the TeV
emission disappear completely for several thousand years.
In reality, where the SNR develops asymmetrically and the pulsar has a proper motion,
these drastic changes are presumably washed out to some degree, leading to a more continuous behaviour.
Still, the collision of PWN bubble and reverse shock heavily depends on
the evolution of the whole system and its interaction with the surroundings,
making such evolved PWNe very diverse, non-uniform objects
\citep[see also][]{okkie_arache_2009}.

This non-uniformity becomes even more pronounced if the pulsar, owing to its proper
motion or a tilted crushing of the nebula, spatially leaves the main PWN bubble or even
the SNR. In that case, which is called the relic stage, the pulsar can form a
local plasma bubble while the old nebula from its younger period still remains,
typically as an IC-dominated PWN due to its much lower magnetisation.

\subsection{Modelling}\label{subsec:theory_modelling}

The interpretation of the data we present and of the log-linear trends we fit to the
evolution plots require a comparison to what can be expected in theory with the basic
concepts outlined above. To do so, we built a simpified, time-dependent
model for the evolution of the VHE electron population and TeV emission of
PWNe. We deliberately opted for a simple model because we do not need it to
contain detailed parameters that our TeV data does not allow us to investigate. 


The model we describe in \aref{app:modelling} assumes a time-dependent
injection of electrons with a fixed power-law spectrum\footnote{A
spectral break at lower injection energies is generally necessary to model low-energy data, but
since this does not impact the TeV regime, and we therefore cannot constrain
it with the data presented in this paper, we focus on the VHE part with a
single power law.}, but decreasing total power
according to \eref{eq:edot}. Following analytical formulae for the expansion, the
cooling from synchrotron, adiabatic, inverse Compton, and escape losses is
applied to the electron population as a function of time. The respective
characteristic age $\age$ is always tracked as well to compare the
model correctly to data. The photon emission is calculated for each time
step from the electron population, including the full Klein-Nishina formula. 

The strategy for the comparison of PWN data and theory is to define
the parameters of the model such that it reflects both the average trend of
PWN evolution (\model) and the scatter of individual
wind nebulae around that average expectation (\scatter). 
This means that, unlike other works, we do not model individual objects in
their particular
multi-wavelength context. Instead, we try to find out what the typical
evolution is and what the typical variations need to be in order to produce the
picture we obtain for the whole population. The band of the \scatter\ can
therefore be interpreted as the area where a synthesised population would be
found (in the absence of detection selection effects).

As it turns out in the following,
we succeeded in finding such a model describing the
evolution that a typical PWN in a typical, dense spiral-arm surrounding undergoes.
Since this one model implies an evolution curve for every observable we
consider, both along $\age$ and $\edot_0$, a good leverage on its absolute
parameters is given.
Starting from the \model, the parameters are
varied with the aim to realistically reproduce the
scatter of measured PWN observables. This way, the scatter of observables
itself is exploited as another observable, with the large number of curves
leading again to a good handle on the scatter.

It should be noted though that intrinsic (physical) and analytical
(mathematical)
correlations between parameters are neglected in the \scatter. For instance,
the scatter ranges of $\edot_0$ and $\tau_0$, strongly restricted by
\fref{fig:edot_age} (left), may be larger if the two quantities were
anti-correlated such that high-$\edot_0$ pulsars always tend to have a lower
$\tau_0$; this is physically plausible because the two quantities are related through
the pulsar birth period and
magnetic field. 
On the mathematical side, $\edot_0$, $\eta$, the energy injection
range and the background photon
density are all parameters with which the TeV luminosity scales in an
almost linear way.
In our \scatter, we
deal with this redundancy by only varying $\edot_0$, but similar results can be achieved if one of the other
factors is varied instead. See also \sref{subsec:modelling_caveats}\ for
this and other caveats of the model.

\section{Properties of TeV pulsar wind nebulae} \label{sec:pwn_properties}

\begin{figure*}
\includegraphics[width=0.494\textwidth]{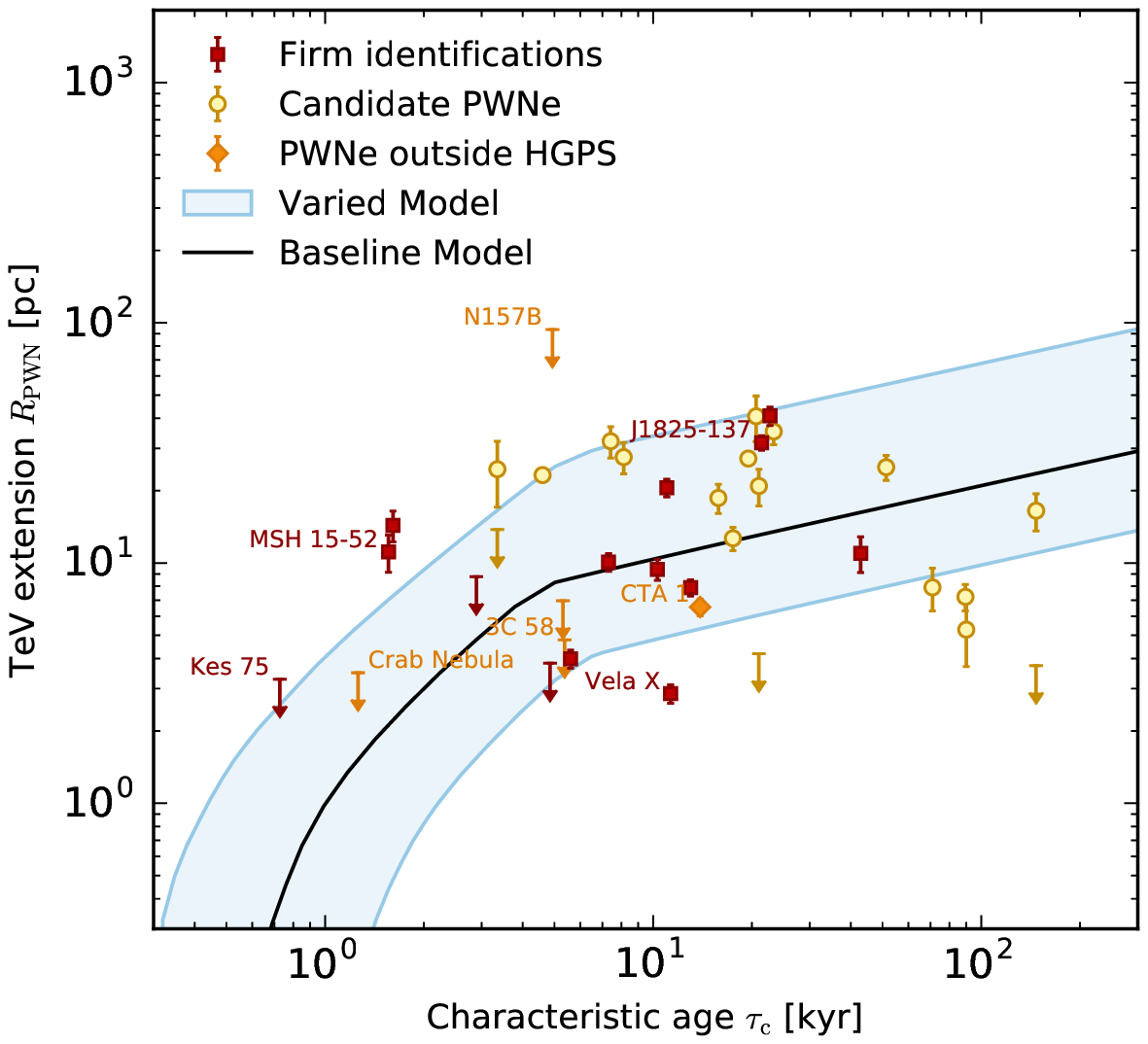}\includegraphics[width=0.494\textwidth]{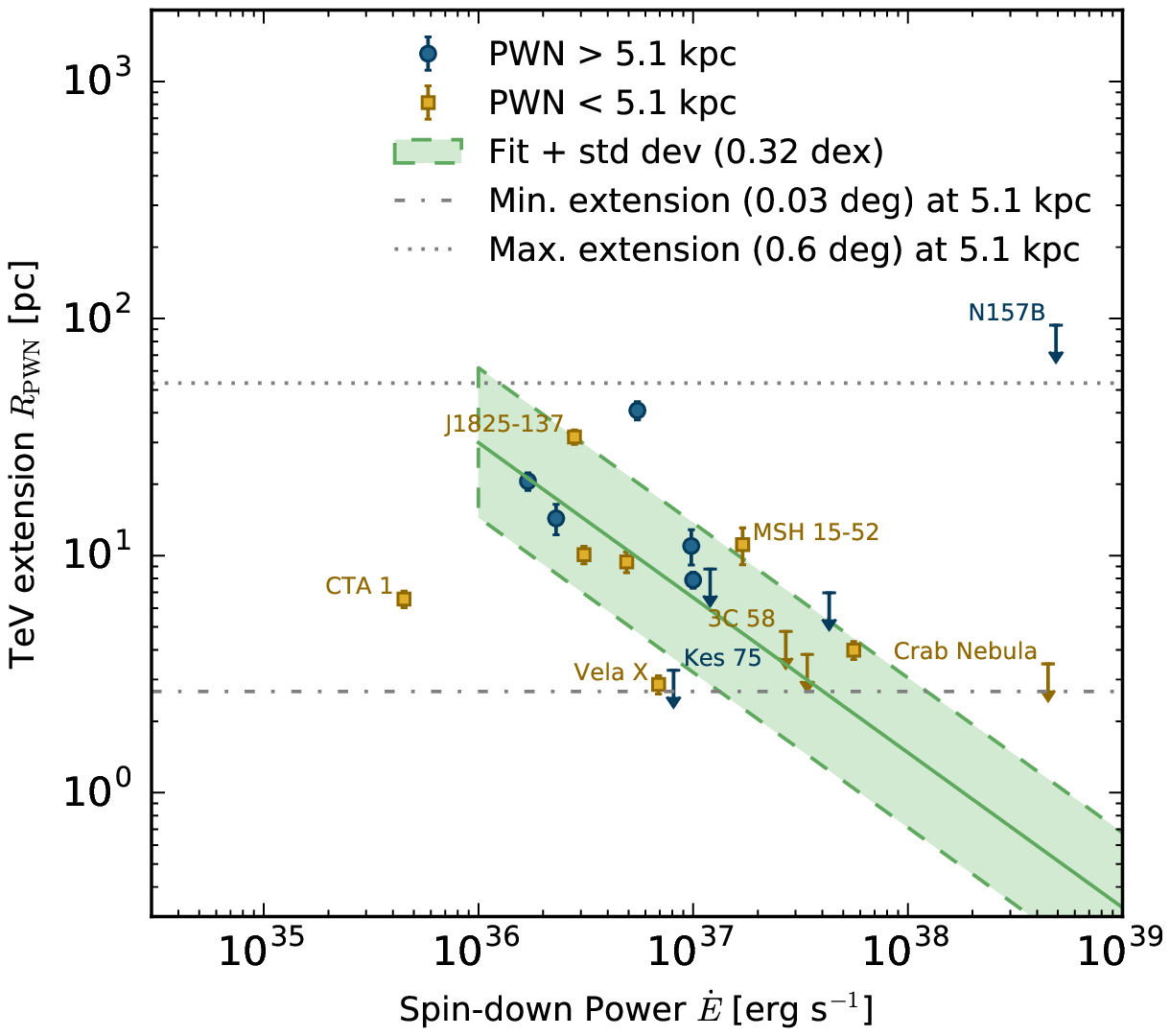}
\caption{
Left: PWN extension evolution with time, in comparison to the modelling 
considered in this work. Right: PWN extension evolution with $\edot$, as
fitted in the $R\tin{PWN}(\edot)$ column of \tref{tab:asurv_results} for
pulsar wind nebulae with $\edot>10^{36}\ergs$ (see \sref{subsubsec:extension}).
The shaded range shows the fit range and standard deviation $\sigma\tin{\lgt R}$.
$1\eh{dex}$
refers to an order of magnitude and is the unit of the logspace defined
$\sigma_{\lgt Y}$. For
clarity, this plot excludes PWN candidates and
divides the sample into nearby and far pulsar wind nebulae to illustrate the potential selection or reconstruction bias
(see text). The dot-dashed and dotted lines indicate the systematic minimum of
$0.03\dg$ and the maximum measured extension in the HGPS of $0.6\dg$, respectively, which are both projected to the
average PWN distance of $5.1\kpc$.  
\label{fig:extension1}}
\end{figure*}

In this section we present and discuss the distributions and correlations of TeV wind nebulae and
their respective pulsars. For each topic we describe what we present,
discuss potential biases, and then interpret what we find, using the modelling
described in \aref{app:modelling} where needed and appropriate. The
presented plots serve to evaluate the plausibility of our current candidate sample (\sref{sec:candidate_rating}) and may prove useful in investigating future PWN candidates. 

\subsection{Fitting and statistical treatment of uncertainties}\label{subsec:errors}

The properties of PWN are intrinsically scattered (see
\sref{subsec:dynamical_theory}) and 
all observables are calculated using a distance estimation based on the dispersion measure of the
pulsar and a model of the Galactic free electron distribution,
whose uncertainty is not statistically  well described.
Consequently, the probability density functions (p.d.f.) of our
observables (size, luminosity, and offset) for 
a given $\age$ or $\edot$ are dominated by the scatter of intrinsic properties
and errors in the distance estimation and not by
our statistical uncertainty.

As a consequence, in the cases where we pursue a fit of observables
with the aim of testing the significance of a correlation or extracting an estimator function,
we follow the approach put forward by 
\citet{vink_efficiency_fit} and \citet{possenti_luminosity_fit}.
They performed
a least-squares fit of the
respective observable with residuals calculated in common logarithmic space.
The fit function is a (log-)linear function,
expressed generally as
\begin{equation} \label{eq:y_pwn}
\lgt Y_{\mathrm{est}}  =  p_0 + p_1 \lgt X.
\end{equation}
In order not to be restricted to detected objects but also to include the valuable limits
from pulsars without VHE emission,  
we use the \asurv\ code
\citep{asurv} for the minimisation. It allows us to apply statistical methods to test for the existence of a
correlation, such as \cox, or to perform a multivariate
regression including limits \citep[see][for an
overview on the statistics inside \asurv]{censored_data_analysis}. Besides the parameters $p_i$ of our
function, \asurv\ also determines the variation $\sigma_{\lgt Y}$ that the
data are scattered with.

Owing to the existing selection biases and the uncertain p.d.f. shapes involved, the derived
estimator function might not always approximate a virtual true evolution
function, but rather evaluate the unweighted average trend of the examined
data points. Table \ref{tab:asurv_results} summarises the fit results that are
referred to in the following paragraphs. The $p$-values are taken from \cox,
which is a regression method for data with upper limits.  This model was originally
developed for biostatistical applications, where it is extensively used. As described in
\citet{censored_data_analysis}, Section III, the model provides an equivalent
$\chi^2$ for the null hypothesis (no correlation), which can be transformed to
a $p$-value. For the linear regressions and parameter determinations, the
expectation maximisation (EM) algorithm is used, which is an iterative least-squares method that allows
for the inclusion of limits \citep[][Section IV]{censored_data_analysis}. 

\subsection{Morphological properties}

\begin{figure*}
\includegraphics[width=0.494\textwidth]{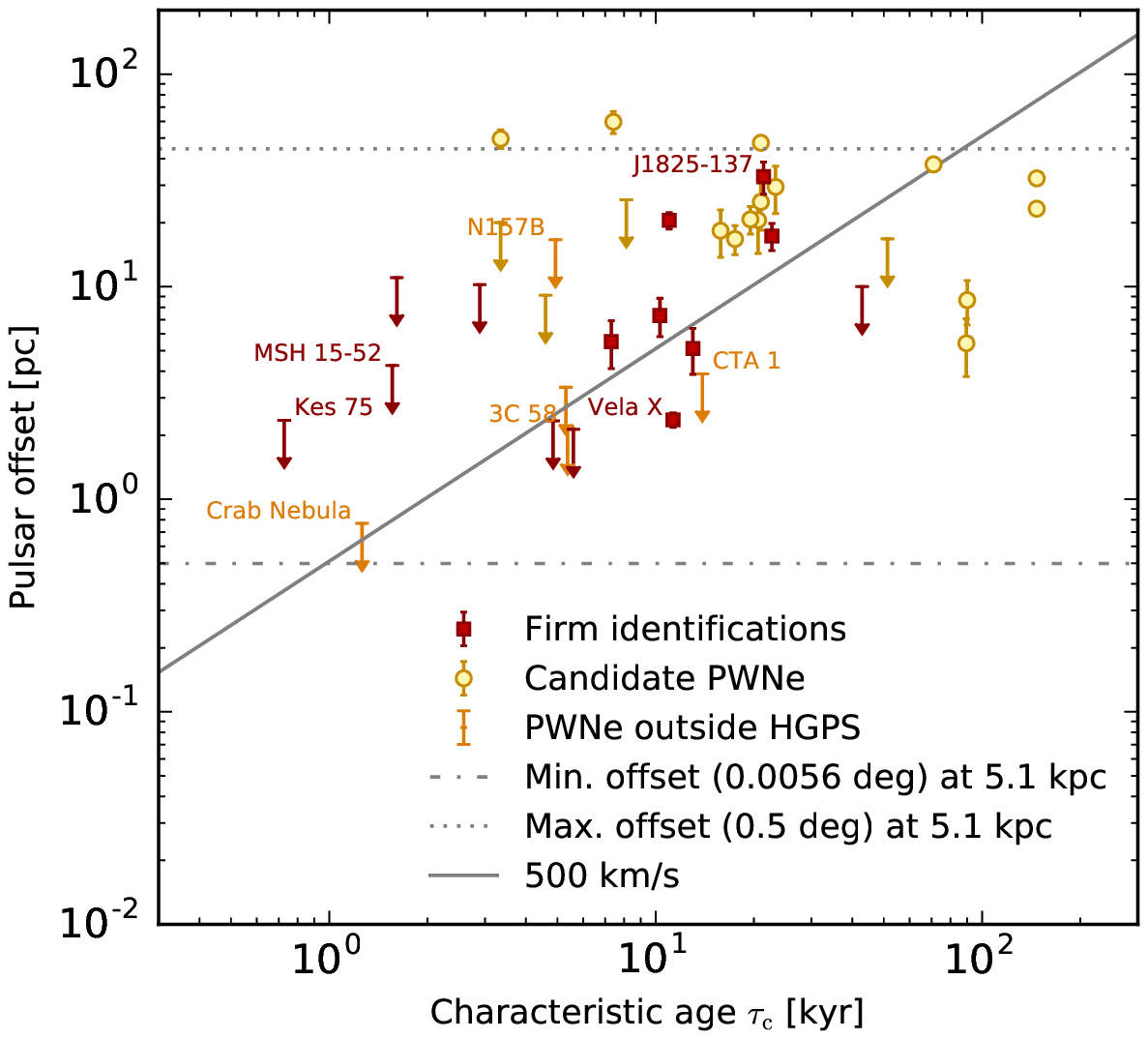}\includegraphics[width=0.494\textwidth]{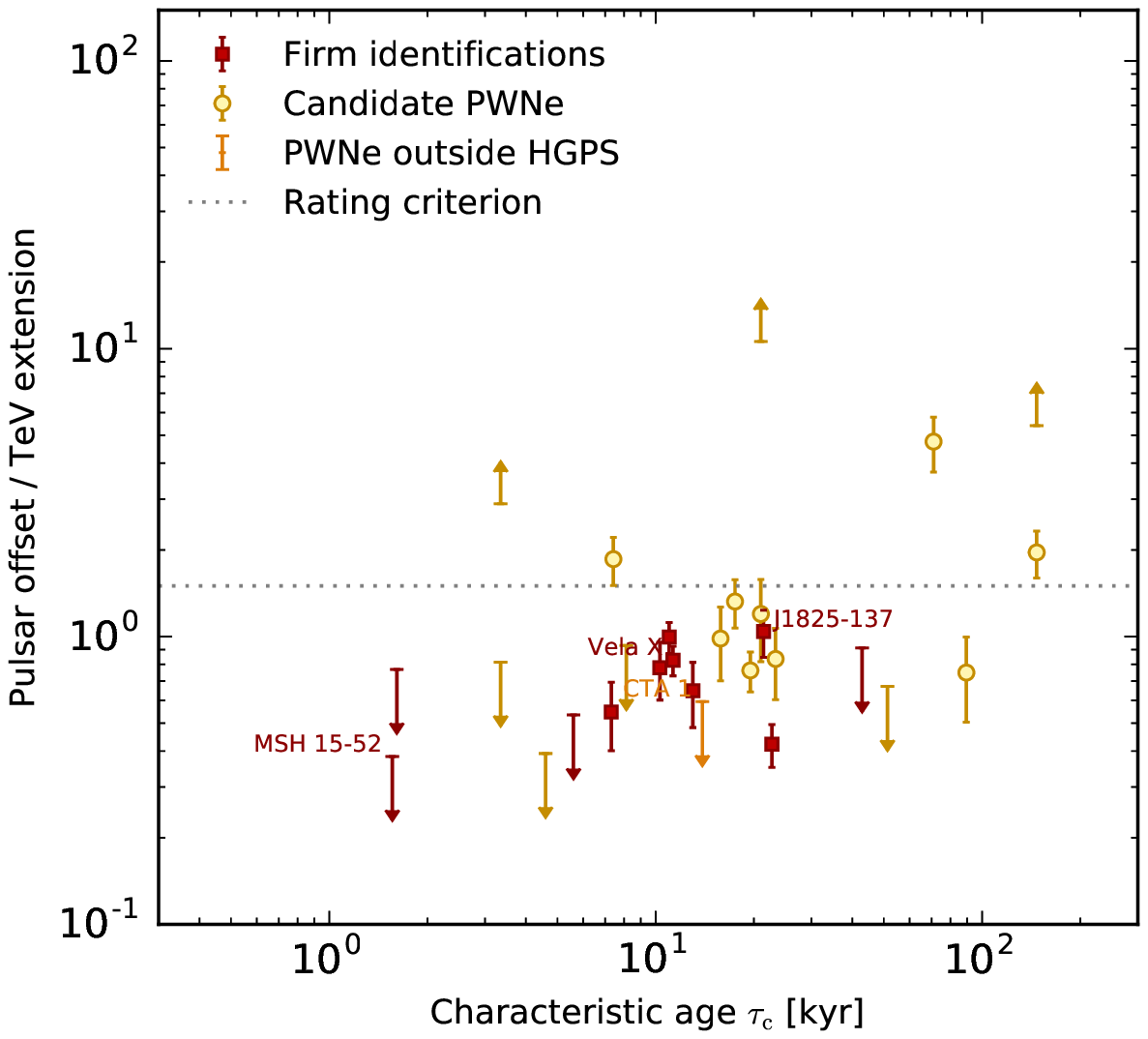}
\caption{
Left: Spatial offset of pulsar and TeV wind nebula as a function of
pulsar characteristic age. The dashed and dotted lines indicate the systematic
minimum of about $0.015\dg$ and the association criterion of $0.5\dg$, which are both
projected to the average PWN distance of $5.1\kpc$. The solid line shows the 
offset one would expect from pulsar motion only (assuming a large
pulsar velocity of 500 km/s). Right: Time evolution of
the containment ratio. Since the pulsar motion can be
assumed to be constant and the expansion decelerates, one expects the
containment fraction to increase and eventually pass unity after some tens of
kiloyears. The dotted horizontal line shows the rating criterion 
(offset/extension $<1.5$) applied in the post-selection of candidates
(\sref{sec:candidate_rating}).
\label{fig:offset}}
\end{figure*}

The morphological parameters provided by the HGPS catalogue are source
position and extension. As a pulsar and its PWN evolve, the PWN is thought to become increasingly extended and
offset from the pulsar position (see \sref{sec:pwn_theory}). This basic evolutionary behaviour can be found
unmistakably in Figs. \ref{fig:extension1} and \ref{fig:offset} (left).

\subsubsection{Extension} \label{subsubsec:extension}

Figure \ref{fig:extension1} (left) shows the evolution of PWN extension as a function of characteristic age $\age$.
%
%
We can determine extensions beyond a systematic
minimum of around $0.03\dg$ and at least up to the observed extension of
\velax, at around $0.6\dg$ (see
\sref{sec:galaxy}). As shown in
\fref{fig:extension_dist}, most known pulsars lie at distances that therefore allow for
the measurement of PWN extensions between $3$ to
$60\pc$. In \fref{fig:extension1} (right), where the extensions are plotted
against pulsar spin-down, far and close-by systems are
distinguished. This elucidates our ability to resolve far and near systems and shows the
plain correlation of size with $\edot$.

A caveat is that there is a selection bias from the fact that extension estimates
or limits are only available for sources that are detected. Systems that
are too faint or too large to be detected with our sensitivity and
FOV are missing in the PWN sample. Since we cover a wide range of
different 
distances, sources that are large and bright, or faint and
small, can still be represented to some level in the sample.
However, if there is a state in which PWNe are faint and large at the same time, it might
be that they cannot be detected at any distance. From
the current understanding of PWN theory, this can be the case for PWNe of ages
beyond few tens or hundreds of kiloyears, so the study presented here has to
be taken with some caution in that regime. To unbias the sample in the fitting
procedure below, we apply a cut of $\edot>10^{36}\ergs$, beyond which the likeliness of
detection is reasonably high and the detected objects can be considered representative for
their stage of evolution.

A measurement bias we may have is that 
the limited FOV might truncate the tails of the source for very extended sources. This effect
was suggested by \citet{argo_extended} as an explanation for the differences between some IACT spectra and the results of the air shower detector ARGO.
We cannot entirely verify or falsify this claim here, but since
only few sources approach the critical regime beyond $1\dg$, it
is presumably a minor effect in this study.

A possible physics bias that might enhance the effect seen in \fref{fig:extension1}
(right) is that close-by objects are on average located farther away from the
Galactic centre and therefore in less dense surroundings than far
objects. This might influence the average dynamical evolution they experience.

Fitting the data to check for
correlations with $\age$ or $\edot$ yields the results shown in
\tref{tab:asurv_results}. The low $p$-values and non-zero $p_{1,2}$ confirm,
on $2$--$3$ standard deviation confidence levels, that
the extension increases along the evolution of a PWN, i.e. with falling
$\edot$ and increasing $\age$. A more general 2D fit of $R\tin{PWN}(P, \pdot)$
does not lead to a significant improvement of the fit, nor a lower $p$-value.
The parametrisation of $R\tin{PWN}(\edot)$ is shown in in
\fref{fig:extension1} (right) to show that it is indeed suitable for predicting
the extensions of the detected young PWNe ($\edot>10^{36}\ergs$) reasonably
well. The only PWN below $10^{36}\ergs$, CTA~1, does not follow the
extrapolation of that trend and appears to be dynamically different from the
rest of the
population.

The relation $R\sim\age^{0.55 \pm 0.23}$ can
be compared to the \model\ in
\fref{fig:extension1} (left), which assumes the canonical $R\sim t^{1.2}$ and
$t^{0.3}$, at early and late times,
respectively, and thus encloses the measured value well. The
conversion between true age and $\age$ according to \eref{eq:age} is taken
into account in the displayed model curves. 

Comparing the data with the model, the initial and fast free expansion can
accommodate
the non-detections
of extensions of very young pulsar wind nebulae, while the slope of $t^{0.3}$ for evolved
PWNe \citep{reynolds84} is roughly consistent with the comparably small
extensions of the few older PWNe in the sample. One has to keep in mind that
the curve at high
ages is more an upper limit than a prediction because a potential crushing (as a
sudden decrease in size 
after the free expansion) is not included in the model.
The absolute scale of the curves is a free parameter of
the model, but later turns out to be constrained by the surface brightness
values we measure (\sref{subsubsec:surf_br}).

In conclusion, the fact
that TeV pulsar wind nebulae generally grow with time until an age of few tens of
kiloyears is clear and supported by the \asurv\
fits. There are, however, few pulsar systems older than that  to place stringent
constraints on the model at later evolution stages.

\subsubsection{PSR-TeV offset}\label{subsubsec:offset}

An offset between a pulsar and its TeV wind nebula can be caused by a combination of pulsar
proper motion, asymmetric crushing of the PWN by the surrounding SNR and,
possibly, by asymmetric pulsar outflow. \citet{hobbs2005} determined
the mean 2D speed for non-millisecond pulsars to be $307\pm 47\kms$,
and the velocity distributions were found to be compatible with a Maxwellian
distribution. Other works, such as \citet{psr_velocity_distribution}, suggest
a more complex distribution and high-velocity outliers, but it is clear that the
bulk of the pulsars have 2D velocities of less than $500\kms$. In \fref{fig:offset} (left),
the offset against characteristic age is compared to a 2D velocity of $500\eh{km/s}$.
The true age of young pulsars can be less than $\age$, in which case those
points, shown in true age, may even move to the left, and thus enlarge the
distance to the $500\eh{km/s}$ line. 
To give an idea of
which offsets can be detected, lines for the maximum offset implied by our
angular selection criterion and systematic minimum resolution are also
shown for the mean PWN
distance of
$5.1\kpc$\che{2016/06/09}.

The \asurv\ fits (\tref{tab:asurv_results}) suggest that the trend of
increasing offset with rising $\age$ and falling $\edot$ is statistically manifest in the
data. What is interesting beyond this general increase is that 5 of 9 pulsars with ages beyond $7\kyr$
in \fref{fig:offset} (left)
are more offset from their PWN than expected
from mere pulsar motion. At these ages, PWNe presumably are beyond their free expansion
phase and have started interaction with the SNR reverse shock or surrounding
medium. While the velocity distribution of pulsars can have outliers that
are significantly faster than average, it is unlikely that such a high
fraction of the
high-$\edot$ pulsars with
TeV-detected nebulae are so fast ($1000\kms$ or more would be required). This suggests that the asymmetric evolution of the PWN,
caused by interaction with the reverse shock and/or asymmetric surrounding
medium \citep{blondin01}, 
is in fact the dominant offset mechanism for middle-aged wind nebulae. Further
support for this conclusion comes from the very few measured pulsar transverse velocity
vectors that are currently available in the ATNF catalogue for our PWN sample (e.g.\ for \velax\ and
\jxviii). These vectors do not consistently point away from the PWN, as one
would expect from a pulsar motion dominated offset.

\subsubsection{Containment}

Containment of a pulsar in its TeV wind nebula, although not strictly binding in the
relic stage, is often taken as an argument to
claim the PWN nature of an object.
We define the containment ratio as the PWN offset divided
by the PWN extension radius. Given the offset
and extension evolution discussed above,
a pulsar is not expected to leave its (then relic) wind nebula before some
tens of kiloyears; yet the ratio should increase and approach unity at some
point, unless the relative
movement is in the direction of the line of sight.
 \figref{fig:offset} (right) shows the evolution of the containment
ratio with characteristic age.

An additional caveat to mention for this quantity is that no upper or lower limit can be
calculated if both offset and extension are already limits, which is the case
for 7\che{2016/06/09} of the 19\che{2016/06/09} firmly identified objects in our
sample. Another selection bias concerns the identification itself. Good reasons, such as observations at other
wavelengths, are required to argue for a non-contained association of a pulsar with a TeV object;
for old systems, however, these MWL data are very difficult to acquire since the
synchrotron component has become very faint. This bias can be regarded
as intrinsic to the decomposition of old, ``dissolving" PWNe, whose remains become inevitably
difficult to associate with the pulsar as time passes.

In \figref{fig:offset} (right), most young pulsars are well contained in their
nebulae, but there are a few older pulsar wind nebulae that were firmly associated to a pulsar close to or slightly beyond their (1-$\sigma$ Gaussian) extension radius.

\begin{figure*}
\includegraphics[width=0.494\textwidth]{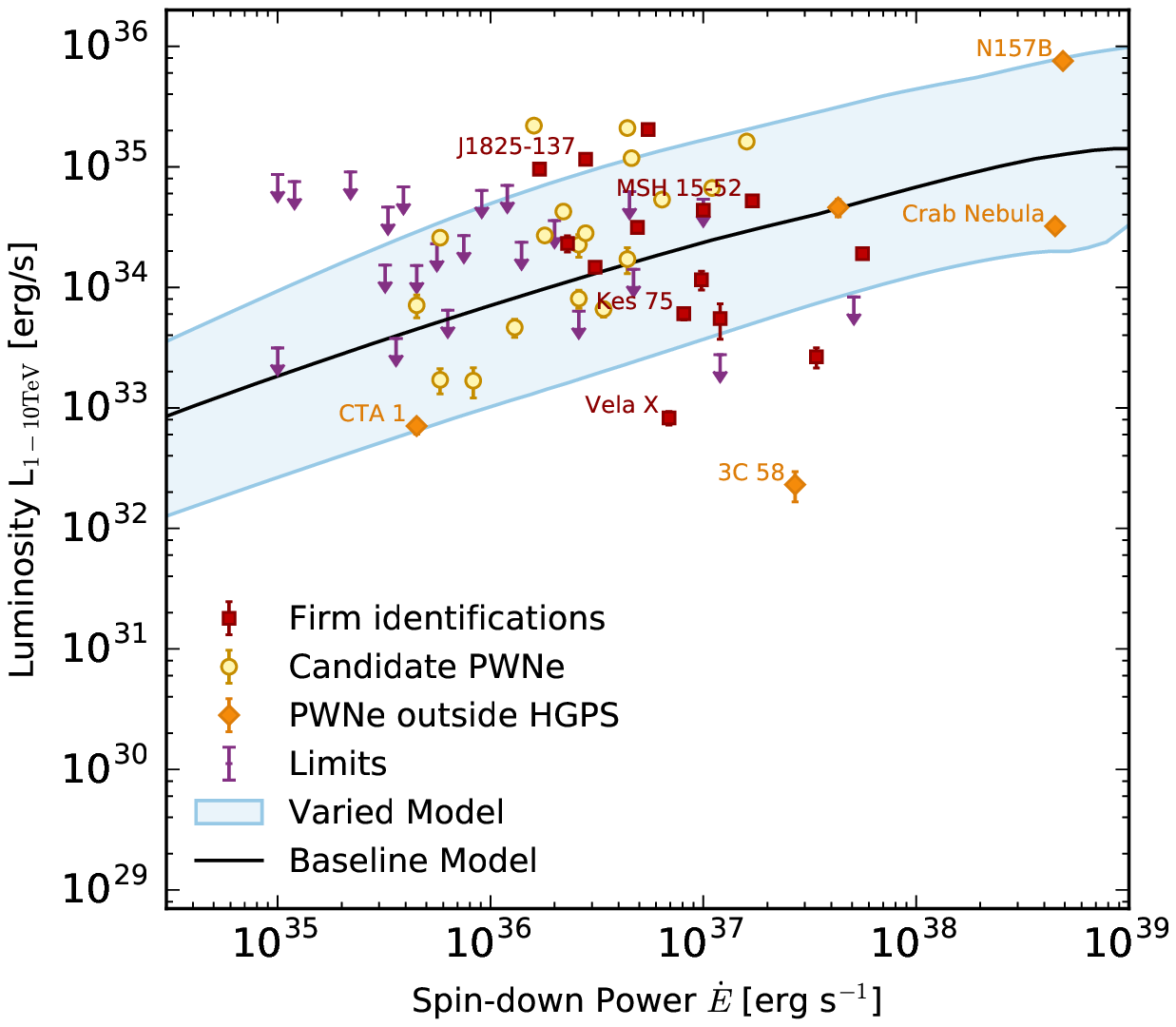}\includegraphics[width=0.494\textwidth]{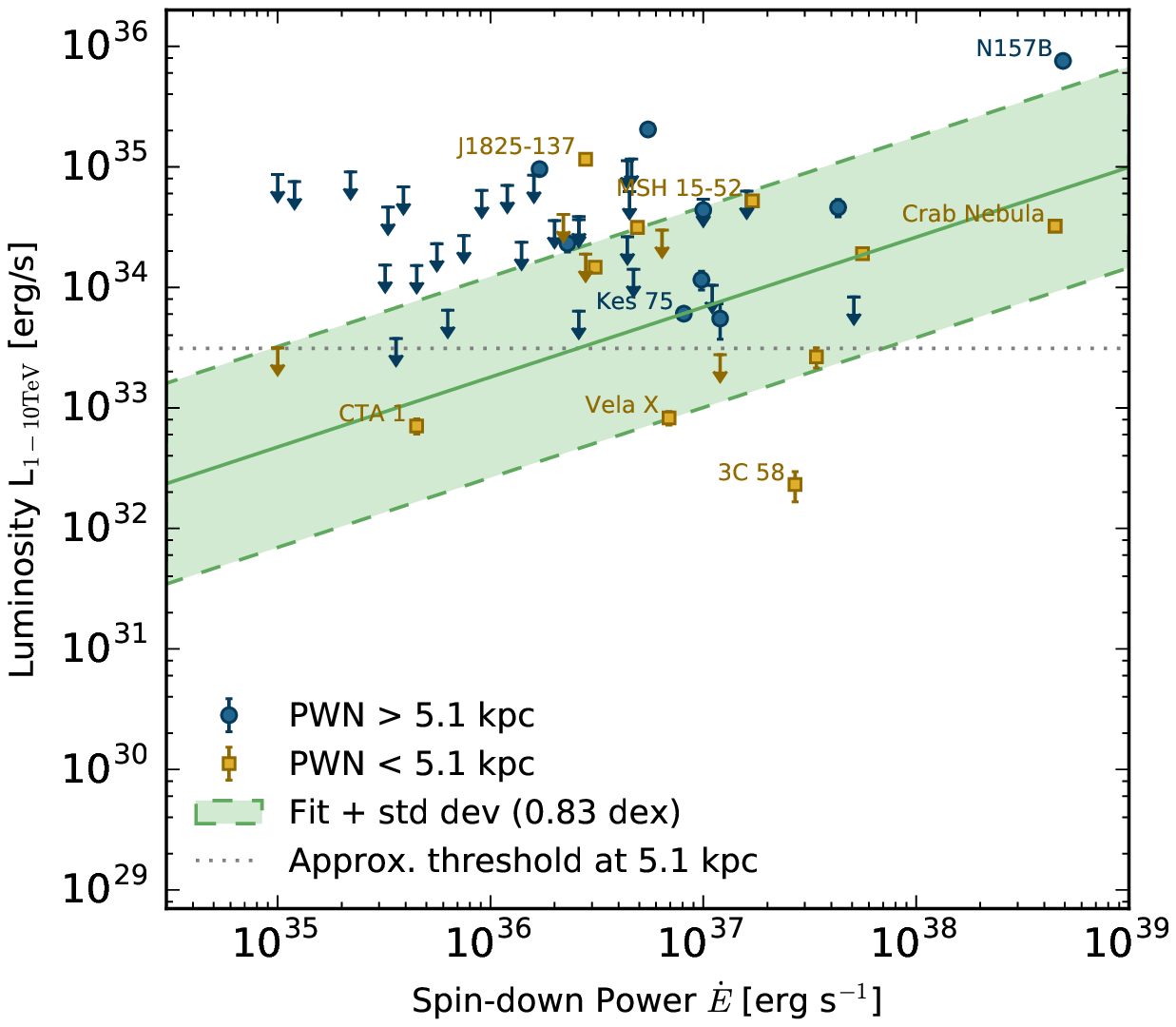}
\caption{
Left: Relation of TeV luminosity and
pulsar $\edot$. Right: Same as \fref{fig:extension1}, but for luminosity and including
the upper limits from the left panel and 11 upper limits calculated for
pulsars of candidate PWNe (see text). 
\label{fig:luminosity}}
\end{figure*}

\begin{table*}
\centering
\caption{List of \asurv\ fit results.}
\label{tab:asurv_results}
\tiny
\centering
\begin{tabular}{lllllllll}
\hline\hline
                     & $R\tin{PWN}(\edot)$ & $R\tin{PWN}(\age)$ & $\lumi(\edot)$ & $\lumi(\age)$  & $S(\edot)$     & $d\tin{P-P}(\varepsilon\tin{TeV})$ & $d\tin{P-P}(\edot)$, & $d\tin{P-P}(\age)$ \\
\hline
$p$-value            & $0.012$             & $0.047$            & $0.010$        & $0.13$         & $0.0013$       & $0.0004$                           & $0.035$              & $0.0086$ \\
$\sigma\tin{\lgt Y}$ & $0.32$              & $0.39$             & $0.83$ & $0.91$         & $0.28$         & $0.18$                             & $0.49$               & $0.42$ \\
$p_0$                & $ 1.48\pm0.20$      & $0.38\pm0.22$      & $33.22\pm0.27$ & $34.1\pm0.4$   & $30.62\pm0.13$ & $1.97\pm0.16$                      & $1.07\pm0.25$        & $-0.9\pm0.5$ \\
$p_1$                & $-0.65\pm0.20$      & $0.55\pm0.23$      & $0.59\pm0.21$  & $-0.46\pm0.36$ & $0.81\pm0.14$  & $0.52\pm0.07$                      & $-0.75\pm0.29$       & $1.4\pm0.5$  \\
\hline                                   
\end{tabular}
\tablefoot{\eref{eq:y_pwn} is applied with one or two
parameters. The $p$-value is calculated after \cox. The fit used (within
\asurv)
is the ``\emalg''. $P$ is given in $0.1\eh{s}$, $\pdot$
in $10^{-13}\eh{s\, s^{-1}}$,
$\edot$ in $10^{36}\ergs$, and $\age$ in $\kyr$. $R\tin{PWN}$ is given in
$\pc$, $\lumi$ in $\ergs$, $S$ in $\ergs\pc^2$. The 2D pulsar-PWN offset
$d\tin{P-P}$ is given in parsecs, and $\varepsilon\tin{TeV} = \lumi/\edot$ is the
apparent TeV efficiency.\che{2016/06/09}}
\end{table*}

\subsection{Luminosity, limits, and derived parameters}\label{subsec:luminosity}

\subsubsection{Luminosity}\label{subsubsec:luminosity}

\begin{figure*}
\includegraphics[width=0.494\textwidth]{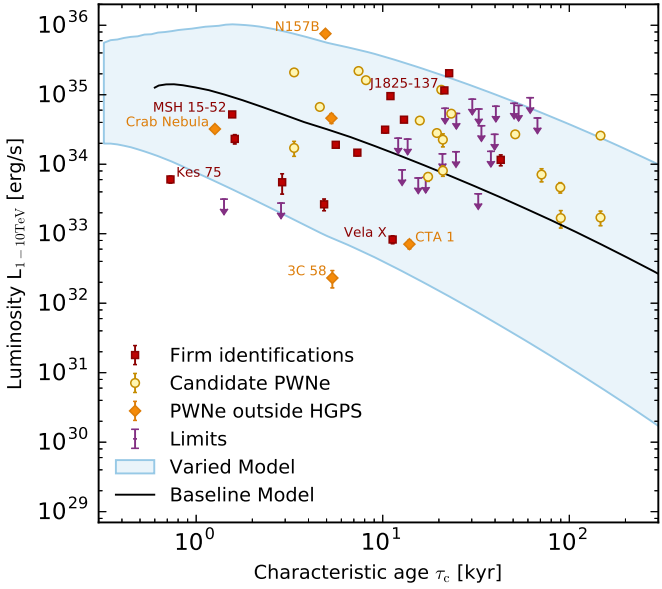}\includegraphics[width=0.494\textwidth]{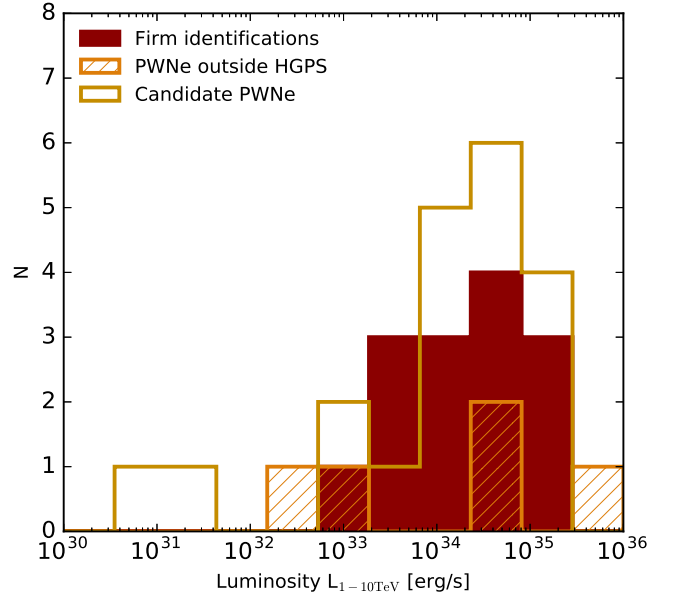}
\caption{
Left: Evolution of TeV luminosity with characteristic age. Right: Distribution
of $1$--$10\tev$ luminosity for PWNe and
 PWN candidates treated in this work. \label{fig:lum_age_hist}}
\end{figure*}

From \sref{sec:pwn_theory} and in our model, the TeV luminosity of pulsar wind
nebulae
is expected to
rise quickly within the first few hundred years and decay slowly over many
thousands of years.
Figure~\ref{fig:luminosity} (left) shows the evolution of luminosity with pulsar
spin-down power and \fref{fig:lum_age_hist} (left) the evolution with
characteristic age.

Figure~\ref{fig:lum_age_hist} (right) indicates the distribution of luminosities. The
average detection threshold of energy flux between $1$ and
$10\tev$ is at around
$10^{-12}\ergs\eh{cm^{-2}}$\che{2016/06/09}\citep{hgps_catalog_paper}, which is equivalent to a luminosity threshold of
$3\ttt{33}\ergs$\che{2016/06/09} at the mean PWN distance of
$5.1\kpc$\che{2016/06/09}.
In this work, we reduce the selection bias present in previous studies by
involving flux upper limits for all eligible pulsars with $\edot >
10^{35}\ergs$. As explained in \sref{subsec:hgps_data_extraction} and
\sref{subsec:candidate_selection}, about one-third of these high-$\edot$ pulsars in the ATNF
catalogue can be expected to have an extension\  small enough from which it is possible to extract a
meaningful limit. So again, PWNe that are very large, presumably
with ages beyond a few tens of kiloyears (below $\sim 10^{36}\ergs$), might be truncated from our data set. 
Figure~\ref{fig:luminosity}
(right) is the equivalent of \fref{fig:extension1} (right), showing the
luminosities and limits in two bands of distance.
The expected extensions and derived limits are listed in \tref{tab:withLIMd2}. 
In the fit below we add further flux limits calculated for the pulsars associated with the candidate PWNe, applying the
same calculation method as for the limits in \tref{tab:withLIMd2}. This adds
11\che{2016/06/09} further valid limits, which are also included in
\fref{fig:luminosity} (right).
This flux limit can actually be below
the flux of the candidate, for instance if the candidate is more
extended than predicted by the model (e.g. in the case of HESS
J1023$-$575).

The primary feature of the data is a mild but stable correlation of luminosity
with pulsar spin-down\footnote{The $p$-value without N157B is still $0.06$, so the
correlation does not only depend on this one source.}.
The \asurv\ fit suggests a relation of
$\lumi\sim\edot^{0.59\pm0.21}$ (see \tref{tab:asurv_results}). The model supports 
this, indicating a power index of around $0.5$. The slow but steady
decay, combined with the growing extension, is what hampers a TeV detection once the pulsar spin-down power falls below $\sim 10^{36}\ergs$. 
This decay could not be observed
in other works before \citep[e.g.][]{mattana_2009,kargaltsev_2010} owing to the missing upper
limits\footnote{The $p$-value for the fit of $\lumi(\edot)$ without the limits
is $0.31$.}. 
Figure \ref{fig:luminosity} (right) shows the result of the fitted
parametrisation $\lumi(\edot)$ derived from our data.

In contrast, the $\lumi$ over $\age$ (\fref{fig:lum_age_hist}, left) is scattered widely and
a correlation is statistically not clear (see \tref{tab:asurv_results}). This, however, matches the broad scatter suggested by
the \scatter\ (shaded area). Apparently, $\edot$ is the better variable to characterise
the evolutionary state of the PWN luminosity.

\subsubsection{Apparent TeV efficiency}
\label{subsubsec:efficiency}

The TeV efficiency, conventionally defined as $\varepsilon\tin{TeV}=\lumi/\edot$, is not the real present
efficiency of a PWN because $\lumi$ is a result of the whole injection history,
whereas $\edot$ characterises the present outflow of the pulsar. Therefore, TeV pulsar wind nebulae can
in principle have TeV efficiencies exceeding unity.

\begin{figure*}
\includegraphics[width=0.494\textwidth]{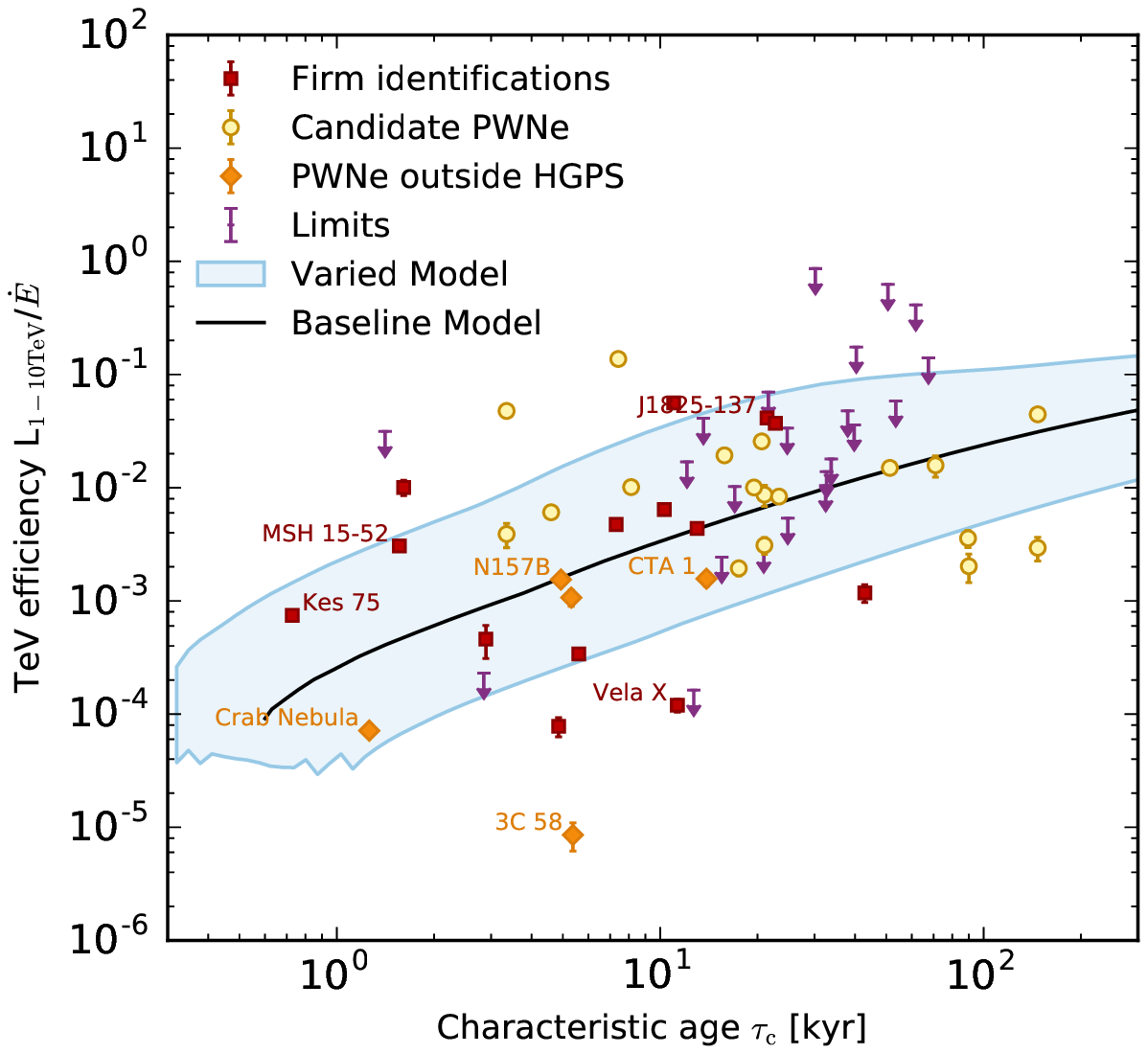}\includegraphics[width=0.494\textwidth]{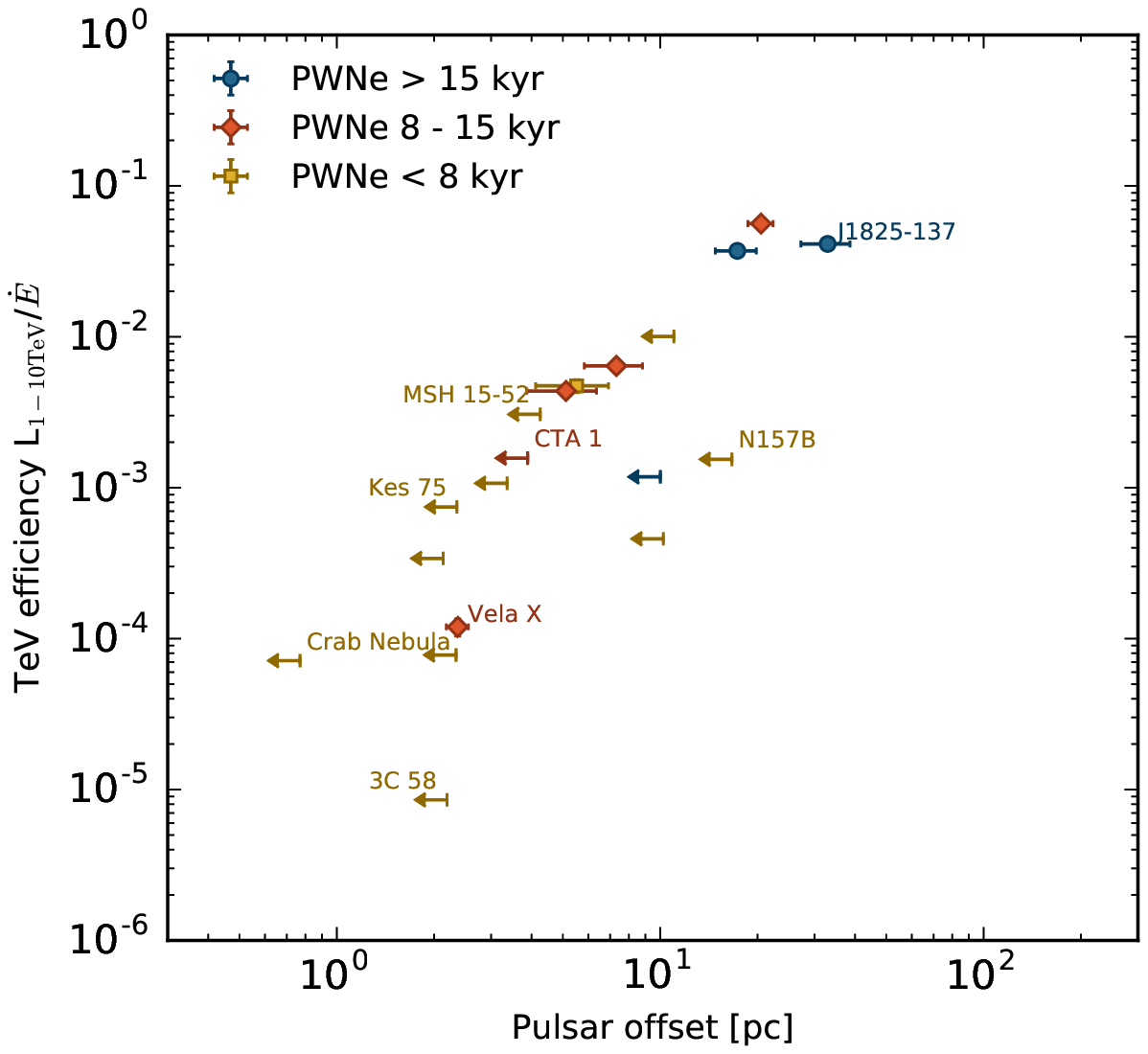}
\caption{
Left: Evolution of TeV efficiency ($\lumi/\edot$) with pulsar characteristic
age. Right: TeV efficiency as a function of pulsar
offset, plotted for pulsars of different age groups. High-offset systems tend to be
more TeV-efficient than low-offset systems.
\label{fig:efficiency2}}
\end{figure*}

Figure~\ref{fig:efficiency2} (left) shows the evolution of the efficiency with the
pulsar characteristic age.
Interestingly, the efficiency seems to be scattered more than suggested
by the \scatter, unlike in \fref{fig:lum_age_hist} (left).
To shed light on the cause of this it is illustrative to plot TeV efficiency
versus the PSR-PWN offset for different groups of characteristic age
(\fref{fig:efficiency2}, right). With the sample of detected PWNe,
a relatively clear correlation can be confirmed also in the \asurv\ fit
(\tref{tab:asurv_results}). Apparently, all low-efficiency
PWNe are found at low offsets from their pulsar and all high-efficiency
wind nebulae have larger offsets. To some level, this correlation is trivial
because both efficiency and offset increase with time. After subdividing the
sample into different age groups, however, it becomes clear that the plot does
not only sort by age; instead, even for PWNe with similar ages, efficiency and
offset are correlated and the age groups overlap each other.
In the plot, a bias might occur because low-efficiency, high-offset systems
may be difficult to identify, but the absence of high-efficiency, low-offset
systems must be genuine. A second systematic effect may be that both
efficiency and offset depend on the PSR distance estimation $d$ (or its square),
so if there were a strong bias in $d$, it would also appear as a trend in the
plot (though hardly at scales of more than a factor of 10).

This is a mild indication that a pronounced offset, as may be induced by SNR reverse shock crushing,
comes with a high TeV efficiency, while systems that interact less with their
surroundings remain fainter overall. One cannot disentangle at this point whether the
crushing itself heats up the plasma bubble or whether the correlation is
indirect because denser environments might provide both crushing material and
a higher IC target photon density. More case studies in the future might
clarify the situation here.

\subsubsection{Surface brightness}
\label{subsubsec:surf_br}

All of the TeV quantities discussed so far rely on the knowledge of the
distance to a given pulsar system, which in many
cases, however, is not very well constrained observationally. A quantity that is
independent of the distance is the TeV surface brightness, defined as
\begin{equation}
S = \frac{\lumi}{4\pi R\ind{PWN}^2} \approx \frac{F_{1-10\,\mathrm{TeV}}}{\sigma^2}
,\end{equation}
where $R\ind{PWN}$ is the physical PWN radius (in pc), $\sigma$ is its angular
extent as seen from Earth, and $F_{1-10\mathrm{TeV}}$ is the integral energy flux
between $1\tev$ and $10\tev$ measured at Earth.

\begin{figure*}
\includegraphics[width=0.494\textwidth]{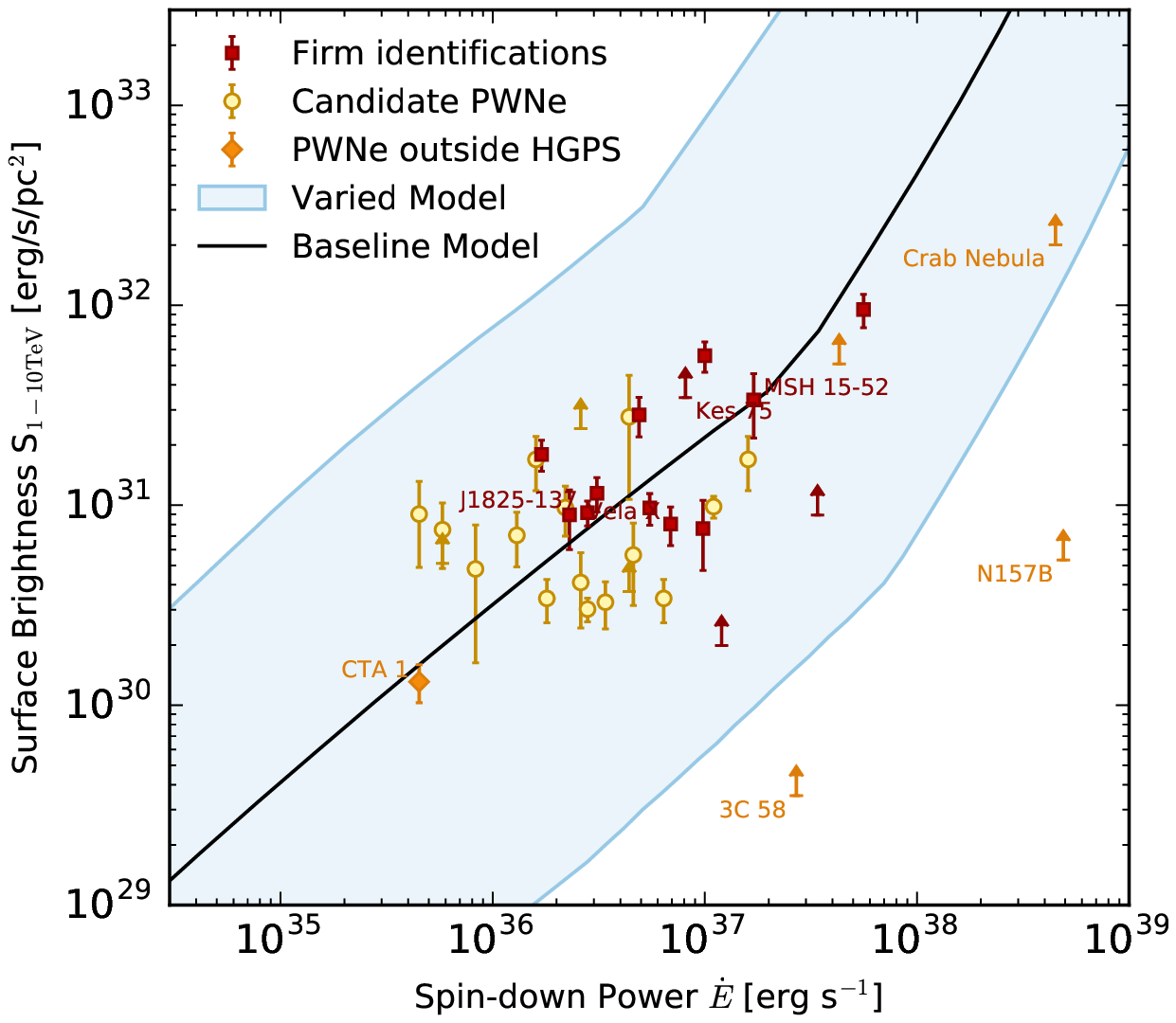}\includegraphics[width=0.494\textwidth]{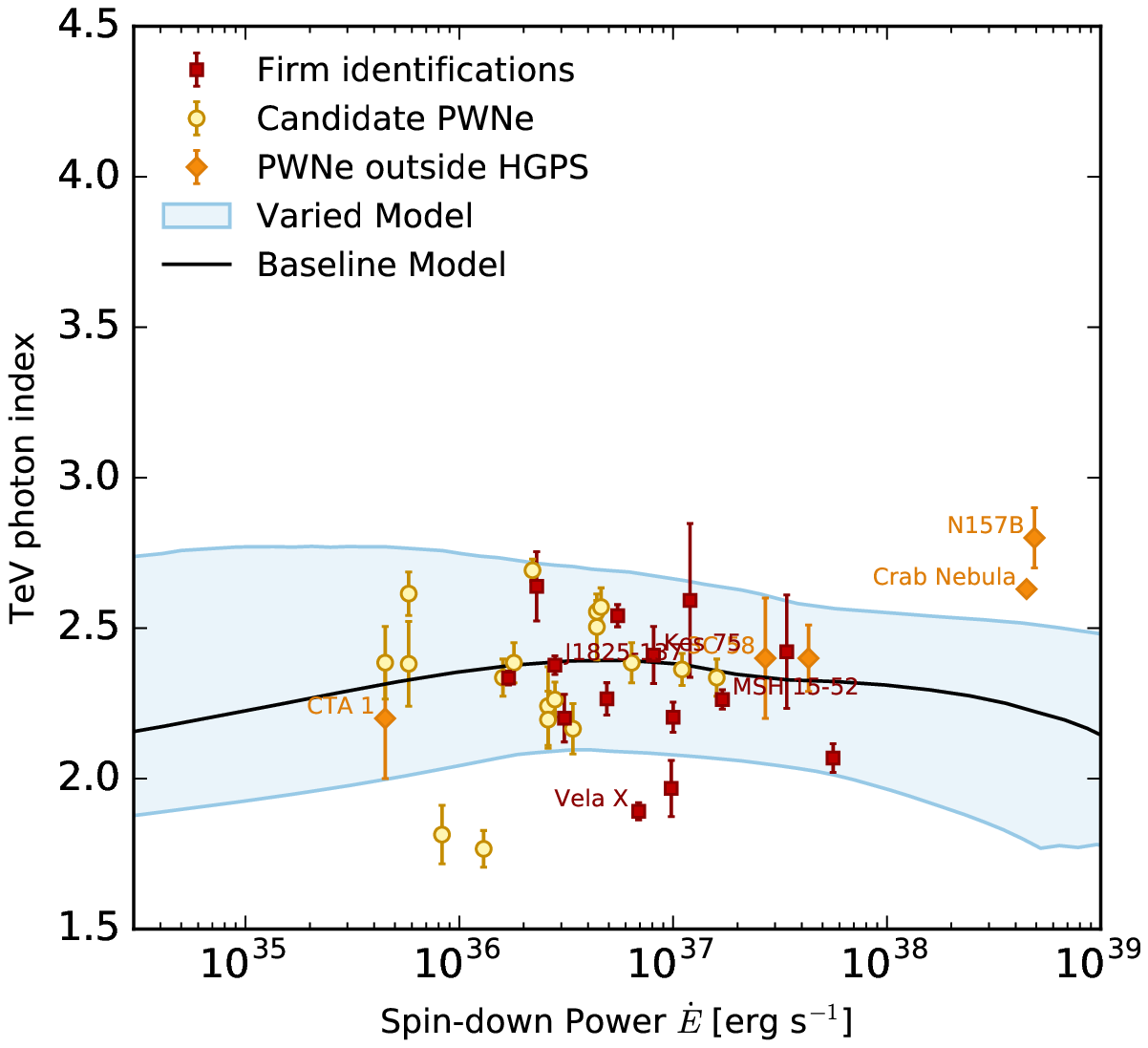}
\caption{
Left: Dependence of surface brightness $S$ (see text) on pulsar spin-down
$\edot$. 
Right: TeV photon index over pulsar spin-down $\edot$ for all detected PWNe
and candidates.
\label{fig:surf_br_index}}
\end{figure*}

Figure~\ref{fig:surf_br_index} (left) shows the dependence of surface brightness on the pulsar's
$\edot$. Like the extension, $S$ can only be calculated for detected systems
and therefore suffers a selection bias expected to become more important with decreasing $\edot$.
Below a spin-down power of $10^{36}\ergs$, the data sample is truncated
at low surface brightness values.

As seen in the \asurv\ fit values in \tref{tab:asurv_results}, a comparably strong 
correlation is found, confirming the above findings of a decreasing luminosity
and increasing extension of ageing pulsar wind nebulae. The measured power-law
relation of $S\sim \edot^{0.81\pm 0.14}$
matches what the model suggests ($\sim 0.9$ for the part where $\edot<2\ttt{37}\ergs$).
We find that the surface brightness gives a strong handle on
the self-consistency of the model because it links the dynamical evolution (i.e.
the extension) to the spectral evolution (i.e. the flux). That is, the scales
of the extension and luminosity evolutions cannot be adjusted independently;
they must lead to a consistent surface brightness scale.

An interesting feature to note is that the scatter suggested by the \scatter\ seems to be much larger
than what is found in the data ($\sigma\tin{\lgt S}\sim 0.3$). This might indicate that flux and extension are not as
independent as implied by a free variation of the respective model
parameters. Another effect might be the missing systematic scatter of
$S$ from the
distance measurements. If the scatter of luminosity and extension measurements were
dominated by the errors on the distance, the \scatter\ shown here would
implicitly include that scatter, and therefore overestimate the actual source-intrinsic scatter. This in turn would lead to a spread of the predicted surface brightness evolution that is too large.

\subsubsection{Photon index}
\label{subsubsec:photonindex}

The average photon index in our sample (firm identifications) is $\sim 2.3$,
and about half\che{2016/06/09} of the PWN indices
deviate significantly from that. Figure~\ref{fig:surf_br_index} (right) shows
the relation of photon index and pulsar $\edot$.

A selection bias  can be expected because non-detections do not appear in the plot and very soft
spectrum
sources are more difficult to detect than hard spectrum sources.

The general range of measured indices ($1.9 \ldots 2.8$) is in accordance with
the model; most of the firm identifications lie in the predicted range of the
\scatter\ or have error bars that are compatible with this model. The precise index is a product of the lepton spectral energy
distribution, in particular of elderly cooled
electrons (see \fref{fig:sed_evolution}, right) and the IC target photon fields, the
combination of which on average seems to be appropriate in our model.
The two exceptions are the Crab nebula and N157B, for which TeV emission is likely
dominated either by IC scattering off their own synchrotron radiation (SSC, for
Crab), or dominated by a very high surrounding photon field (N157B). These
special features are not incorporated in our generic model. 

The peak of the IC emission does not have a clear tendency in our model,
although a mild trend for an increasing peak position seems
to be manifest in \fref{fig:sed_evolution} (left) beyond ages of few kiloyears. 
Also, such a trend is not generally agreed on between different modelling codes. The
MILAGRO and HAWC observations of the ancient Geminga PWN indicate a multi-TeV
nebula \citep{milagro_fermilist, hawc_geminga}, presumably with a high-peaking spectrum, despite its age of $\sim 300\kyr$.
The data discussed in this paper do not allow for a clear statement
here, but show that the trend, if present, is weak.

\begin{figure*}
\includegraphics[width=0.32\textwidth]{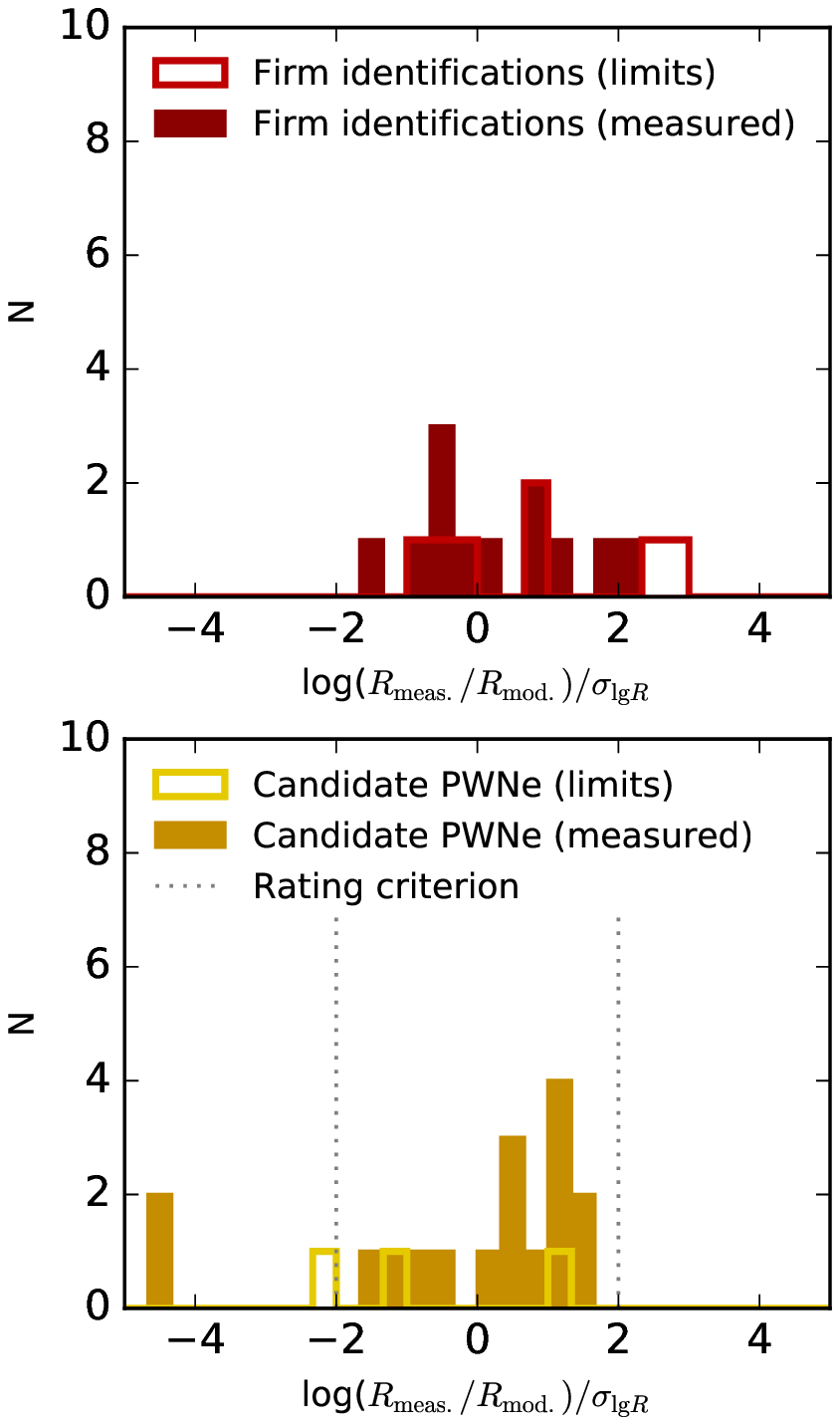}\includegraphics[width=0.32\textwidth]{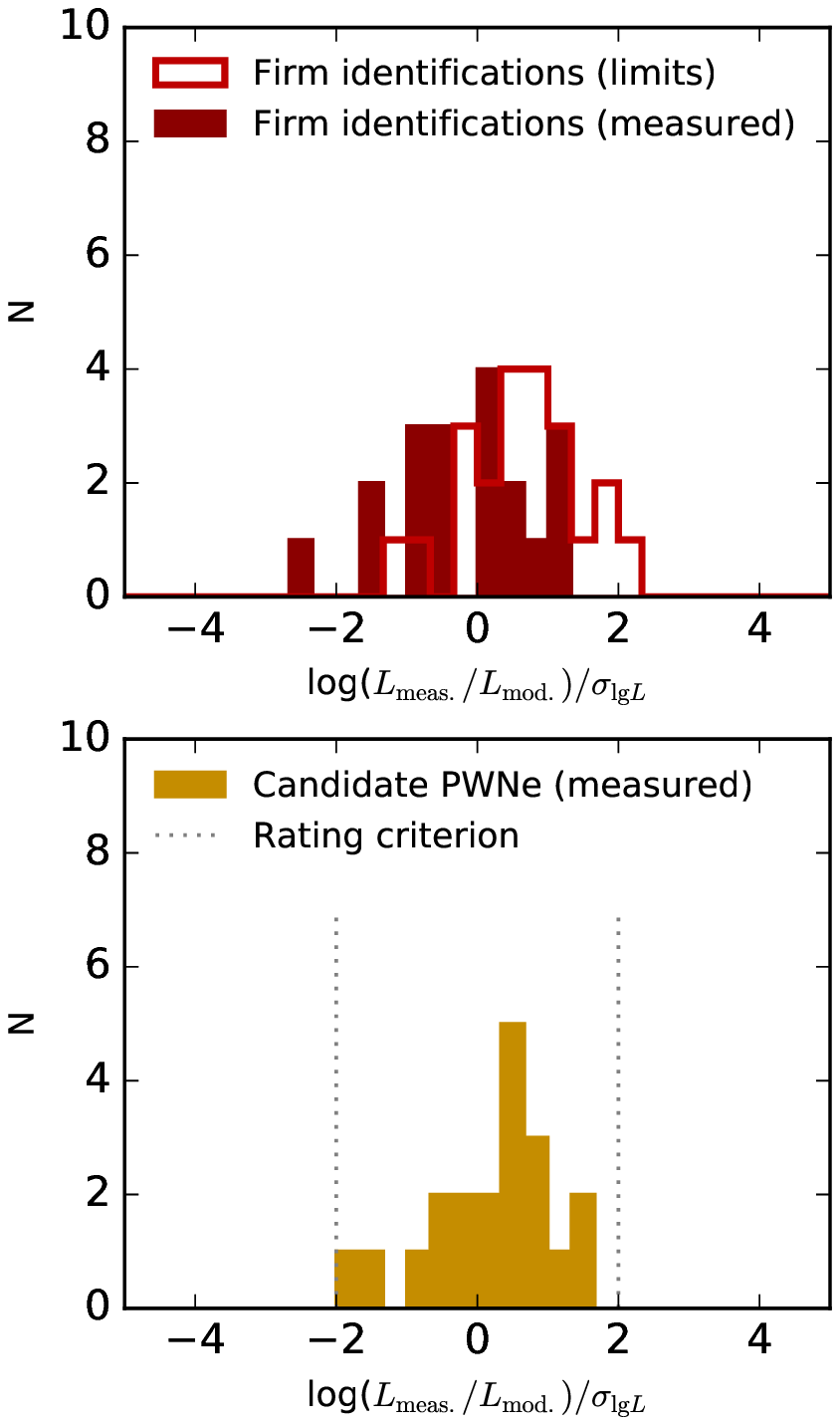}\includegraphics[width=0.32\textwidth]{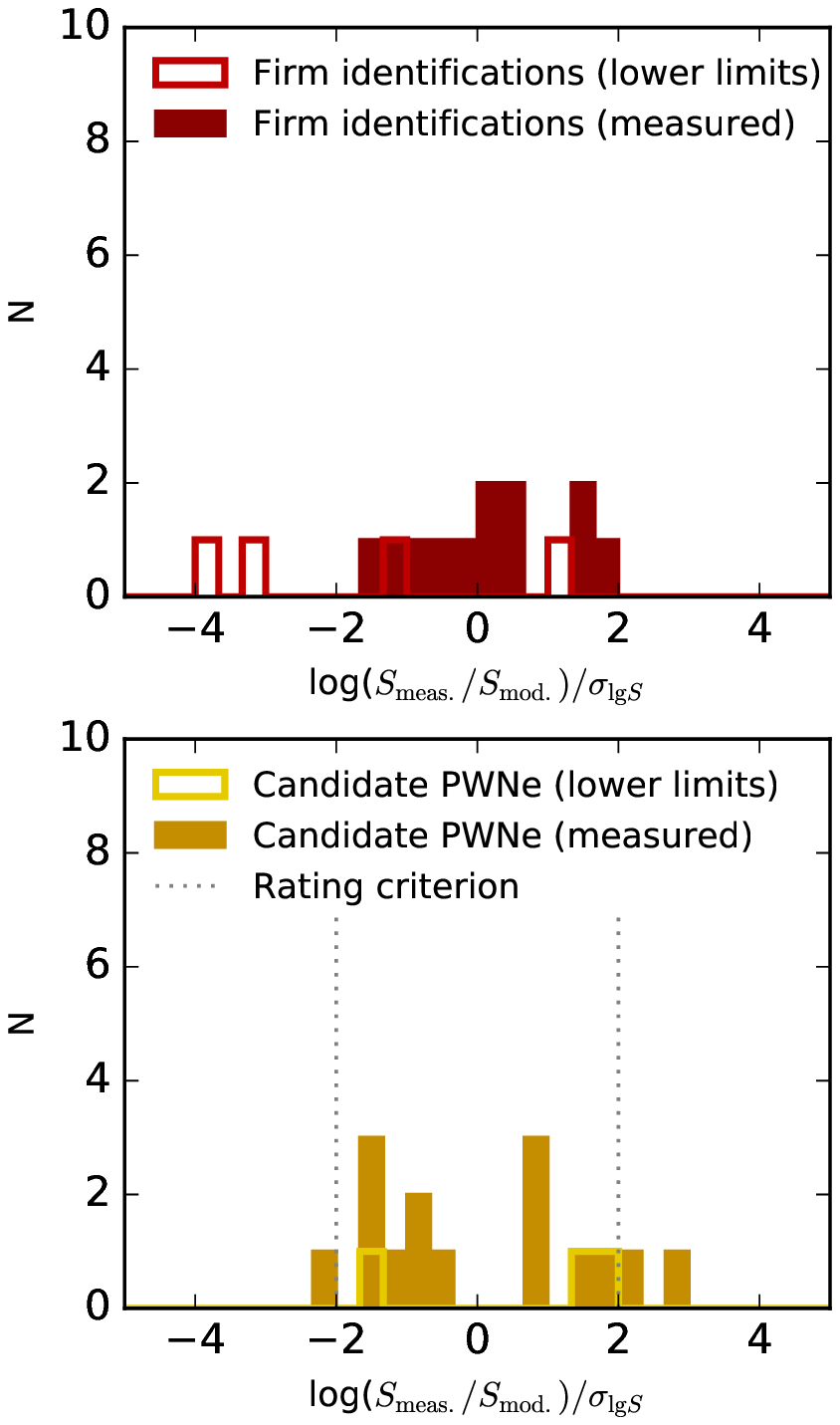}
\caption{
Common logarithmic residuals of rating criteria 2--4, using the standard
deviations $\sigma\tin{\lgt Y}$ explained in \sref{sec:candidate_rating}. Left: Extension with respect to
the model shown in \fref{fig:extension1} (left). Middle: The same
for luminosity (\fref{fig:luminosity}, left) Right: The same
for surface brightness (\fref{fig:surf_br_index}, left).
In all cases, limits are shown separately as outlined histograms.
\label{fig:pulls1}}
\end{figure*}

\section{Review of pulsar wind nebula candidates}
\label{sec:candidate_rating}

In the previous section, the preselection candidate PWNe defined in
\sref{subsec:candidate_selection} were shown along with the firmly
associated PWNe and with average model expectations. Some of the
candidates very consistently lie among other PWNe and close to the model
prediction, while others do not.
In order to compare the candidates among each other, in this section we apply uniform post-selection (``rating") criteria to all of them.

It is important to note that such a rating only evaluates the
plausibility of a given candidate in the context of firmly identified
PWNe or of our model (which is adjusted to the PWNe). Therefore,
a badly rated candidate may
either be an atypical PWN, or an object that contains a PWN alongside a second source (such as a stellar
cluster or SNR), or no PWN at all. Arguments from observations at other
wavelengths are ignored in this uniform approach here since they are not available for all
candidates. Consequently, our rating evaluates the plausibility of a PWN
candidate by how
normal the TeV properties of the PWN candidates are.

We evaluate four criteria: three are comparisons to the model evolution and one concerns the containment of the
pulsar inside the PWN.
Specifically, we apply the following criteria: \che{2016/06/09}

\begin{enumerate}
\item \textbf{Containment ratio} (\fref{fig:offset}, right): The pulsar offset
should be $<1.5$ extension
radii.  
\item \textbf{TeV extension versus age} (\fref{fig:pulls1}, left):
Log-residual from
model (\fref{fig:extension1}, left) should be within $2$ standard deviations, using the measured $\sigma\tin{\lgt R} =
0.39$ (\tref{tab:asurv_results}). 
\item \textbf{TeV luminosity versus pulsar spin-down} (\fref{fig:pulls1}, middle):
Log-residual from
model (\fref{fig:luminosity}, left) should be within $2$ standard deviations,
using the measured $\sigma\tin{\lgt L} =
0.83$ (\tref{tab:asurv_results}). 
\item \textbf{Surface brightness versus pulsar spin-down} (\fref{fig:pulls1}, right):
Log-residual from
model (\fref{fig:surf_br_index}, left) should be within $2$ standard deviations,
using the measured $\sigma\tin{\lgt S} =
0.30$ (\tref{tab:asurv_results}). 
\end{enumerate}

Table~\ref{tab:CAPd2} shows the ratings of the considered candidates. There
are 20\che{2016/06/09} PSR-TeV pairs, in which there are two TeV double associations
(one HGPS source qualifying for two pulsars) and four PSR double associations
(one PSR qualifying for two HGPS sources).

Ten\che{2016/07/22}\ of
the candidate PSR-TeV pairs fulfill all criteria and seem to be plausible TeV pulsar
wind nebula associations.
%
%
All\che{2016/06/09}\ of these candidate pairs have already been
discussed as possible TeV PWNe, namely
HESS~J1616$-$508 and HESS~J1804$-$216 \citep[both in][]{survey2006},  
HESS~J1809$-$193 and HESS~J1718$-$385 \citep[both in][]{two_new_candidates},  
HESS~J1857+026 \citep{j1857},  
HESS~J1908+063 \citep[aka MGRO~J1908+06; e.g.][]{j1908, veritas_j1908},
HESS~J1640$-$465 \citep[PWN hypothesis disfavoured, though]{j1640},  
HESS~J1708$-$443 \citep{j1706},  
HESS~J1023$-$575 \citep[coinciding with massive stellar cluster Westerlund
2,][]{westerlund2}, and HESS~J1018$-$589B \citep[the extended additional
component close to the binary HESS~J1018$-$589A,][]{j1018}.

%
Of the ten\che{2016/07/22} disfavoured candidates, one is an alternative
association for the above strong candidate HESS~J1616$-$508, disfavoured due to its offset.
Similarly, PSR~J1811$-$1925 is a second pulsar in the area of
HESS~J1809$-$193, already argued in \citet{two_new_candidates} to be the less
likely counterpart of the two pulsars that can be considered.
HESS~J1026$-$582 was previously hypothesised to be a PWN, but receives an unfavourable rating due to its
pulsar offset, although the HGPS analysis may not be optimal to reveal the
morphology of this hard-spectrum source \citep{westerlund2}.

The two sources HESS~J1745$-$303 and J1746$-$308, both associated with the very nearby
old PSR~B1742$-$30, are a special case. The pulsar is a factor of 10--100 older
than most other PWNe discussed here, so the extrapolation performed for the rating
cannot be considered to be very robust. In fact, these two objects could not be represented in most of the
figures because they are too far off the axis ranges. They obtain a bad rating mostly because they
are both too underluminous and too small for their age. It could well be,
though, that HESS~J1746$-$308 is a late-phase PWN, created locally near the pulsar after
the main relic PWN bubble has become very faint and/or has dissolved. The predicted
size of the PWN according to our model would be $32\pc$ or $9\dg$ in the sky,
which is impossible to detect with state-of-the-art IACT analysis methods.

In conclusion, about half of the PWN candidates evaluated in this work are viable
PWNe, judging by their TeV and pulsar properties in relation to the population
as such. The number of disfavoured
candidates (10) matches well with the expectation of $\sim10$ chance coincidences
evaluated in \sref{subsec:psr_pwn_correlations}. Hence, it seems plausible 
that most of the ten\che{2016/06/09} high-rated candidates are indeed genuine pulsar wind
nebulae. If this were the case, a total of 25\che{2016/06/09} in
78\che{2016/06/09} HGPS sources would be pulsar wind nebulae (including
G0.9+0.1 here).

\section{Conclusions} \label{sec:discussion}

In this work we subsume and examine the population of TeV pulsar wind nebulae
found to date. The census presents 14\che{2016/06/09} objects reanalysed in the HGPS
catalogue pipeline, which we consider to be firmly identified PWNe, and five\che{2016/06/09} more objects found
outside that catalogue range or pipeline. In addition to those, we conclude that there
are ten\che{2016/06/09} strong further candidates in the HGPS data.
Most of the PWNe are located in the bright and dense Crux Scutum arm of the
inner Milky Way. 
A spatial correlation study 
confirmed the picture
drawn in earlier studies, namely that only
young, energetic pulsars grow TeV pulsar wind nebulae that are bright enough for
detection with presently available Cherenkov telescopes.
For the first time, flux upper limits for undetected PWNe were given
around 22\che{2016/06/09} pulsars with a spin-down power beyond
$10^{35}\ergs$\che{2016/06/09} and with expected apparent extensions (plus
offsets) below
$0.6\dg$ in the sky.

Of the 17\che{2016/06/09} most energetic ATNF pulsars, with a spin-down power
$\edot \geq 10^{37}\ergs$, 11\che{2016/06/09} have either an identified TeV
wind nebula (9\che{2016/06/09}) or candidate (2\che{2016/06/09}) featured in the
present study. Of the remaining 6\che{2016/06/09},
\begin{itemize}
        \item 3\che{2016/06/09} are included in the flux upper limits in \tref{tab:withLIMd2};
        \item 3 are out of the range of the HGPS:
        \begin{itemize}
           \item PSR J2022+3842: SNR~G076.9+01.0, contains an X-ray PWN; not
reported in TeV
           \item PSR J2229+6114: Boomerang, contains an X-ray PWN; detected by MILAGRO and VERITAS, but of
unclear nature in TeV
           \item J0540$-$6919: In the Large Magellanic Cloud; a limit is given
in \citet{hess_lmc_science}. Converting the limit to luminosity yields
$\lumi<5.7\ttt{34}\ergs$, which is compatible with the predicted
$3.3\ttt{34}\ergs$\che{2016/06/09} that can be taken from $\lumi(\edot)$ in \tref{tab:asurv_results}.
        \end{itemize} 
\end{itemize} 
In summary, only 5 of the 17 highest-$\edot$ pulsars remain without a detected
potential counterpart in the TeV band.

Figures~\ref{fig:extension1} to \ref{fig:surf_br_index} showed a variety of
trends between pulsar and TeV wind nebula parameters, and consistently
compared them to a simple one-zone time-dependent emission model of the TeV
emission with a varied range of model input parameters. The
main conclusion of this work is that for several observables, a trend was found in
the data and the trends suggested by our
model are consistent with these findings. With only a moderate variation of the
model input parameters, we can mimic the spreads of the observables, although
the precise value of the parameter ranges is subject to the model caveats
discussed in \sref{subsec:theory_modelling}.
Our first-order understanding of the evolution of TeV pulsar wind
nebulae with ages up to several tens of kiloyears therefore seems to be compatible with what the whole population of detected and
undetected PWNe suggests.

Using the flux limits for undetected PWNe, we find evidence that the TeV luminosity
of PWNe decays with time while they expand in (angular)
size, preventing the detection of those whose pulsar has dropped 
below $\sim 10^{36}\ergs$ (roughly corresponding to several tens of kiloyears). 
This was implicitly known before from the mere non-detection of
old pulsar wind nebulae, but for the first time could be put into a
quantitative perspective here, both by fitting data and limits, and by comparing the
data to model predictions. The power-law relation between TeV luminosity and
pulsar spin-down power could be estimated as $\lumi\sim\edot^{0.58\pm0.21}$,
in consistency with the model that suggests a power index of around $0.5$.

Another feature that was discussed on some individual objects before 
\citep[e.g.][]{hess_j1825_detection, crushing2015} is the ``crushing"
of PWNe, which can be exerted by the inward-bound reverse front of the
supernova shock wave. For SNRs developing asymmetrically, for instance due to
an inhomogeneous surrounding medium (ISM), this crushing may result in considerable
distortion and displacement of the wind nebula. Put to a population-scoped graph
(\fref{fig:offset}, left), it becomes clear that pulsar proper motions
are insufficient to
explain the large offsets observed, which may instead
be due to reverse shock interaction being
a dominant and frequent cause of pulsar-PWN offset in middle-aged systems
\citep[see also][]{okkie_arache_2009}.
Furthermore, the offset appears to relate to high efficiency
(\fref{fig:efficiency2}, right), suggesting that the PWN
either gains energy and brightness through the process that causes the offset or that dense
surroundings amplify both the IC luminosity and the offset between pulsar and
wind nebula. While the evidence for this at present is not very strong, following
up with expanded future studies is certainly worthwhile. 

The expansion of PWNe with time (i.e. rising characteristic age and falling
pulsar spin-down) could also be shown to be evident in the data. The fitted
relation $R\sim\age^{0.55 \pm 0.23}$ suggests an average expansion coefficient
in between those expected theoretically ($1.2$ and $0.3$). The data set is not
comprehensive enough to do a fit with two power laws, but appears to be
consistent with the model. Notably, and in coherence with what was discussed
already in
\citet{aharonian1997}, this expansion is not so clear in X-rays, where the synchrotron
emission always remains very local because it only traces the young particles in areas of high magnetic field
relatively close to the pulsar. Most of the old objects ($>30\kyr)$ in \citet{kargaltsev2013}
are therefore smaller than $1\pc$ in their bright core emission. On the other
hand, in a limited sample of eight PWNe, \citet{bamba} have reported the existence of an additional 
extended and expanding X-ray emission component, which might be the emission
from the particles we see in TeV.

An interesting relation was found between the PWN surface brightness and 
pulsar $\edot$ (\fref{fig:surf_br_index}, left). What stands out is not
only the correlation itself, but also its relatively low scatter. This might either
suggest that luminosity and extension are more correlated than reproduced in
our model (such that a high-luminosity outlier is always balanced by an
accordingly large extension), or it is an indication that the large scatter in
all the other plots is dominated by the distance uncertainty, which is
cancelled out in the surface brightness parameter. If this latter were true,
it would mean that PWNe in fact evolve even more uniformly than suggested by
our \scatter.

The evolution trend of the photon index remains an open issue in this study. Neither the data
nor the model are particularly clear about it for the young to middle-aged PWNe we investigated. A more sensitive
data set -- as expected from CTA \citep{cta} -- will reduce the uncertainties
on spectral indices and reduce the selection bias by detecting more
soft-spectrum PWNe.

Since both the \hgps\ and the ATNF pulsar database only cover a fraction of the
Milky Way, depending on TeV and pulsar brightnesses, this study suffers from
several selection biases discussed throughout the text. For TeV-bright,
high-$\edot$, young pulsar systems ($>10^{36}\ergs$) we achieve a relatively
good coverage, whereas for systems beyond some tens of kiloyears of age we likely
miss many sources. In the plots discussing flux-related quantities, this is
partly compensated by the inclusion of flux limits, allowing for statements
that consider non-detections. For extension- and position-related quantities,
however, we can only rely on the detected cases.
It requires a full
population synthesis study to judge whether some of the correlations are genuine or
include side effects of other correlations or selection biases. Our plots and
fits are meant to draw
attention to where correlations may lurk and we encourage further work on this matter beyond the scope of this
paper.

One presumably very influential parameter ignored in this study is the density
of matter and background light at the position of each pulsar. It is likely
due to such circumstances that \velax, \threec,\ and CTA~1 are so faint, and \lmc\ (in the
Large Magellanic Cloud) is so
bright. In the scope of a population synthesis study, one could use a
specific Milky Way model to ``calibrate" the calorimetric objects that
TeV pulsar wind nebulae are assumed to be.

On the modelling side, we are able to describe the trends and scatter of the
TeV properties of the present
PWN population with a relatively simple time-dependent modelling. Its $12$ free parameters
($7$ of which were varied for the \scatter) were well below the $4\times
19$\che{2016/06/09} observed
parameters that the firmly identified PWNe provided\footnote{All
plotted parameters were derived from the four parameters $P$, $\pdot$, $\lumi$,
and the PWN extension; the TeV offset was not dealt with in the modelling.}.
It is remarkable that the adaptive parameters need to be varied in a
fairly small range, compared to what one may fathom
from the modelling literature\footnote{Even considering only the four papers mentioned in
\sref{subsec:injection}, $\edot_0$ and $\tau_0$ vary there by factors of $250$
and $6$, respectively, compared to $10$ and $1.4$ in our work (see
\tref{tab:model-params}).}, while still producing sufficient scatter in
the predicted observables (even excluding distance uncertainties and target
photon densities as additional factors). Whether this indicates that the underlying 
variations of the individual PWN parameters are indeed small, or
whether this is because the parameters are (anti-)correlated (see
\sref{subsec:theory_modelling}), cannot be clarified in this work. This requires a
deeper physical model of the pulsars and possibly a multidimensional
likelihood fit to correctly quantify all correlations and
identify the true distributions of its parameters.

In the CTA era, most of the PWNe that will be detected in addition to the
now assessed population will be middle-aged and old systems that are too
faint or too extended to be detected with current instruments.
Also, SKA \citep{ska} will enlarge
the sample of pulsars detected in our Galaxy.
To gain new insights from studying these systems, a solid and publicly
available modelling code is needed that includes the
difficult reverse shock interaction phase of a PWN in a reproducible
way. This may help to understand the effect and influence of the amount of
crushing and pulsar offset of the PWN, which is likely an 
influential factor of later PWN evolution.

On the analysis side, it would be beneficial to (i) improve the angular
resolution and get to smaller scales of extension, (ii) find ways to
reliably disentangle overlapping sources and their spectra, and (iii) aim for
detecting objects larger than the IACT camera FOV. The latter
is also of interest because pulsar systems in our Galactic neighbourhood, at few
hundred parsecs from Earth, are considered plausible candidates to strongly contribute
to the cosmic-ray electron and positron fluxes at Earth \citep[e.g.][]{ams_pulsars}.
The CTA cameras
will provide us with a larger FOV \citep{cta}, which improves the capability
of mapping out close-by PWNe. Detecting TeV objects even larger than
that FOV will require better modelling and/or
treatment of the cosmic-ray background event distribution and its systematics
\citep[e.g.][]{spengler, generalized_likelihood}.
In parallel, more generalised analysis
packages with wholistic likelihood approaches \citep{ctools} might help us to unriddle sources that occult each other in
the densely populated arms of the Galaxy.

\section*{Acknowledgements}

We would like to thank Rolf B{\"u}hler for fruitful discussions on the PWN and
Crab subjects. This work made extensive use of the ATNF Pulsar Catalogue
\citep{atnf}, the Python
packages {\tt NumPy}/{\tt SciPy} \citep{numpy, scipy}, and {\tt Matplotlib}
\citep{matplotlib}. The TeVCat Online Catalog for TeV astronomy \citep{tevcat}
was a helpful resource in our literature research.
Blackstars
\ding{72} were used in memory of
David Bowie, starman waiting
in the sky.

The support of the Namibian authorities and of the University of Namibia in facilitating the construction and operation of H.E.S.S. is gratefully acknowledged, as is the support by the German Ministry for Education and Research (BMBF), the Max Planck Society, the German Research Foundation (DFG), the French Ministry for Research, the CNRS-IN2P3 and the Astroparticle Interdisciplinary Programme of the CNRS, the U.K. Science and Technology Facilities Council (STFC), the IPNP of the Charles University, the Czech Science Foundation, the Polish Ministry of Science and Higher Education, the South African Department of Science and Technology and National Research Foundation, the University of Namibia, the Innsbruck University, the Austrian Science Fund (FWF), and the Austrian Federal Ministry for Science, Research and Economy, and by the University of Adelaide and the Australian Research Council. We appreciate the excellent work of the technical support staff in Berlin, Durham, Hamburg, Heidelberg, Palaiseau, Paris, Saclay, and in Namibia in the construction and operation of the equipment. This work benefitted from services provided by the H.E.S.S. Virtual Organisation, supported by the national resource providers of the EGI Federation.

\appendix

\section{Basic modelling of TeV pulsar wind nebulae}
\label{app:modelling}

In the interpretation of the TeV characteristics of the PWN population described in this 
paper we have made use of a time-dependent one-zone model. It allows us
to trace the evolution of the VHE lepton population, and hence the radiative output of a PWN, 
based on a few general assumptions.
The specific model we adopt here
was 
introduced by \citet{mayer_model}, but extended and improved for this work.
Its essential traits are outlined in the following.

\subsection{Spin-down evolution and energy conversion into energetic leptons}
\label{subsec:model-spindown}

The model allows us to calculate the evolution of the non-thermal emission of a PWN
in discrete time steps with an adaptive step size $\delta t$.
In each step, the amount of spin-down energy converted into relativistic 
electrons and positrons is given as
\begin{equation}
\Delta E_\mathrm{p}(t) = \eta \int_t^{t+\delta t} \edot(t')\mathrm{d}t' ,
\end{equation}
where the spin-down evolution $\edot(t)$ of the pulsar
(\eref{eq:edot}, p.~\pageref{eq:edot}) is 
characterised by the braking index $n$, the initial
spin-down timescale $\tau_0$, and the initial spin-down $\edot_0$. The lepton
conversion efficiency $\eta$ can be adjusted to account for
additional cooling effects, but in this work is set to $1$.
This neglects the sub-percent fraction of magnetic energy release that
should technically be missing in the particle outflow, but is negligible here.

While it is possible to transfer the dependency on $\tau_0$ to one on $P_0$ using 
\begin{equation}\label{eq:tau0}
\tau_0 = \frac{2 \age}{n-1} \, \left( \frac{P_0}{P} \right)^{n-1},
\end{equation}
in this work we take $\tau_0$ as the free parameter.

\subsection{Lepton injection spectrum} \label{subsec:model-injection}

For the energy spectrum  of leptons freshly injected into the nebula we assume
the following power-law shape:
\begin{equation}
\frac{\mathrm{d}N_\mathrm{inj}}{\mathrm{d}E}(E,t)=\Phi_0(t)
\left(\frac{E}{1\,\mathrm{TeV}}\right)^{-\beta}
\end{equation}
with a power-law index $\beta$.
$\Phi_0(t)$ can be calculated imposing
\begin{equation}
\Delta E_\mathrm{p}(t) \stackrel{!}{=} 
\int_{E_{\mathrm{min}}}^{E_{\mathrm{max}}}
\frac{\mathrm{d}N_\mathrm{inj}}{\mathrm{d}E}(E,t)\,\mathrm{d}E\,\mathrm{.}
\end{equation}
The lepton energies needed to deliver the relevant X-ray
and gamma-ray energies cover a range of
$E_{\mathrm{min}}$ to $E_{\mathrm{max}}$. Varying the
boundary energies essentially changes the number of particles contained in the 
IC-relevant energy range, and thus the efficiency, but does not fundamentally
change the relative evolution of observables. A low-energy break in the
injection spectrum is often applied in literature (e.g. \citet{torres2014}),
but only impacts the lower ends of the emission spectra. We omit it here
because it neither influences, nor is constrained by our data.

\subsection{Cooling mechanisms} \label{subsec:model-cooling}

Cooling is approximated as
 \begin{equation}
 \frac{\mathrm{d}N_\mathrm{cooled}}{\mathrm{d}E}(E,t)
 =\frac{\mathrm{d}N}{\mathrm{d}E}(E,t-\delta t)\cdot
 \exp\left(-\frac{\delta t}{\tau_{\mathrm{eff}}(E,t)}\right)
,\end{equation}
with an effective cooling timescale
\begin{equation}
\tau_\mathrm{eff}^{-1} = \tau_\mathrm{syn}^{-1} 
+ \tau_\mathrm{esc}^{-1} + \tau_\mathrm{ad}^{-1},
\end{equation}
which comprises synchrotron, escape, and adiabatic losses.
This strategy, as well as the expressions for the first two terms,
are adopted from \citet{zhang_2008}
\begin{eqnarray}
\tau_{\mathrm{syn}}(E,t) & = & 
12.5\cdot\left[\frac{B(t)}{10\,\mathrm{\mu G}}\right]^{-2}          
   \cdot\left[\frac{E}{10\tev}\right]^{-1}\kyr \\
\tau_{\mathrm{esc}}(E,t)  & = & 
34\cdot\left[\frac{B(t)}{10\,\mathrm{\mu G}}\right]         
    \cdot\left[\frac{E}{10\tev}\right]^{-1}  \nonumber \\         
    & & \qquad \cdot\left[\frac{R(t)}{1\,\pc}\right]^{2}\kyr.
\end{eqnarray}
Here, $R(t)$ and $B(t)$ describe the time evolution of the
PWN radius and the magnetic field strength inside the PWN
(cf. \sref{subsec:model-dynamics} below). 
The timescale for adiabatic losses, 
$\tau_\mathrm{ad} = -\frac{E}{\dot{E_\mathrm{p}}}$,
is governed by the expansion of the nebula and can be calculated 
\citep[following][]{deJager92} from
\begin{equation}
\frac{\mathrm{d}E_\mathrm{ad}}{\mathrm{d}t} =
 -\frac{E}{3}\nabla \vec{v}_\mathrm{\bot}(R)=\dot{E}_\mathrm{p},
\end{equation}
with $\vec{v}_\mathrm{\bot}(R)$ being the radial component of the particle
velocity. In general, its divergence can be calculated to
\begin{align}
\nabla\vec{v}_\mathrm{\bot}(R) &=\frac{1}{R^{2}}\cdot\frac{\partial
(R^{2}\vec{v}_\mathrm{\bot})}{\partial R}\\
&=\frac{1}{R(t)^{2}}\cdot\frac{\partial (R(t)^{2}
\vec{v}_\mathrm{\bot}(t))}{\partial t}\cdot\frac{\partial t}{\partial R}
\end{align}
making use of the radial evolution function $R(t)$ given in the next section.
In addition to the above formulation we take into account losses originating from
inverse Compton (IC) emission. This is achieved by subtracting the IC emissivity in each
time step dependent on the electron energy (see~\sref{subsec:model-radiation} for further details on the IC emissivity).

\subsection{Dynamical evolution} \label{subsec:model-dynamics}

In order to take into account that the growth rate of a PWN strongly 
depends on its evolutionary state, the model builds on analytical studies of the 
development of PWNe inside their SNR environment \citep[e.g.][]{chevalier77, 
reynolds84}. The time evolution implemented in the model comprises 
three
\footnote{The original version of the model presented in
\citet{mayer_model} does not incorporate a free expansion phase and 
 uses only two evolutionary stages.} 
distinct phases, which define the expansion behaviour of the PWN according to 
the age of the system in terms of the spin-down timescale $\tau_0$ and the 
reverse-shock interaction time $t_\mathrm{rs}$. Usually, the reverse-shock 
passage and the subsequent reverberations are expected to occur at a time 
$t_\mathrm{rs} > \tau_0$. For this case, the following relations have been 
derived in the aforementioned works:
\begin{equation}
R(t) \quad \propto \quad
        \begin{cases}
                t^{6/5}         &       \text{for $t \leqslant \tau_0$} \\
                t               &       \text{for $\tau_0 < t \leqslant
t_\mathrm{rs}$} \\
                t^{3/10}        &       \text{for $t > t_\mathrm{rs}$}.
        \end{cases}
\end{equation}
In the (supposably much less common) opposite case, $t_\mathrm{rs} < \tau_0$, 
the time evolution of $R_\mathrm{PWN}$ is modified to
\begin{equation}
R(t) \quad \propto \quad
        \begin{cases}
                t^{6/5}                 &       \text{for $t \leqslant
t_\mathrm{rs}$} \\
                t^{11/15}       &       \text{for $t_\mathrm{rs} < t \leqslant
\tau_0$} \\
                t^{3/10}                &       \text{for $t > \tau_0$}.
        \end{cases}
\end{equation}

As a simplification, the crushing of the PWN by the
SNR reverse shock is not modelled here. 
Such crushing presumably reduces the radius between free expansion and reverse
shock interaction phase.

The magnetic field evolution is adapted from \citet{zhang_2008} as
\begin{equation} \label{eq:b-field}
B(t) =\frac{B_0}{1+\left(\frac{t}{\tau_0}\right)^{\alpha}}+B_\mathrm{ISM},
\end{equation}
assuming a constant and homogeneous ISM contribution of $3\,\mathrm{\mu G}$, 
and adopting an index of $\alpha=0.6$ in order to satisfy the conservation of 
magnetic flux.

\subsection{Time-dependent lepton energy distribution and radiative processes}
\label{subsec:model-radiation}

The framework laid out in the previous sections allows us to calculate
the energy distribution of the leptons contained in the PWN at any given 
time. More specifically, the number of leptons with energy $E$ residing in
the nebula at a time $t + \delta t$ is determined by the balance of 
freshly injected leptons and those cooled out of the respective energy 
interval,
\begin{equation}
\frac{\mathrm{d}N}{\mathrm{d}E}(E,t+\delta t) = 
\frac{\mathrm{d}N_\mathrm{cooled}}{\mathrm{d}E}(E,t) +  
\frac{\mathrm{d}N_\mathrm{inj}}{\mathrm{d}E}(E,t+\delta t) .\label{eq:amount}
\end{equation}
The iterative evaluation of \eref{eq:amount} then yields the lepton energy distribution
as a function of time. The time binning is adjusted adaptively to
guarantee high precision at a still reasonable computation cost (see
\citet{mayer_diploma}, Sect.~5.2.2. and Fig.~5.8 for details on this).

From the lepton distribution, the photon population arising from
synchrotron emission and inverse Compton scattering as the most important
processes can be obtained. The physics of these processes is described in the
comprehensive review article by \cite{blumenthal70}, which we follow in the
implementation of the radiation mechanisms within our model. The target
photon fields required as an input for IC scattering are CMB, starlight, and infrared
photons. While the uniform CMB component is modelled as a black-body spectrum with an
energy density of 0.26\,eV\,cm$^{-3}$ and temperature of 2.7\,K, the starlight and infrared components can be adopted from the  {\sc Galprop} code \citep{galprop}. In order to derive a representative radiation field composition for the baseline model, the {\sc Galprop} fields at the positions of all firmly identified PWNe were averaged, using the mean temperature and energy densities as input for the respective black-body spectra.
Following this set-up, the energy densities of the starlight and infrared
fields are 1.92\,eV\,cm$^{-3}$ and 1.19\,eV\,cm$^{-3}$, respectively.
The temperatures at the spectral peaks are 107\,K for the infrared and 7906\,K
for the starlight field component.

\subsection{Results of the time-dependent modelling}
\label{subsec:model-results}
        
In summary, the model takes the parameters listed in \tref{tab:model-params}.
The table contains two compilations of parameters: the first one
states the values used for the \model, which is\ depicted as a black line throughout the
population plots in this paper; the second one gives the ranges of parameters
we used to mimic the intrinsic spread of the PWN properties. The PWN
evolution implied by our \model\ is listed in \tref{tab:model-evolution}.

The considerations that went into the choice of the model parameters and
ranges are the following:
\begin{itemize}
\item We want to mimic a typical PWN in a typical (dense spiral arm) surrounding. 
For this reason, we do not give objects like \velax, \threec, or \ctaone\ too much
consideration in the adjustment of the parameters. This can make the model
differ from the fit results, which take all objects into account.
\item $n$: The braking index defines the slope of the pulsar trajectory on
\fref{fig:edot_age}, which has to be $\sim 3$--$4$ to match the measured pulsar
population. The theoretical
expectation is that $n=3$ if the energy loss is dominated by magnetic dipole
radiation,
whereas a spin-down dominated by gravitational
radiation leads to a longer energy release through $n=5$
\citep[e.g.][]{yue_xu_zhu}. By contrast, the few direct measurements of
braking indices presently available lie in the range of
$0.9$--$2.8$ \citep[for a compilation see][]{braking_indices}, indicating a
much faster spin-down decay. In this study, we set it to the canonical $n=3$.
\item $\tau_0$, $\edot_0$: These parameters define the starting point of the
pulsar trajectory on \fref{fig:edot_age} and the total energy budget of the
pulsar (see \aref{app:formulae_derivation} and \fref{fig:edot_age}). With the chosen combination and
the canonical $n=3$, the pulsar energy outflow evolves along the path where
ATNF pulsars are actually found and starts with a total energy of
$\edot_0\tau_0 = 3.1\ttt{49}\erg$.
\item $B_0$: The initial B-field is chosen such that, using \eref{eq:b-field}
for its decay,
Crab-like young PWNe have (present) fields $B\tin{PWN} \sim 100\eh{\mu G}$, while older objects
at some point arrive at few tens of $\mu$G or less. This is consistent
with the ranges found in other modelling works, such as
\citet{torres2014} and \citet{zhang_2008}, in which this scale is set with the goal of
producing realistic X-ray luminosities.
\item $t_{\mathrm{rs}}$, $R_{5}$: These parameters determine the dynamical
evolution and are set such that the PWN extension trajectory evolves roughly through
the middle of the firmly identified PWNe. They also have strong influence on
the surface brightness plot \fref{fig:surf_br_index} (left), which interlinks
them with the luminosity related parameters. 
\item $B\tin{ISM}$: Set to the canonical $3\eh{\mu G}$.
\item $\eta$: The lepton efficiency can account for a substantial fraction of
energy going into magnetic fields or hadron acceleration, neither of which we
assume to be large. Hence, we set $\eta=1$.
\item $\alpha$: Set to $0.6$ in order to satisfy the conservation of
magnetic flux. 
\item $\beta$: An injection index of $2$ is a typical value found to lead to good agreement
with observed spectral indices here and in other works. The variation we
induce produces a realistic variation of gamma-ray photon indices.
\item  $E_{\mathrm{min}}$, $E_{\mathrm{max}}$: These energy bounds mainly
determine the ranges of the synchrotron and IC photon spectra, and therefore also the
amount of photons produced specifically in the $1$--$10\tev$ band considered
here. They are not constrained by our plots beyond this efficiency variation
they can provoke.
\end{itemize}

\begin{figure*}[!tb]
\includegraphics[width=0.494\textwidth]{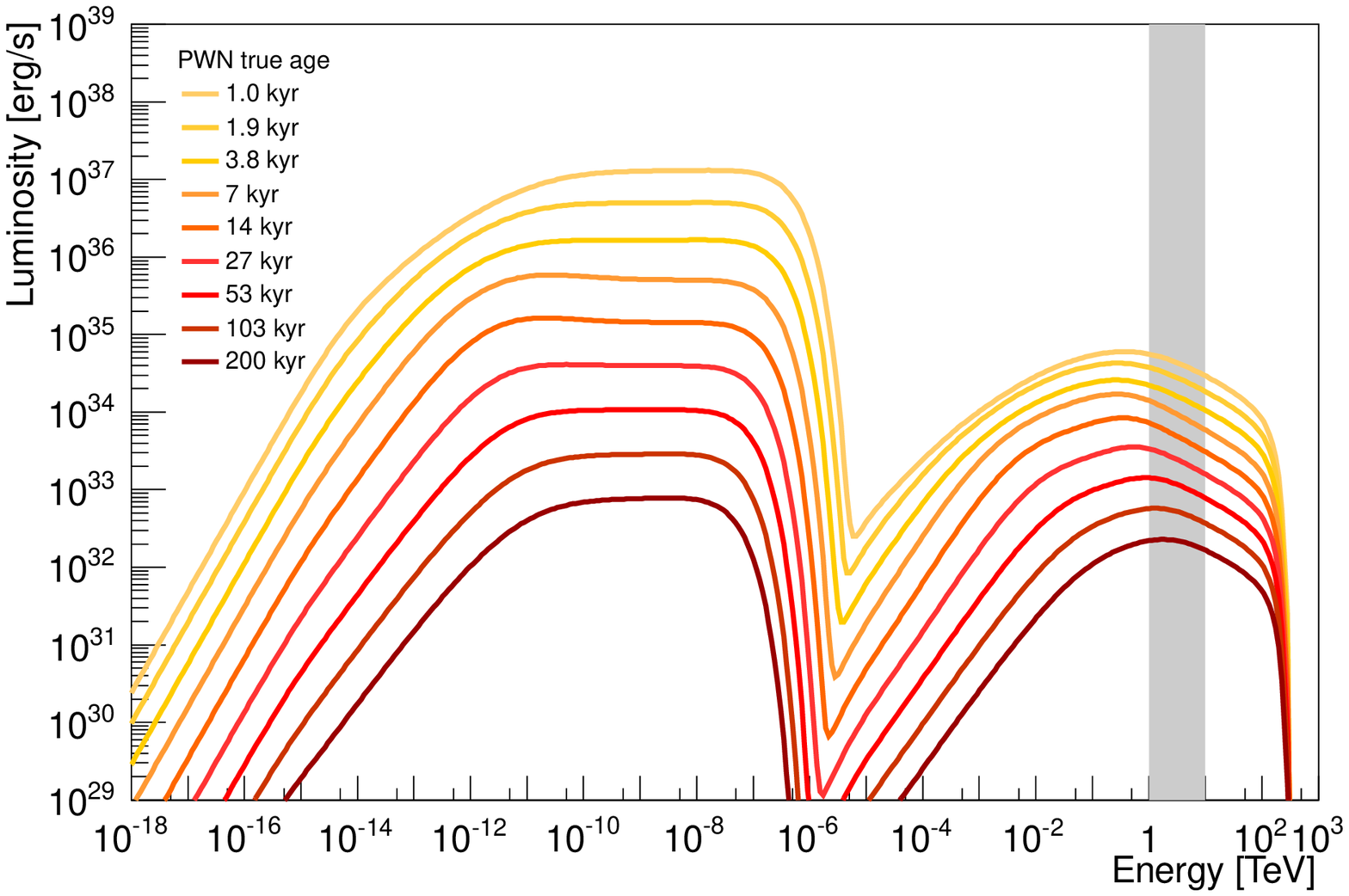}\includegraphics[width=0.494\textwidth]{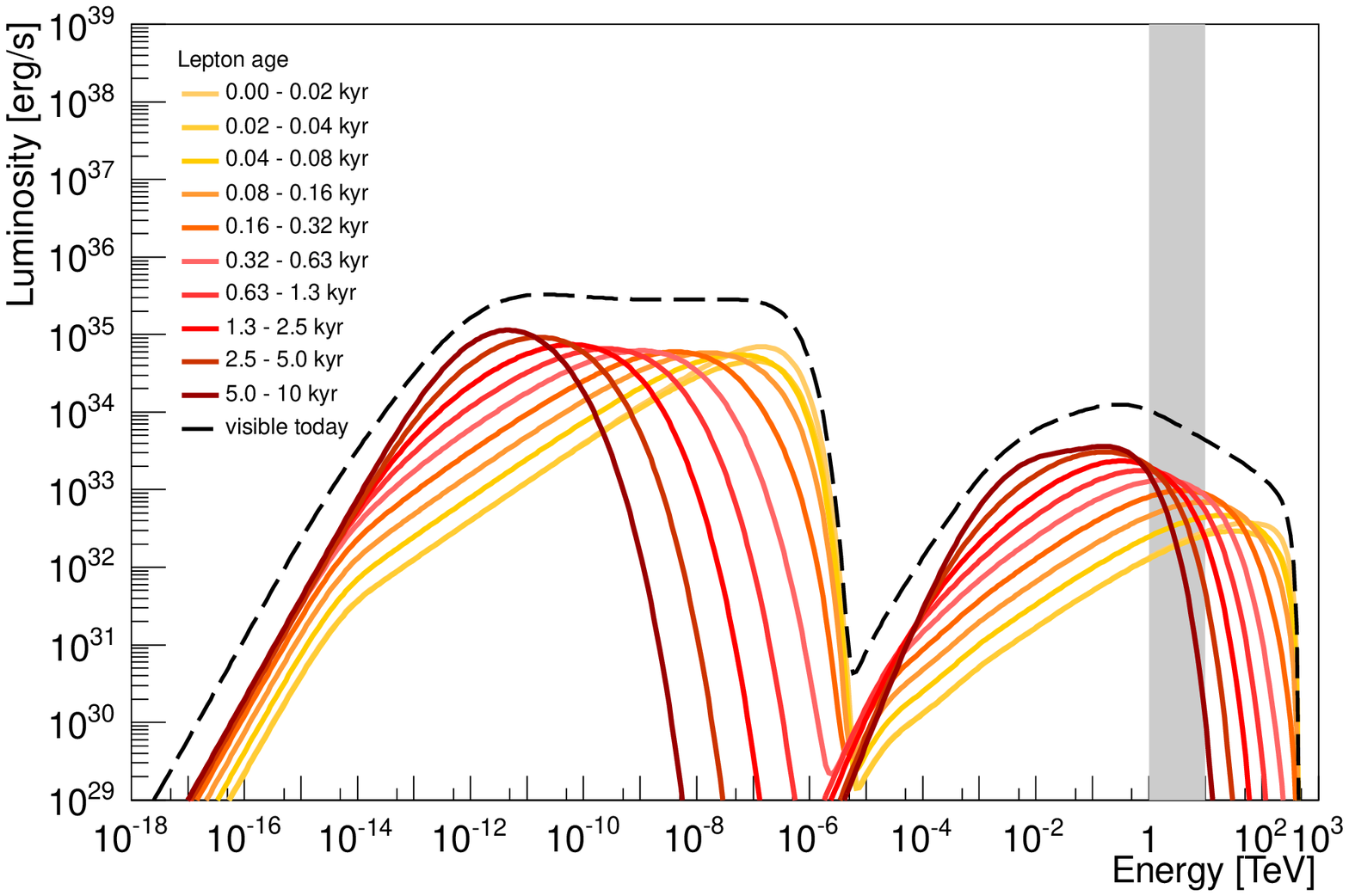}
\caption{Modelled spectral energy distribution (SED) of a generic PWN with
parameters according to \model\ given in \tref{tab:model-params}. See
\sref{subsec:modelling_caveats} for caveats of the SEDs.
Left: Time evolution of the SED, ranging from $1\kyr$ to $200\kyr$.
Right: Decomposition of the SED of a 
middle-aged PWN ($10\kyr$; black dashed curve) into contributions
by leptons from various injection epochs (coloured lines).
The grey-shaded bands indicate the energy range of $1$--$10\tev$ explored in this
paper.
\label{fig:sed_evolution}
}
\end{figure*}

The parameter set of the \model\ was also used to construct the spectral
energy distributions (SEDs) shown in \fref{fig:sed_evolution}. 
These sample SEDs illustrate the time evolution of the radiative output of a generic PWN
according to the presented model. \\
The set of model curves in \fref{fig:sed_evolution} (left) 
traces the various evolutionary stages of the energy flux at PWN ages 
ranging between $0.5\kyr$ and $150\kyr$, calculated
in equidistant steps on a logarithmic timescale. Even though both the 
synchrotron and IC contributions obviously undergo significant 
development with increasing age of the system, the decline of the synchrotron
energy flux (due to its strong dependence on the decaying magnetic field
strength) is more pronounced than that of the IC component.\\ 
Figure~\ref{fig:sed_evolution} (right) depicts the SED of a generic
middle-aged PWN decomposed into numerous contributions from
individual epochs. The dominance of the very youngest leptons in
producing the synchrotron component (most notably the X-ray part) 
is manifest in this plot.
By contrast, accumulated leptons from various ages contribute to
the IC radiation, in particular in the TeV energy band.

\newcommand{\mydots}{\ensuremath{\,\ldots}\;}

\begin{table*}
\caption{Overview of parameters used for the modelling and the calculation of
\scatter\ ranges\che{2016/06/08}}
\label{tab:model-params}
\centering
\tiny
\begin{tabular}{lllcc}
\hline\hline
\multicolumn{3}{c}{Parameter description} & \multicolumn{2}{c}{Parameter
values} \\
 & & & \model\ & \scatter \\
\hline
Braking index                             & $n$                &                 &  3.0  &  2.5  \mydots  3.5 \\
Initial spin-down power                   & $\edot_0$          & ($10^{39}\ergs$)&  2.0  &  1.0  \mydots  4.0 \\
Initial spin-down timescale               & $\tau_0$           & ($\kyr$)        &  0.5  &  0.32 \mydots  0.77 \\
Initial magn. field strength              & $B_0$              & ($\mu$G)        &  200  &  110  \mydots  270 \\
Reverse shock interaction timescale       & $t_{\mathrm{rs}}$  & ($\kyr$)        &  4.0  &  4.0  \mydots  8.0  \\
PWN radius at $t = 3\kyr$                 & $R_{3}$            & (pc)            &  6.0  &  3.0  \mydots  12.0 \\
Adopted const. ISM magn. field strength   & $B\tin{ISM}$       & ($\mu$G)        &  3.0  &  3.0 \\
Lepton conversion efficiency              & $\eta$             &                 &  1.0  &  1.0 \\
Index of magn. field evolution            & $\alpha$           &                 &  0.6  &  0.6 \\
Index of lepton injection spectrum        & $\beta$            &                 &  2.0  &  1.75 \mydots 2.25 \\
Lower bound of lepton energy distribution & $E_{\mathrm{min}}$ & (TeV)           &  0.03 &  0.03 \\
Upper bound of lepton energy distribution & $E_{\mathrm{max}}$ & (TeV)           &  300  &  300 \\
\hline                                   
\end{tabular}
\end{table*}

\begin{table*}
\caption{Evolution of a PWN in our \model.}
\label{tab:model-evolution}
\centering
\tiny
\begin{tabular}{lllllll}
\hline\hline
\multicolumn{3}{c}{Pulsar} & \multicolumn{4}{c}{PWN} \\
$t$ & $\age$ & $\edot$          & $B\tin{PWN}$ & $R\tin{PWN}$ & $\lumi$          & $\Gamma$ \\
(kyr)      & (kyr)  & ($10^{38}\ergs$) & ($\mu$G)      & (pc)           & ($10^{33}\ergs$)   &  \\
\hline
0.10 & 0.60 &  1.39$\times 10^{39}$ & 148 & 0.142 & 1.27$\times 10^{35}$ & 2.08 \\
0.14 & 0.63 &  1.23$\times 10^{39}$ & 140 & 0.207 & 1.36$\times 10^{35}$ & 2.11 \\
0.19 & 0.69 &  1.05$\times 10^{39}$ & 131 & 0.316 & 1.41$\times 10^{35}$ & 2.14 \\
0.26 & 0.76 &  8.63$\times 10^{38}$ & 122 & 0.458 & 1.42$\times 10^{35}$ & 2.17 \\
0.36 & 0.85 &  6.78$\times 10^{38}$ & 113 & 0.665 & 1.37$\times 10^{35}$ & 2.19 \\
0.49 & 0.99 &  5.07$\times 10^{38}$ & 103 & 0.971 & 1.28$\times 10^{35}$ & 2.22 \\
0.67 & 1.17 &  3.61$\times 10^{38}$ & 94.0 & 1.34 & 1.16$\times 10^{35}$ & 2.25 \\
0.92 & 1.42 &  2.44$\times 10^{38}$ & 84.6 & 1.84 & 1.01$\times 10^{35}$ & 2.28 \\
1.27 & 1.77 &  1.58$\times 10^{38}$ & 75.6 & 2.54 & 8.36$\times 10^{34}$ & 2.30 \\
1.74 & 2.24 &  9.82$\times 10^{37}$ & 67.0 & 3.49 & 6.72$\times 10^{34}$ & 2.31 \\
2.40 & 2.89 &  5.89$\times 10^{37}$ & 59.0 & 4.79 & 5.27$\times 10^{34}$ & 2.32 \\
3.29 & 3.79 &  3.44$\times 10^{37}$ & 51.7 & 6.58 & 4.04$\times 10^{34}$ & 2.33 \\
4.52 & 5.02 &  1.96$\times 10^{37}$ & 45.0 & 8.30 & 3.19$\times 10^{34}$ & 2.35 \\
6.21 & 6.71 &  1.10$\times 10^{37}$ & 39.0 & 9.13 & 2.46$\times 10^{34}$ & 2.38 \\
8.53 & 9.03 &  6.05$\times 10^{36}$ & 33.7 & 10.0 & 1.84$\times 10^{34}$ & 2.39 \\
11.7 & 12.2 &  3.30$\times 10^{36}$ & 29.1 & 11.0 & 1.35$\times 10^{34}$ & 2.39 \\
16.1 & 16.6 &  1.79$\times 10^{36}$ & 25.1 & 12.2 & 9.71$\times 10^{33}$ & 2.38 \\
22.1 & 22.6 &  9.63$\times 10^{35}$ & 21.6 & 13.4 & 6.92$\times 10^{33}$ & 2.35 \\
30.4 & 30.9 &  5.16$\times 10^{35}$ & 18.6 & 14.7 & 4.87$\times 10^{33}$ & 2.32 \\
41.8 & 42.2 &  2.76$\times 10^{35}$ & 16.1 & 16.2 & 3.39$\times 10^{33}$ & 2.29 \\
57.4 & 57.9 &  1.47$\times 10^{35}$ & 13.9 & 17.8 & 2.32$\times 10^{33}$ & 2.25 \\
78.8 & 79.3 &  7.84$\times 10^{34}$ & 12.1 & 19.6 & 1.58$\times 10^{33}$ & 2.21 \\
108 & 109 &  4.17$\times 10^{34}$ & 10.6 & 21.5 & 1.06$\times 10^{33}$ & 2.17 \\
149 & 149 &  2.21$\times 10^{34}$ & 9.33 & 23.7 & 6.96$\times 10^{32}$ & 2.14 \\
204 & 205 &  1.17$\times 10^{34}$ & 8.26 & 26.0 & 4.53$\times 10^{32}$ & 2.11 \\
281 & 281 &  6.23$\times 10^{33}$ & 7.37 & 28.6 & 2.91$\times 10^{32}$ & 2.08 \\
386 & 386 &  3.30$\times 10^{33}$ & 6.62 & 31.5 & 1.83$\times 10^{32}$ & 2.05 \\
530 & 530 &  1.75$\times 10^{33}$ & 6.00 & 34.6 & 1.14$\times 10^{32}$ & 2.03 \\
728 & 728 &  9.29$\times 10^{32}$ & 5.49 & 38.1 & 7.03$\times 10^{31}$ & 2.00 \\
1000 & 1000 &  4.92$\times 10^{32}$ & 5.06 & 41.9 & 4.26$\times 10^{31}$ & 1.98 \\
\hline                                   
\end{tabular}
\tablefoot{$t$ is the true age of the pulsar, $\age$ its characteristic age, $\edot$
its spin-down luminosity. $B\tin{PWN}$ is the magnetic field in the PWN,
$R\tin{PWN}$ the PWN radius, $\lumi$ the TeV luminosity, and $\Gamma$ the
gamma-ray index between $1$ and $10\tev$.}
\end{table*}

\subsection{Caveats} \label{subsec:modelling_caveats}

As already emphasised in \sref{subsec:theory_modelling} and
\ref{subsec:model-injection}, the aim of this model is to serve for the
interpretation of the TeV data we have. Spectral breaks, potential
reverberation compressions, and other aspects that cannot be judged with the
present data are therefore omitted on purpose. The multi-wavelength spectra it
predicts, though found in the right order of magnitude, may therefore not be very
accurate at energies other than the TeV regime.

Another caveat to note is the correlation of parameters. We vary only 7 of the
12 parameters (the target photon field could additionally be regarded as a 13th
parameter), but the variations in the model can of course also be achieved by
varying more of the parameters by a smaller magnitude. A variation of
$\edot$ is for instance
indistinguishable, from the point of view of the TeV properties, from a variation of lepton efficiency. So
the variation solution we found leads to a sensible range in predicted
observable ranges, but is not unique. Similarly, a correlation of two
parameters can mean that larger variations are possible, such as in the example
described in \sref{subsec:theory_modelling}. 

\section{Derivation of basic formulae around the relation of $\edot$ and $\age$}\label{app:formulae_derivation}

Since the following relations are relatively fundamental to the energy input
evolution of PWNe, but still rather hard to find in recent literature, we
briefly want to wrap up what Eqs.~\ref{eq:edot}--\ref{eq:edotvsage} and
\ref{eq:tau0} are derived from.

As pointed out by \citet{gunn_dipole_radiation}, the energy loss rate of a
rotating magnetic dipole depends on the angular velocity  $\Omega$ as 
\begin{equation}\label{edot_omega_dipole}
\edot = -k'\,\Omega^4.
\end{equation}
Since the angular momentum loss rate is
\begin{equation}
\dot{J} = \frac{\edot}{\Omega}=k'\,\Omega^3
\end{equation}
it follows that the velocity loss rate is
\begin{equation}
\dot{\Omega}=\frac{\dot{J}}{I}=-k\,\Omega^3
,\end{equation}
where $I$ is the neutron star moment of inertia.
To generalise this relation for the non-dipole
case, the index $3$ is replaced by the braking index $n$,
\begin{equation}\label{eq:omega}
\dot{\Omega}=\frac{\dot{J}}{I}=-k\Omega^n,
\end{equation}
which turns \eref{edot_omega_dipole} into 
\begin{equation}\label{eq:edot_omega}
\edot = -k'\,\Omega^{n+1}.
\end{equation}
The general solution of the differential equation \eref{eq:omega}
can be written as
\begin{equation}
\Omega(t)=\Omega_0\left(1+\frac{t}{\tau_0}\right)^{-\frac{1}{n-1}}.
\end{equation}
Using \eref{eq:edot_omega}, and $P=2\pi/\Omega$, and differentiating $P$ 
one obtains
\begin{eqnarray} \label{eq:edot_appendix}
\edot(t)   =& \edot_0\left(1+\frac{t}{\tau_0}\right)^{-\frac{n+1}{n-1}} \\
P(t)       =& P_0\left(1+\frac{t}{\tau_0}\right)^{\frac{1}{n-1}}
\label{eq:p_evol}\\
\pdot(t)   =& \frac{P_0}{\tau_0
(n-1)}\left(1+\frac{t}{\tau_0}\right)^{-\frac{n-2}{n-1}}. \label{eq:pdot_evol}
\end{eqnarray}
The canonical formulae to calculate $\edot$ and $\age$ from $P$ and $\pdot$
 then yield 
\begin{eqnarray}
\edot(t)       = & 4\pi^2 I\frac{\pdot(t)}{P(t)^3} & =  \frac{4\pi^2
I}{\tau_0\,P_0^2\,(n-1)}\left(1+\frac{t}{\tau_0}\right)^{-\frac{n+1}{n-1}} \\
\age(t)   \equiv & \frac{P(t)}{2\pdot(t)}          & =  \frac{n-1}{2}(t+\tau_0)
\end{eqnarray}
(cp.~the notation in \cite{gs}, Eqs.~5 and 6). Note that neither of these expressions
 relies on the dipole hypothesis of $n=3$. 
At the birth of the pulsar,
$t=0$, $\age$ is $\tau_0\,(n-1)/2$ and increases steadily.

For the relation of $\pdot$ and $P$, Eqs.~\ref{eq:p_evol} and \ref{eq:pdot_evol}
furthermore imply
\begin{equation} \label{eq:pdot_vs_p}
\pdot(P) = \frac{P_0}{\tau_0} \frac{1}{n-1} \left(\frac{P}{P_0}\right)^{2-n}
,\end{equation}
which can be taken to discuss plausible braking indices directly from
\fref{fig:edot_age} (right).

In order to see what happens if $\edot$ is plotted against $\age$, one has to
resolve the dependency on $t$ to arrive at
\begin{equation}
\edot(\age)=\edot_0
\left(\frac{2}{n-1}\frac{\age}{\tau_0}\right)^{-\frac{n+1}{n-1}}.
\end{equation}
Clearly, the evolution curve of a pulsar on the $\edot$-$\age$ diagram starts
at a point $[\tau_0\,(n-1)/2, \edot_0]$, which depends on $\tau_0$, $n$, and
$\edot_0$, but the slope of the power law is only dictated
by the braking index $n$. This index is not predetermined to be $3$ by the way $\age$ is
constructed.

Assuming that \eref{eq:edot_appendix} describes the energy outflow of the
pulsar throughout its lifetime, one can calculate the energy deposited up to a
certain time as follows:
\begin{eqnarray}
E\tin{dep}(t) & = & \int_0^t \edot(t') \mathrm{d}t' \\
              & = & \frac{n-1}{2}\, \edot_0\, \tau_0 - \edot(t)\,\age(t)
\end{eqnarray}
For $t\rightarrow\infty$, $\edot(t)\,\age(t)$ vanishes, so the first term represents the total energy budget that is emitted and, using \eref{eq:edot_appendix}, can be made equivalent to
$I\,\Omega_0^2/2$, the total rotational energy of the pulsar. Unfortunately,
$n$, $\edot_0$, and $\tau_0$ are three unknown initial
properties of the pulsar, so it cannot be measured. Unlike that, the second term $\edot\,\age$, which represents the
present budget of rotational energy, can be calculated from the measured $P$
and $\pdot$. The ordinary (low-aged) pulsar with the maximum present budget of energy is PSR~J0537$-$6910
in \lmc, with $7.6\ttt{49}\erg$, which is a lower limit to the maximal initial
rotational energies that can be reached.

\bibliographystyle{aa}
\bibliography{pwnpop}

\end{document}

%% file: table2d2.tex
\begin{table*}
\centering
\tiny
\caption{\label{tab:rePAPd2}HGPS sources considered as firmly identified pulsar wind nebulae in this paper.}
\begin{tabular}{llllllllll}
\hline\hline
HGPS name & ATNF name & Canonical name & $\lg\edot$ & $\tau_\mathrm{c}$ & $d$ & PSR offset & $\Gamma$ & $R_{\mathrm{PWN}}$ & $L_{\mathrm{1-10\,TeV}}$ \\
 &  &  &  & (kyr) & (kpc) & (pc) &  & (pc) & ($10^{33}\,\mathrm{erg\,s^{-1}}$) \\
J1813$-$178$^{[1]}$ & J1813$-$1749 &  & 37.75 & 5.60 & 4.70 & $<2$ & $2.07 \pm 0.05$ & $4.0 \pm 0.3$ & $19.0 \pm 1.5$ \\
J1833$-$105 & J1833$-$1034 & G21.5$-$0.9$^{[2]}$ & 37.53 & 4.85 & 4.10 & $<2$ & $2.42 \pm 0.19$ & $<4$ & $2.6 \pm 0.5$ \\
J1514$-$591 & B1509$-$58 & MSH 15$-$52$^{[3]}$ & 37.23 & 1.56 & 4.40 & $<4$ & $2.26 \pm 0.03$ & $11.1 \pm 2.0$ & $52.1 \pm 1.8$ \\
J1930+188 & J1930+1852 & G54.1+0.3$^{[4]}$ & 37.08 & 2.89 & 7.00 & $<10$ & $2.6 \pm 0.3$ & $<9$ & $5.5 \pm 1.8$ \\
J1420$-$607 & J1420$-$6048 & Kookaburra (K2)$^{[5]}$ & 37.00 & 13.0 & 5.61 & $5.1 \pm 1.2$ & $2.20 \pm 0.05$ & $7.9 \pm 0.6$ & $44 \pm 3$ \\
J1849$-$000 & J1849$-$0001 & IGR J18490$-$0000$^{[6]}$ & 36.99 & 42.9 & 7.00 & $<10$ & $1.97 \pm 0.09$ & $11.0 \pm 1.9$ & $12 \pm 2$ \\
J1846$-$029 & J1846$-$0258 & Kes 75$^{[2]}$ & 36.91 & 0.728 & 5.80 & $<2$ & $2.41 \pm 0.09$ & $<3$ & $6.0 \pm 0.7$ \\
J0835$-$455 & B0833$-$45 & Vela X$^{[7]}$ & 36.84 & 11.3 & 0.280 & $2.37 \pm 0.18$ & $1.89 \pm 0.03$ & $2.9 \pm 0.3$ & $0.83 \pm 0.11^{*}$ \\
J1837$-$069$^{[8]}$ & J1838$-$0655 &  & 36.74 & 22.7 & 6.60 & $17 \pm 3$ & $2.54 \pm 0.04$ & $41 \pm 4$ & $204 \pm 8$ \\
J1418$-$609 & J1418$-$6058 & Kookaburra (Rabbit)$^{[5]}$ & 36.69 & 10.3 & 5.00 & $7.3 \pm 1.5$ & $2.26 \pm 0.05$ & $9.4 \pm 0.9$ & $31 \pm 3$ \\
J1356$-$645$^{[9]}$ & J1357$-$6429 &  & 36.49 & 7.31 & 2.50 & $5.5 \pm 1.4$ & $2.20 \pm 0.08$ & $10.1 \pm 0.9$ & $14.7 \pm 1.4$ \\
J1825$-$137$^{[10]}$ & B1823$-$13 &  & 36.45 & 21.4 & 3.93 & $33 \pm 6$ & $2.38 \pm 0.03$ & $32 \pm 2$ & $116 \pm 4$ \\
J1119$-$614 & J1119$-$6127 & G292.2$-$0.5$^{[11]}$ & 36.36 & 1.61 & 8.40 & $<11$ & $2.64 \pm 0.12$ & $14 \pm 2$ & $23 \pm 4$ \\
J1303$-$631$^{[12]}$ & J1301$-$6305 &  & 36.23 & 11.0 & 6.65 & $20.5 \pm 1.8$ & $2.33 \pm 0.02$ & $20.6 \pm 1.7$ & $96 \pm 5$ \\
\hline
\end{tabular}
\tablefoot{The sources are sorted by decreasing $\edot$. $\lg\edot$ stands for
$\log_{10}(\dot{E}/\mathrm{erg\,s^{-1}})$, $\age$ is the pulsar characteristic
age, $d$ is the pulsar distance, $R\tin{PWN}$ is the 1-sigma Gaussian
extension and $\lumi$ is the TeV luminosity. The pulsar
distances are printed uniformly here, but their uncertainties might often be larger
or not available; see ATNF
Catalogue references (\url{http://www.atnf.csiro.au/people/pulsar/psrcat/})
for detailed information. The limits are $2$-sigma limits (see
\sref{subsec:hgps_data_extraction}).\newline
$^{*}$The luminosity of Vela X
is calculated as described in \sref{subsec:hgps_data_extraction}.}
\tablebib{Previous publications on these sources: $[1]$ \citet{funk_1813}; $[2]$ \citet{hess_kes75_g21}; $[3]$ \citet{hess_msh}; $[4]$ \citet{veritas_g54}; $[5]$ \citet{kookaburra}; $[6]$ \citet{igr_terrier}; $[7]$ \citet{hess_velax_discovery}; $[8]$ \citet{psr_1837}; $[9]$ \citet{renaud_newpwne_2008}; $[10]$ \citet{hess_j1825_detection}; $[11]$ \citet{fermi_pwne}; $[12]$ \citet{hess_j1303_identification}.}
\end{table*}

%% file: table1d2.tex
\begin{table*}
\centering
\tiny
\caption{\label{tab:rePORGd2}Pulsar wind nebulae outside the HGPS catalogue.}
\begin{tabular}{lllllllll}
\hline\hline
Canonical name & ATNF name & $\lg\edot$ & $\tau_\mathrm{c}$ & $d$ & PSR offset & $\Gamma$ & $R_{\mathrm{PWN}}$ & $L_{\mathrm{1-10\,TeV}}$ \\
 &  &  & (kyr) & (kpc) & (pc) &  & (pc) & ($10^{33}\,\mathrm{erg\,s^{-1}}$)   \\
N157B$^{[1]}$ & J0537$-$6910 & 38.69 & 4.93 & 53.7 & $<22$ & $2.80 \pm 0.10$ & $<94$ & $760 \pm 80$    \\
Crab Nebula$^{[2]}$ & B0531+21 & 38.65 & 1.26 & 2.00 & $<0.8$ & $2.63 \pm 0.02$ & $<3$ & $32.1 \pm 0.7$    \\
G0.9+0.1$^{[3]}$ & J1747$-$2809 & 37.63 & 5.31 & 13.3 & $<3$ & $2.40 \pm 0.11$ & $<7$ & $46 \pm 7$    \\
3C58$^{[4]}$ & J0205+6449 & 37.43 & 5.37 & 2.00 & $<2$ & $2.4 \pm 0.2$ & $<5$ & $0.23 \pm 0.06$    \\
CTA 1$^{[5]}$ & J0007+7303 & 35.65 & 13.9 & 1.40 & $<4$ & $2.2 \pm 0.2$ & $6.6 \pm 0.5$ & $0.71 \pm 0.10$    \\
\hline
\end{tabular}
\tablefoot{See 	\tref{tab:rePAPd2} for the explanation of the columns.
G0.9+0.1 is listed in the catalogue, but not treated in the HGPS analysis
pipeline, so we treat it as an HGPS-external result. Offset limits were calculated as for the HGPS
(see \sref{subsec:hgps_data_extraction}). In the case of N157B and 3C58,
$2\sigma\tin{PSF}$ was used as conservative extension limit since no value is
given in the respective papers.}
\tablebib{$[1]$ \citet{hess_n157b, hess_lmc_science}; $[2]$ extension limit: \citet{hegra_crab}, flux: \citet{hess_crab}; $[3]$ \citet{g09}; $[4]$ \citet{3c58_magic}; $[5]$ \citet{veritas_cta1}.}
\end{table*}

%% file: table3d2.tex
\begin{table*}
\centering
\tiny
\caption{\label{tab:CAPd2}Candidate pulsar wind nebulae from the pre-selection.}
\begin{tabular}{lllllllllp{0.001cm}p{0.001cm}p{0.001cm}p{0.001cm}}
\hline\hline
HGPS name & ATNF name & $\lg\edot$ & $\tau_\mathrm{c}$ & $d$ & PSR offset & $\Gamma$ & $R_{\mathrm{PWN}}$ & $L_{\mathrm{1-10\,TeV}}$ & \multicolumn{4}{c}{Rating} \\
 &  &  & (kyr) & (kpc) & (pc) &  & (pc) & ($10^{33}\,\mathrm{erg\,s^{-1}}$) &  1 & 2 & 3 & 4\\
J1616$-$508 (1) & J1617$-$5055 & 37.20 & 8.13 & 6.82 & $<26$ & $2.34 \pm 0.06$ & $28 \pm 4$ & $162 \pm 9$& \ding{72} & \ding{72} & \ding{72} & \ding{72}  \\
J1023$-$575 & J1023$-$5746 & 37.04 & 4.60 & 8.00 & $<9$ & $2.36 \pm 0.05$ & $23.2 \pm 1.2$ & $67 \pm 5$& \ding{72} & \ding{72} & \ding{72} & \ding{72}  \\
J1809$-$193 (1) & J1811$-$1925 & 36.81 & 23.3 & 5.00 & $29 \pm 7$ & $2.38 \pm 0.07$ & $35 \pm 4$ & $53 \pm 3$& \ding{72} & \ding{72} & \ding{72} & $\lightning$  \\
J1857+026 & J1856+0245 & 36.66 & 20.6 & 9.01 & $21 \pm 6$ & $2.57 \pm 0.06$ & $41 \pm 9$ & $118 \pm 13$& \ding{72} & \ding{72} & \ding{72} & \ding{72}  \\
J1640$-$465 & J1640$-$4631 (1) & 36.64 & 3.35 & 12.8 & $<20$ & $2.55 \pm 0.04$ & $25 \pm 8$ & $210 \pm 12$& \ding{72} & \ding{72} & \ding{72} & \ding{72}  \\
J1641$-$462 & J1640$-$4631 (2) & 36.64 & 3.35 & 12.8 & $50 \pm 5$ & $2.50 \pm 0.11$ & $<14$ & $17 \pm 4$& $\lightning$ & $\star$ & \ding{72} & $\star$  \\
J1708$-$443 & B1706$-$44 & 36.53 & 17.5 & 2.60 & $17 \pm 3$ & $2.17 \pm 0.08$ & $12.7 \pm 1.4$ & $6.6 \pm 0.9$& \ding{72} & \ding{72} & \ding{72} & \ding{72}  \\
J1908+063 & J1907+0602 & 36.45 & 19.5 & 3.21 & $21 \pm 3$ & $2.26 \pm 0.06$ & $27.2 \pm 1.5$ & $28 \pm 2$& \ding{72} & \ding{72} & \ding{72} & \ding{72}  \\
J1018$-$589A & J1016$-$5857 (1) & 36.41 & 21.0 & 8.00 & $47.5 \pm 1.6$ & $2.24 \pm 0.13$ & $<4$ & $8.1 \pm 1.4$& $\lightning$ & $\star$ & \ding{72} & $\star$  \\
J1018$-$589B & J1016$-$5857 (2) & 36.41 & 21.0 & 8.00 & $25 \pm 7$ & $2.20 \pm 0.09$ & $21 \pm 4$ & $23 \pm 5$& \ding{72} & \ding{72} & \ding{72} & \ding{72}  \\
J1804$-$216 & B1800$-$21 & 36.34 & 15.8 & 4.40 & $18 \pm 5$ & $2.69 \pm 0.04$ & $19 \pm 3$ & $42.5 \pm 2.0$& \ding{72} & \ding{72} & \ding{72} & \ding{72}  \\
J1809$-$193 (2) & J1809$-$1917 & 36.26 & 51.3 & 3.55 & $<17$ & $2.38 \pm 0.07$ & $25 \pm 3$ & $26.9 \pm 1.5$& \ding{72} & \ding{72} & \ding{72} & \ding{72}  \\
J1616$-$508 (2) & B1610$-$50 & 36.20 & 7.42 & 7.94 & $60 \pm 7$ & $2.34 \pm 0.06$ & $32 \pm 5$ & $220 \pm 12$& $\lightning$ & \ding{72} & \ding{72} & \ding{72}  \\
J1718$-$385 & J1718$-$3825 & 36.11 & 89.5 & 3.60 & $5.4 \pm 1.6$ & $1.77 \pm 0.06$ & $7.2 \pm 0.9$ & $4.6 \pm 0.8$& \ding{72} & \ding{72} & \ding{72} & \ding{72}  \\
J1026$-$582 & J1028$-$5819 & 35.92 & 90.0 & 2.33 & $9 \pm 2$ & $1.81 \pm 0.10$ & $5.3 \pm 1.6$ & $1.7 \pm 0.5$& $\lightning$ & \ding{72} & \ding{72} & \ding{72}  \\
J1832$-$085 & B1830$-$08 (1) & 35.76 & 147 & 4.50 & $23.3 \pm 1.5$ & $2.38 \pm 0.14$ & $<4$ & $1.7 \pm 0.4$& $\lightning$ & $\lightning$ & \ding{72} & $\star$  \\
J1834$-$087 & B1830$-$08 (2) & 35.76 & 147 & 4.50 & $32.3 \pm 1.9$ & $2.61 \pm 0.07$ & $17 \pm 3$ & $25.8 \pm 2.0$& $\lightning$ & \ding{72} & \ding{72} & $\lightning$  \\
J1858+020 & J1857+0143 & 35.65 & 71.0 & 5.75 & $38 \pm 3$ & $2.39 \pm 0.12$ & $7.9 \pm 1.6$ & $7.1 \pm 1.5$& $\lightning$ & \ding{72} & \ding{72} & $\lightning$  \\
J1745$-$303 & B1742$-$30 (1) & 33.93 & 546 & 0.200 & $1.42 \pm 0.15$ & $2.57 \pm 0.06$ & $0.62 \pm 0.07$ & $0.014 \pm 0.003$& $\lightning$ & $\lightning$ & \ding{72} & $\lightning$  \\
J1746$-$308 & B1742$-$30 (2) & 33.93 & 546 & 0.200 & $<1.1$ & $3.3 \pm 0.2$ & $0.56 \pm 0.12$ & $0.009 \pm 0.003$& $\star$ & $\lightning$ & \ding{72} & $\lightning$  \\
\hline
\end{tabular}
\tablefoot{See \tref{tab:rePAPd2} for the explanation of the columns. In the
rating columns (1: PSR containment, 2: extension, 3: luminosity, 4: surface
brightness, see \sref{sec:candidate_rating}), a big star \ding{72} denotes a
quantity that fulfills its requirement, a small star $\star$
denotes a compatible limit, a lightning symbol $\lightning$ denotes a limit or
measurement in conflict with the requirement (see \sref{sec:candidate_rating}). Numbers in brackets indicate double associations.}
\end{table*}

%% file: table4d2.tex
\begin{table*}
\centering
\tiny
\caption{\label{tab:withLIMd2}Flux and luminosity upper limits (95\%~CL) for regions around pulsars without detected PWN.}
\begin{tabular}{lllllllll}
\hline\hline
ATNF name & $\lg\edot$ & $\tau_\mathrm{c}$ & $d$ & $\theta_{\mathrm{pred}}$ & $\theta_{\mathrm{int}}$ & Significance & $F_{\mathrm{>1\,TeV}}$ & $L_{\mathrm{1-10\,TeV}}$ \\
 &  & (kyr) & (kpc) & (deg) & (deg) & ($\sigma$) & ($10^{-12}\,$cm$^{-2}\,$s$^{-1}$) & ($10^{33}\,\mathrm{erg\,s^{-1}}$) \\
J1400$-$6325 & 37.71 & 12.7 & 7.00 & 0.150 & 0.2 & 1.4 & $<0.41$ & $<8.3$ \\
J1124$-$5916 & 37.08 & 2.85 & 5.00 & 0.137 & 0.2 & 1.0 & $<0.27$ & $<2.8$ \\
J1410$-$6132 & 37.00 & 24.8 & 15.6 & 0.127 & 0.2 & 2.8 & $<0.53$ & $<54$ \\
J1935+2025 & 36.67 & 20.9 & 6.21 & 0.29 & 0.4 & 1.9 & $<0.88$ & $<14$ \\
J1112$-$6103 & 36.65 & 32.7 & 12.2 & 0.21 & 0.4 & 3.7 & $<1.0$ & $<62$ \\
J1801$-$2451 & 36.41 & 15.5 & 5.22 & 0.30 & 0.4 & 1.1 & $<0.56$ & $<6.3$ \\
J1837$-$0604 & 36.30 & 33.8 & 6.41 & 0.42 & 0.4 & 9.5 & $<2.1$ & $<36$ \\
J1341$-$6220 & 36.15 & 12.1 & 11.1 & 0.129 & 0.2 & 2.6 & $<0.46$ & $<24$ \\
J1055$-$6028 & 36.08 & 53.5 & 15.5 & 0.25 & 0.4 & 1.1 & $<0.70$ & $<70$ \\
J1934+2352 & 35.96 & 21.6 & 11.6 & 0.175 & 0.2 & 1.6 & $<1.1$ & $<64$ \\
J1932+2220 & 35.88 & 39.8 & 10.9 & 0.29 & 0.4 & -0.9 & $<0.55$ & $<27$ \\
J1702$-$4310 & 35.80 & 17.0 & 5.14 & 0.35 & 0.4 & 0.9 & $<0.59$ & $<6.5$ \\
J1413$-$6141 & 35.75 & 13.6 & 10.1 & 0.161 & 0.2 & 2.8 & $<0.54$ & $<23$ \\
J1909+0749 & 35.65 & 24.7 & 9.48 & 0.24 & 0.4 & 0.9 & $<0.41$ & $<15$ \\
J1815$-$1738 & 35.59 & 40.4 & 8.78 & 0.37 & 0.4 & 8.9 & $<2.1$ & $<68$ \\
J1646$-$4346 & 35.56 & 32.5 & 5.79 & 0.48 & 0.4 & -2.0 & $<0.27$ & $<3.8$ \\
J1850$-$0026 & 35.52 & 67.5 & 11.1 & 0.44 & 0.4 & 3.7 & $<0.91$ & $<46$ \\
J1907+0918 & 35.51 & 38.0 & 7.79 & 0.40 & 0.4 & 2.7 & $<0.61$ & $<15$ \\
J1406$-$6121 & 35.34 & 61.7 & 8.15 & 0.56 & 0.4 & 4.4 & $<3.3$ & $<91$ \\
J1412$-$6145 & 35.08 & 50.6 & 7.82 & 0.51 & 0.4 & 5.0 & $<3.0$ & $<75$ \\
J1550$-$5418 & 35.00 & 1.41 & 4.00 & 0.29 & 0.4 & 1.0 & $<0.47$ & $<3.1$ \\
J1841$-$0524 & 35.00 & 30.2 & 5.34 & 0.53 & 0.4 & 20.9 & $<7.3$ & $<86$ \\
\hline
\end{tabular}
\tablefoot{In addition to the table variables explained in \tref{tab:rePAPd2}, $\theta_{\mathrm{pred}}$ is the predicted PWN extension (including offset), $\theta_{\mathrm{int}}$ is the correlation radius of the map where the limit is taken from, and $F_{\mathrm{>1\,TeV}}$ is the actual flux limit (see \sref{subsec:hgps_data_extraction} for details). In the cases of high significance, the pulsar coincides with a TeV source that is not considered the PWN.}
\end{table*}